\RequirePackage{ifpdf}
\ifpdf 
\documentclass[pdftex]{sigma}
\else
\documentclass{sigma}
\fi

\usepackage{bbm}

\newtheorem{summary}{Summary}
\numberwithin{equation}{section}

\newcommand{\cA}{\mathcal{A}}
\newcommand{\cB}{\mathcal{B}}
\newcommand{\cC}{\mathcal{C}}
\newcommand{\cF}{\mathcal{F}}
\newcommand{\cH}{\mathcal{H}}
\newcommand{\cM}{\mathcal{M}}
\newcommand{\cP}{\mathcal{P}}
\newcommand{\cS}{\mathcal{S}}
\newcommand{\cT}{\mathcal{T}}
\newcommand{\cW}{\mathcal{W}}
\newcommand{\dC}{{\mathbbm{C}}}
\newcommand{\dF}{{\mathbbm{F}}}
\newcommand{\dR}{{\mathbbm{R}}}
\newcommand{\dM}{{\mathbbm{M}}}
\newcommand{\dN}{{\mathbbm{N}}}
\newcommand{\1}{{\mathbbm{1}}}
\newcommand{\ex}[1]{\operatorname{exp}\{ #1 \}}
\newcommand{\wh}[1]{{\widehat{#1}}}
\newcommand{\ovl}[1]{\overline{#1}}
\newcommand{\betr}[1]{\left| #1 \right|}
\newcommand{\norm}[1]{\left\| #1 \right\|}
\newcommand{\dual}[2]{\langle #1;#2\rangle}
\newcommand{\skal}[2]{(#1|#2)}
\newcommand{\klr}[1]{\left( #1 \right)}

\begin{document}
\allowdisplaybreaks

\renewcommand{\PaperNumber}{047}

\FirstPageHeading

\renewcommand{\thefootnote}{$\star$}

\ShortArticleName{Field-Theoretic Weyl Deformation Quantization of
Enlarged Poisson Algebras}

\ArticleName{Field-Theoretic Weyl Deformation Quantization\\
                                        of Enlarged Poisson Algebras\footnote{This paper is a
contribution to the Special Issue on Deformation Quantization. The
full collection is available at
\href{http://www.emis.de/journals/SIGMA/Deformation_Quantization.html}{http://www.emis.de/journals/SIGMA/Deformation\_{}Quantization.html}}}

\Author{Reinhard HONEGGER, Alfred RIECKERS and Lothar SCHLAFER}

\AuthorNameForHeading{R. Honegger, A. Rieckers and L. Schlafer}

\Address{Institut f\"ur Theoretische Physik, Universit\"at T\"ubingen,\\
Auf der Morgenstelle 14, D--72076 T\"ubingen, Germany}
\Email{\href{mailto:reinhard.honegger@uni-tuebingen.de}{reinhard.honegger@uni-tuebingen.de},
        \href{mailto:alfred.rieckers@uni-tuebingen.de}{alfred.rieckers@uni-tuebingen.de},\\
\hspace*{14mm}\href{mailto:lothar.schlafer@uni-tuebingen.de}{lothar.schlafer@uni-tuebingen.de}}

\ArticleDates{Received December 20, 2007, in f\/inal form May 06,
2008; Published online May 29, 2008}

\Abstract{$C^*$-algebraic Weyl quantization is extended by
allowing also degenerate pre-symplectic forms for the Weyl
relations with inf\/initely many degrees of freedom, and by
starting out from enlarged classical Poisson algebras. A powerful
tool is found in the construction of Poisson algebras and
non-commutative twisted Banach-$*$-algebras on the stage of
measures on the not locally compact test function space. Already
within this frame strict deformation quantization is obtained, but
in terms of Banach-$*$-algebras instead of $C^*$-algebras. Fourier
transformation and representation theory of the measure
Banach-$*$-algebras are combined with the theory of continuous
projective group representations to arrive at the genuine
$C^*$-algebraic strict deformation quantization in the sense of
Rief\/fel and Landsman. Weyl quantization is recognized to depend
in the f\/irst step functorially on the (in general) inf\/inite
dimensional, pre-symplectic test function space; but in the second
step one has to select a family of representations, indexed by the
deformation parameter $\hbar$. The latter ambiguity is in the
present investigation connected with the choice of a folium of
states, a structure, which does not necessarily require a Hilbert
space representation.}


\Classification{46L65; 47L90; 81R15}

\section{Introduction}
\label{WQ-s1}

In the present investigation we elaborate a kind of f\/ield
theoretic Weyl quantization, which generalizes the more popular
strategies in a threefold manner: We admit inf\/initely many
degrees of freedom (by using inf\/inite dimensional test function
spaces $E$), we formulate the Weyl relations in terms of a
possibly degenerate pre-symplectic form $\sigma$, and we start out
for quantization from rather large classical Poisson algebras. An
appropriate choice of the latter is basic for a rigorous
quantization scheme, for which we join the ideas of the so-called
\emph{strict deformation quantization}. Because of the mentioned
generalizations, we have however need for some technical
pecularities, some of which we want to motivate in the f\/inite
dimensional case.

\subsection[Integration extension in  finite dimensions]{Integration extension in  f\/inite dimensions}
\label{WQ-s1mot}

Let us outline our approach to Weyl quantization at hand of the
test function space $E:=\dR^d\times\dR^d$, $d\in\dN$, which
contains the coordinate tuples $f=(u,v)$ with $u,v\in\dR^d$. The
symplectic form $\sigma$ is given by
\begin{gather*}
\sigma((u,v),(u',v')):= u\cdot v' - v\cdot u' , \qquad\forall \,
u,v,u',v'\in \dR^d.
\end{gather*}
The dual space $E'=\dR^d\times\dR^d$ serves as phase space for a
system with $d$ degrees of freedom. The duality relation
\begin{gather*}
F(f)=u\cdot q + v\cdot p, \qquad\forall \, F=[q,p]\in E',\quad
\forall \, f=(u,v)\in E,
\end{gather*}
imitates the smearing of the f\/ield in f\/inite dimensions. (For
smeared f\/ields cf. e.g.
\cite{Emch72,BratteliRobinson2,ReedSimon2}). A~classical Weyl
element, a ``Weyl function'', is given by the periodic phase space
function
\begin{gather}
\label{WQB0:eq:Wc(f)} W_c(f):E'\longrightarrow\dC, \qquad
F=[q,p]\longmapsto W_c(f)[F]=\ex{iF(f)}=\ex{i(u\cdot q+v\cdot
p)},\!\!\!
\end{gather}
for each $f=(u,v)\in E$. According to Weyl  \cite{Weyl28,Weyl31}
one  does not quantize the unbounded coordinate functions but the
Weyl functions and uses the Schr\"{o}dinger representation. So we
do not adress here the question of further representations of the
canonical commutation relations (CCR), inequivalent to the
Schr\"{o}dinger representation (e.g.~\cite{Putnam67,GaPa89,Hon93a}). We only mention von Neumann's
uniqueness result~\cite{vonNeumann31} that the Schr\"{o}dinger
realization of the CCR is the unique (up to unitary equivalence)
irreducible representation, for which the CCR arise from the Weyl
relations.

Denoting in $\mathrm{L}^2(\dR^d)$ the quantized momenta by
$P_k=-i\hbar\frac{\partial}{\partial x_k}$ (depending on
$\hbar\neq0$) and the position operators (multiplication by the
coordinate functions $x_l$) by $Q_l$, one arrives at the Weyl
operators in the Schr\"{o}dinger representation
\begin{gather*}
W_S^\hbar(u,v) :=\ex{i(u\cdot Q+v\cdot P)} =\ex{\tfrac{i}{2}\hbar
u\cdot v}\ex{iu\cdot Q}\ex{iv\cdot P}
\end{gather*}
with $(u,v)=f\in E$. These unitary operators satisfy the
\emph{Weyl relations}
\begin{gather}
\label{WQB0:eq:Weyl-relations} W_S^\hbar(f)W^\hbar_S(g)=
\ex{-\textstyle\frac{i}{2}\hbar\sigma(f,g)} W_S^\hbar(f+g),\qquad
W_S^\hbar(f)^*=W_S^\hbar(-f),\quad \forall\, f,g\in E,
\end{gather}
and act as $(W_S^\hbar(u,v)\psi)(x)= \ex{\frac{i}{2}\hbar u\cdot
v} \ex{iu\cdot x}\psi(x+\hbar v)$, $\forall\, x\in\dR^d$, on
$\psi\in\mathrm{L}^2(\dR^d)$. The smallest $C^*$-algebra
containing all Weyl operators $W_S^\hbar(f)$, $f\in E$, is
$*$-isomorphic to the abstract $C^*$-Weyl algebra
$\cW(E,\hbar\sigma)$ over the symplectic space $(E,\hbar\sigma)$.

The concept of \emph{Weyl quantization} includes a special
operator ordering, implicitely given by the linear, $*$-preserving
quantization map
\begin{gather}
\label{WQB0:eq:QhbarW} Q_\hbar^S\left( \sum\limits_{k=1}^n z_k
W_c(f_k)\right):= \sum\limits_{k=1}^n z_k
W^\hbar_S(f_k),
\end{gather}
with $n\in\dN$, $z_k\in\dC$, and with dif\/ferent $f_k\in E$.

The $C^*$-algebraic completion of this Weyl quantization covers
also certain inf\/inite sums, since
$\sum\limits_{k=1}^\infty\betr{z_k}<\infty$ implies the
convergence of $\sum\limits_{k=1}^\infty z_k W_c(f_k)$ and
$\sum\limits_{k=1}^\infty z_k W^\hbar_S(f_k)$ in the $C^*$-norms
(classically the supremum norm, and quantum mechanically the
operator norm). (There exist of course $C^*$-algebraic
quantization prescriptions dif\/ferent from the indicated
symmetric Weyl ordering cf.~\cite{Rieffel94a,BinzHonRie01a}, and
references therein.)

Because $E\ni f\mapsto W^\hbar_S(f)$ is discontinuous with respect
to the operator norm, an extension of the quantization map
$Q_\hbar^S$ to more general functions than the \emph{almost
periodic} functions of the type $\sum\limits_{k=1}^\infty z_k
W_c(f_k)$ is impossible in terms of the norm. However, we are
dealing with a~representation, the Schr\"{o}dinger representation,
of the abstract Weyl algebra $\cW(E,\hbar\sigma)$, and so there
exist weaker topologies than the norm. And indeed, $E\ni f\mapsto
W^\hbar_S(f)$ is continuous with respect to all weak operator
topologies (weak, $\sigma$-weak, strong, \ldots). This allows for
extending the quantization map $Q_\hbar^S$ to suitable
``continuous linear combinations'' of the periodic functions
$W_c(f)$: Let $\cM(E)$ be the f\/inite, complex, regular Borel
measures on $E$. The Fourier transform $\wh{\mu}[F]:=\int_E
d\mu(f) \ex{iF(f)}$, $F\in E'$, of the measure $\mu\in \cM(E)$ is
a bounded, continuous function $\wh{\mu}:E'\rightarrow\dC$, which
may be considered as the point-wise ``continuous linear
combination''
\begin{gather}
\label{WQB0:eq:mu-LC} \wh{\mu}=\int_E d\mu(f) \;W_c(f)
\end{gather}
of the $W_c(f)$, $f\in E$. So the desired extension of $Q_\hbar^S$
from the original linear combinations in
equation~\eqref{WQB0:eq:QhbarW} to the ``continuous'' ones
from~\eqref{WQB0:eq:mu-LC} is given by
\begin{gather}
\label{WQB0:eq:QhbarW-cont} Q_\hbar^S(\wh{\mu}):=\int_E
d\mu(f)\;Q_\hbar^S(W_c(f)) =\int_E d\mu(f)\; W^\hbar_S(f).
\end{gather}
The integral exists -- as the limit of the Lebesgue partial sums
-- with respect to the weak operator topologies, constituting a
bounded operator in $\mathrm{L}^2(\dR^d)$. Especially, the above
inf\/inite series
\begin{gather}
\label{WQB0:eq:QhbarW2} Q_\hbar^S(\wh{\mu})=
Q_\hbar^S\left(\sum\limits_{k=1}^\infty z_k W_c(f_k)\right)=
\sum\limits_{k=1}^\infty z_k W^\hbar_S(f_k)
\end{gather}
just belong to the discrete measures $\mu=\sum\limits_{k=1}^\infty
z_k\delta(f_k) \in \cM(E)_d$ on $E$ with f\/inite total weights
$\sum\limits_{k=1}^\infty\betr{z_k}<\infty$ (where $\delta(f)$
denotes the point measure at $f\in E$). Therefore, the map in
\eqref{WQB0:eq:QhbarW-cont} provides, already for f\/inite
dimensions, a considerable extension of the $C^*$-algebraic Weyl
quantization by means of weak integration in a representation
space. This extension is comparable to the transition from Fourier
series to Fourier integrals, where the Fourier transformations of
f\/inite measures are continuous bounded functions. (Certain
classes are described at the end of the paper.)

The connection to deformation quantization is gained by observing
that $Q_\hbar^S$ acts injectively on the Fourier transformed
measures. Hence one may def\/ine a~non-commutative product for
a~certain class of phase space functions by setting
\begin{gather}
\label{WQB0:eq:Moyal-prod} \wh{\mu}\cdot_{\!\hbar}\wh{\nu}:=
{Q_\hbar^S}^{-1}(Q_\hbar^S(\wh{\mu})Q_\hbar^S(\wh{\nu})),
\end{gather}
where $Q_\hbar^S(\wh{\mu})Q_\hbar^S(\wh{\nu})$ means the common
operator product.

The essence of quantization may be considered as deforming the
commutative, pointwise product for phase space functions into a
non-commutative, $\hbar$-dependent product like
$\wh{\mu}\cdot_{\!\hbar}\wh{\nu}$. (Explicit formulations are
given in the form of so-called  Moyal products.) This point of
view has certainly provoked many fresh ideas on the quantization
problem. In this connection let us add as a side remark that,
keeping the phase space functions as observables also in the
quantized theory, speaks a bit against the philosophy that the
canonical observables may not ``exist'' before being measured (and
supports perhaps Einstein's side in the Einstein--Bohr debate).

In spite of the merits of the quantized phase space formalism we
concentrate in the present investigation f\/irst on ``quantizing''
measures on the test function space. Before Fourier
transformation, the deformed product~\eqref{WQB0:eq:Moyal-prod} is
just the twisted convolution
\begin{gather*}
\mu\star_{\!\hbar}\nu=\dF^{-1}(\wh{\mu}\cdot_{\!\hbar}\wh{\nu}),
\qquad \forall\,\mu,\nu\in \cM(E),
\end{gather*}
on the measure space $\cM(E)$ with respect to the multiplier
$\ex{-\textstyle\frac{i}{2}\hbar\sigma(f,g)}$ occurring in the
Weyl relations~\eqref{WQB0:eq:Weyl-relations}. (As forerunners may
be mentioned \cite{GarciaBondia88} and even \cite{vonNeumann31}.)

Equipped with a suitable involution and the twisted convolution
product, $(\cM(E),\star_{\!\hbar})$ constitutes a
Banach-$*$-algebra with respect to the total variation norm. This
leads us to a dif\/ferent reading of formula
\eqref{WQB0:eq:QhbarW-cont}: On the right hand side we see the
Weyl operators $W^\hbar_S(f)$ in the Schr{\"o}dinger
representation, which are integrated over the measure $\mu$  to an
operator, which in general is no longer contained in the
represented $C^*$-Weyl algebra. For varying $\mu  \in \cM(E)$ the
resulting operators realize $*$-homomorphically the $*$-algebraic
operations of $(\cM(E),\star_{\!\hbar})$ in terms of the hermitian
conjugation and operator product. Thus the mapping $\mu \mapsto
Q_\hbar^S(\wh{\mu})$ provides us with a so-called regular
representation of the Banach-$*$-algebra
$(\cM(E),\star_{\!\hbar})$.

For such structures there exist much mathematical literature
related to locally compact groups, to which we have to refer
frequently.
 Still another interpretation
of the quantization formula \eqref{WQB0:eq:QhbarW-cont} results
from this f\/ield of mathematics. In virtue of the Weyl
relations~\eqref{WQB0:eq:Weyl-relations}, the mapping $E\ni
f\mapsto W^\hbar_S(f)$ realizes a $\sigma$-strongly continuous
projective unitary representation of the locally compact vector
group $E$. Integration over the group leads to the realm of
twisted group algebras.

In any case, we emphasize the use of the Banach-$*$-algebras
$(\cM(E),\star_{\!\hbar})$. We do this in modif\/ied form also in
the inf\/inite dimensional case, as is sketched in the next
Subsection. Since the extension of the (represented) $C^*$-Weyl
algebra is performed by weak integration we term it
\emph{integration extension}. Related with this are the
\emph{integration type} representations of
$(\cM(E),\star_{\!\hbar})$.

\subsection[Overview on the extension of field theoretic Weyl quantization]{Overview on the extension of f\/ield theoretic Weyl quantization}
\label{WQ-s1ov}

Extending of the ideas of the previous Subsection to inf\/inite
dimensional test function spaces $E$, where already usual
$C^*$-algebraic Weyl quantization over such $E$ of\/fers certain
subtleties (e.g.~\cite{Kastler65,BinzHonRie03b}), causes
additional mathematical dif\/f\/iculties, the treatment of which
forms the subject of the present investigation. Let us at this
place give only some motivation and an overview on the main steps
of reasoning.

Concerning the generalization of the symplectic form, there are,
in fact, many instances of f\/ield theories with pre-symplectic
test function spaces. In the formulation of quantum
electrodynamics (QED) for non-relativistic optical systems the
transversal parts of the canonical f\/ield variables have to be
separated out by means of a Helmholtz--Hodge decomposition (cf.
e.g. \cite{Loudon79,CohenTann89, Schwarz95}). Only these enter the
canonical formalism (in an inf\/inite cavity) and are quantized.
The corresponding Poisson tensor for the total f\/ield in the
Coulomb gauge is thus highly degenerate. In f\/ield theory for
high energy physics, there are degenerate symplectic forms, for
example, in the frame of conformal f\/ield theory (cf.
e.g.~\cite{BuchhMT88}). Quite generally, f\/ield theories with
intrinsic superselection rules require CCR resulting from
degenerate Poisson brackets.

Also in the frame of deformation quantization, stipulated by the
seminal work of \cite{BFFLS78}, diverse attempts were undertaken
to meet the challenges of quantum f\/ield theory (QFT)
(cf.~\cite{Dito92,Dito93,DFredenhagen01b,
SternheimerAlcade89,Fronsdal91}). The various so-called star
exponentials cover a wide range of bounded functions on the
f\/ield phase space.  Dif\/ferent kinds of ordering for unbounded
f\/ield polynomials occur in their series expansions. As pointed
out especially by Dito~\cite{Dito90,Dito92} the requirement that
the star exponential of the Hamiltonian be well def\/ined
restricts the class of suitable deformed products (star products)
to some extent, leaving still open a wide class of possible
deformations. Most of these formulations have not reached full
mathematical rigor, but they give inspiration to extend the
controlled formalism as much as possible and to systematize the
deformations of f\/ield expressions.

Our mentioned previous works on Weyl quantization over
inf\/initely many degrees of freedom
\cite{BinzHonRie03b,BinzHonRie01a} were based on a more
restrictive notion of deformation quantization, termed
\emph{strict} by Rief\/fel and Landsman
\cite{Rieffel93a,Rieffel94a,Rieffel98a,Landsman98a,
Landsman98book}, and outlined brief\/ly in
Subsection~\ref{WQ-s2.1}. Our mathematical techniques avoid,
however, phase space integrals of the quantized products and
commutators, basic for the f\/inite dimensional deformation
quantization (also used in the papers of Rief\/fel and Landsman),
since these are hard to generalize to inf\/initely many variables.

For def\/initeness consider the abstract Weyl relations
\begin{gather}
\label{WQB1:eq:Weyl-relations} W^\hbar(f)W^\hbar(g)=
\ex{-\textstyle\frac{i}{2}\hbar\sigma(f,g)} W^\hbar(f+g),\qquad
W^\hbar(f)^*=W^\hbar(-f),\quad \forall \, f,g\in E,
\end{gather}
where $E$ denotes the inf\/inite dimensional test function space,
equipped with the (real bilinear, antisymmetric) pre-symplectic
form $\sigma$. According to the deformation strategy, $\hbar$ is a
variable parameter, which we let range in~$\dR$. Generally we
assume $\sigma\neq0$, implying (partial) non-commutativity in the
Weyl relations~\eqref{WQB1:eq:Weyl-relations} for the quantum
regime~$\hbar\neq0$.

For each $\hbar\in\dR$ the (f\/inite) linear combinations of Weyl
elements $W^\hbar(f)$, $f\in E$, determine already the abstract
$C^*$-Weyl algebra $\cW(E, \hbar \sigma)$, also in the case of
degenerate $\sigma$, cf.~Subsection~\ref{WQ-s2.2}. This uniqueness
is achieved by demanding the abstract $C^*$-Weyl algebra to
include all intrinsic superselection observables. For mutually
dif\/ferent values of $\hbar$ the associated Weyl relations lead
to dif\/ferent Weyl algebras $\cW(E, \hbar \sigma)$, possessing
dif\/ferent norms. In Subsection~\ref{WQ-s4.3} we investigate
transformations between the $\cW(E, \hbar \sigma)$ with
dif\/ferent $\hbar$. As is recapitulated in
Subsection~\ref{WQ-s2.3}, the $C^*$-norms are so well calibrated
that the Weyl quantization, induced by the mappings
\begin{gather*}
Q_\hbar\left( \sum_{k=1}^\infty z_k W_c(f_k)\right):=
 \sum_{k=1}^\infty z_k W^\hbar(f_k)
\in \cW(E, \hbar \sigma),
\end{gather*}
satisf\/ies all axioms of a strict deformation quantization,
including Rief\/fel's condition. The geo\-metric features behind
the choice of the classical Poisson bracket are touched upon in
Subsection~\ref{WQ-s2.4}. As mentioned for f\/inite dimensions in
Subsection~\ref{WQ-s1mot}, the dual space $E_\tau'$, formed with
respect to some locally convex (vector space) topology $\tau$,
serves as the (f\/lat) phase space manifold of the classical
f\/ield theory. The classical Weyl elements $W^0(f)$, $f\in E$,
may be realized by the continuous, periodic functions $W_c(f)[F]:=
\ex{iF(f)}$ on $E_\tau'\ni F$. We emphasize however the algebraic
universality of this basic mechanical structure and of the whole
quantization method, involving the $\cW(E,\hbar\sigma)$, which
expresses a functorial dependence on~$(E, \sigma)$.

The f\/irst step for the extended theory in Section~\ref{WQ-s5} is
the introduction of the appropriate, universal measure space
$\cM(E)$ in terms of an inductive limit. The Banach-$*$-algebras
of measures $(\cM(E),\star_{\!\hbar})$, with the (deformed)
convolution products
(equation~\eqref{WQ:eq:9-5:deformed-convolution} below) and
endowed with the total variation norm, for each $\hbar\in\dR$, use
all of $\cM(E)$. Certain semi-norms compatible with $\sigma$ are
introduced for constructing the Poisson algebras of measures. The
existence of their moments with respect to a measure imitates
dif\/ferentiability assumptions on the phase space function
obtained from the measure by Fourier transformation.

Let us mention that in \cite{Kastler65} D.~Kastler already
introduced $(\cM(E),\star_{\!\hbar})$ for the construction of a
Boson f\/ield $C^*$-algebra in terms of a  $C^*$-norm completion,
which depends on a Schr\"{o}dinger-like representation. There this
construction has been used, however, for non-degenerate $\sigma$
(and f\/ixed $\hbar\neq0$), only.

For proving the axioms of strict deformation quantization in the
Banach-$*$-type version, the $\hbar$-independence of the variation
norm is a great help.

Section~\ref{WQ-s5xt} treats the Poisson algebras in the more
common phase space formulation. Here, but also in the previous
measure version, the observables of the classical and of the
quantized theory are the same mathematical quantities. Our remarks
in Subsection~\ref{WQ-s5.4} on the pre-symplectic geometry give a
glimpse on the dif\/f\/iculties one has in dealing with
Hamiltonian vector f\/ields in inf\/inite dimensions. The
introduction of restricted tangent and cotangent spaces touches
the problem of the foliation into symplectic leaves. Our
contribution is the smooth adjunction of such geometrical
structures to the quantized f\/ield theory.

From the f\/inite dimensional theory one knows how useful a
rigorous classical approximation to the quantum theory is, both
for technical and interpretational reasons.

The step to the $C^*$-version of strict f\/ield quantization
requires additional mathematical tools. In Section~\ref{WQ-s4}
certain representations of the $(\cM(E),\star_{\!\hbar})$, termed
\emph{integration type representations}, are investigated. They
have the form
\begin{gather}
\label{WQB1:eq:Pi-IntTyp} \Pi_\hbar(\mu)=\int_E d\mu(f)\,
\pi_\hbar(f), \qquad \forall\,\mu\in\cM(E),
\end{gather}
where $E\ni f\mapsto \pi_\hbar(f)$ is a $\sigma$-strongly
continuous, projective, unitary representation of the vector group
$E$ arising from the Weyl relations. The connection to a
representation of the original $C^*$-Weyl algebra
$\cW(E,\hbar\sigma)$, resp.~of $(\cM(E)_d,\star_{\!\hbar})$, is
given by
\begin{gather*}
\pi_\hbar(f)=\Pi_\hbar(W^\hbar(f))=\Pi_\hbar(\delta(f)),\qquad
\forall \, f\in E,
\end{gather*}
and the representation $\Pi_\hbar$ extends from
$(\cM(E)_d,\star_{\!\hbar})$ to all measures $\cM(E)$
via~\eqref{WQB1:eq:Pi-IntTyp}.

The consideration of representations is necessary, because the
$C^*$-norm of the enveloping $C^*$-algebra is gained by a supremum
over the representation norms (cf. Subsection~\ref{WQ-s2.1}).
Regularity of a representation of $\cW(E,\hbar\sigma)$, which
allows for the representation dependent Boson f\/ield operators by
dif\/ferentiating the represented Weyl elements, has to be
strengthened to our concept of $\tau$-continuity. The normal
states of the $\tau$-continuous representations constitute the
$\tau$-continuous folia. They play, quite generally, an important
role for Boson f\/ield theory e.g. for formulating the dynamics
and for a spatial decomposition theory of the non-separable
$C^*$-Weyl algebra $\cW(E,\hbar\sigma)$. Since they are not so
well known we treat them in some detail. In the present context
they are inserted into the integration type extensions
\eqref{WQB1:eq:Pi-IntTyp} of representations (which display not
all of the desired properties if only regular representations are
inserted).

In Section~\ref{WQ-s6} we f\/inally deal with the extended Weyl
quantization in the $C^*$-version. The quantization maps are based
on the integration representations
\begin{gather}
\label{WQB1:eq:Qhbar-Pi} Q_\hbar^\Pi(\wh{\mu}):=\Pi_\hbar(\mu),
\qquad \forall\,\mu\in\cM(E).
\end{gather}
The need of an intermediate step via Hilbert space representations
causes great dif\/f\/iculties when using a degenerate symplectic
form $\sigma$. Now the intrinsic superselection observables may be
partially lost in a representation, causing def\/icits in the
$C^*$-norms. To calibrate these losses we work with so-called
\emph{well-matched} families of ($\tau$-continuous and regular)
representations (indexed by $\hbar$). Under these assumptions,
partial results on the Rief\/fel condition are derived, whereas
the even more popular Dirac and von Neumann conditions provide no
dif\/f\/iculties.

Altogether, the classical correspondence limit is concisely worked
out for a rather comprehensive set of observables. Subclasses of
these are described at the end of Section~\ref{WQ-s6} in terms of
phase space functions. The other way round, our investigation
illustrates that deformation quantization in inf\/inite dimensions
requires adequate topological and measure theoretic notions. Some
are connected here with Hilbert space representations, which may
however be replaced by the choice of state folia, since the latter
serve equally well to def\/ine weak topologies (and integration
and dif\/ferentiation methods) in the observable algebras.

Physically the choice of a folium expresses classical,
macroscopic, i.e.\ collective, aspects of the quantum f\/ield
system. Disjoint folia describe dif\/ferent features of the
macroscopic preparation of the system (as e.g. dif\/ferent
reservoir couplings, like heat baths or weak links to condensed
particles). They lead to additional superselection sectors
(represented also by additional central observables in the weak
closures) to the intrinsic ones, originating from the algebraic
degeneracy of the commutation relations. (Charges are intrinsic,
thermodynamic variables parametrize folia.)

Due to the described extension of the algebraic structure in
comparison to the usual $C^*$-Weyl algebra of Boson f\/ields, one
may prof\/it now from a still wider class of f\/ield operator
realizations than in traditional algebraic QFT. This may be of use
for improving the more heuristic approaches to inf\/inite
dimensional deformation quantization, which employ directly the
f\/ield observables (in contradistinction to the bounded Weyl
elements). Since the unbounded f\/ield operators depend
\emph{essentially} on the representation, resp.~on the chosen weak
topology (e.g. in a GNS representation over a condensed state
arise additional classical f\/ield parts), the classif\/ication of
their deformed products has to take into account, beside -- and in
combination with -- operator ordering, the ef\/fects of weak
topologies.

\section{Preliminary notions and results}
\label{WQ-s2}

Let us f\/irst make some notational remarks. Generally all
occurring topologies are assumed to be Hausdorf\/f. If not
specif\/ied otherwise (bi-)linearity is understood over the
complex f\/ield $\dC$. The linear hull $\mathrm{LH}\{V\}$ denotes
the (f\/inite) complex linear combinations of the elements of the
set~$V$.

We deal exclusively with $*$-algebras $\cA$ (containing the
arbitrary elements $A$ and $B$) over the complex f\/ield $\dC$
with an associative, but possibly non-commutative product. The
$*$-operation is involutive ($(A^*)^*=A$), anti-linear, and
product anti-homomorphic ($(AB)^*=B^*A^*$). An algebra norm is a
vector space norm with $\norm{AB}\leq \norm{A}\norm{B}$. A Banach
algebra is an algebra, which is complete in an algebra norm. A
Banach-$*$-algebra is a Banach algebra with a $*$-operation, which
satisf\/ies $\norm{A^*}=\norm{A}$. A $C^*$-norm $\norm \cdot$ on a
$*$-algebra $\cA$ is an algebra norm satisfying the so-called
$C^*$-norm property $\norm{A^*A}=\norm{A}^2$. A $C^*$-algebra is a
Banach-$*$-algebra the norm of which is a $C^*$-norm.

Under a representation $(\Pi,\cH_\Pi)$ of a $*$-algebra $\cA$ we
understand a $*$-homo\-mor\-phism $\Pi$ from~$\cA$ into the
$C^*$-algebra of all \emph{bounded} operators on the complex
representation Hilbert space~$\cH_\Pi$. If $(\Pi,\cH_\Pi)$
represents a Banach-$*$-algebra, then $\norm{\Pi(A)} \leq
\norm{A}$. If $(\Pi,\cH_\Pi)$ is a faithful representation of a
$C^*$-algebra, then $\norm{\Pi(A)}=\norm{A}$. If $\cA$ is a
Banach-$*$-algebra one introduces a $C^*$-norm by setting
$\norm{A}_{C^*}:= \sup \{\norm{\Pi(A)} \}$, where the supremum
goes over all representations. The enveloping $C^*$-algebra
$\cC(\cA)$ of a Banach-$*$-algebra  $\cA$ is the completion of
$\cA$ in the mentioned $C^*$-norm.

The commutative $C^*$-algebra
$(\operatorname{C}_b(\mathsf{P}),\cdot_{\!0})$, consisting of the
bounded, continuous functions $A:\mathsf{P}\rightarrow\dC$,
$F\mapsto A[F]$ on the topological space $\mathsf{P}$ has the
sup-norm $\norm{\cdot}_0$ as its $C^*$-norm,
$\norm{A}_0=\sup\{\betr{A[F]}\mid F\in\mathsf{P}\}$, and is
equipped with the usual pointwise-def\/ined commutative
$*$-algebraic operations
\begin{gather}
\label{WQ:eq:9-5:*-algebraic-operations-F} (A \cdot_{\!0}
B)[F]:=A[F]B[F],\qquad A^*[F]:=\ovl{A[F]}, \qquad\forall \, F\in
\mathsf{P}.
\end{gather}

\subsection{The notion of strict deformation quantization}
\label{WQ-s2.1}

Bohr's correspondence principle has reached an especially concise
form in terms of strict deformation quantization. Let us specify
this notion.

A Poisson algebra $(\cP,\{\cdot,\cdot\})$ consists of a
commutative $*$-algebra $(\cP,\cdot_{\!0})$ (over $\dC$ by the
above notational remarks) equipped with a Poisson bracket
$\{\cdot,\cdot\}$. The latter is a bilinear mapping
$\{\cdot,\cdot\}:\cP\times\cP\rightarrow\cP$, which is
anticommutative $\{A,B\}=-\{B,A\}$, real $\{A,B\}^*=\{A^*,B^*\}$,
fulf\/ills the Jacobi identity
$\{A,\{B,C\}\}+\{B,\{C,A\}\}+\{C,\{A,B\}\}=0$, and the Leibniz
rule
$\{A,B\cdot_{\!0}C\}=\{A,B\}\cdot_{\!0}C+B\cdot_{\!0}\{A,C\}$.
$\cP$ is assumed to be $\norm{\cdot}_0$-dense in a commutative
$C^*$-algebra $\cA^0$ ($\norm{\cdot}_0$ denotes the $C^*$-norm on
$\cA^0$), where $\cA^0$ is interpreted as the algebra of
observables for the considered classical f\/ield theory.

In so-called ``algebraic quantum f\/ield theory'' a quantum
f\/ield system is characterized in terms of a non-commutative
$C^*$-algebra $\cA^\hbar$, equipped with the $C^*$-norm
$\norm{\cdot}_\hbar$, on which the $\hbar$-scaled commutator be
def\/ined by
\begin{gather}
\label{WQ:eq:9-5:hbar-scaled-commut}
\{A,B\}_\hbar:=\textstyle\frac{i}{\hbar}(AB-BA), \qquad\forall \,
A,B\in\cA^\hbar.
\end{gather}
The quantum system should be def\/ined for so many values of
$\hbar\neq0$ that the classical correspondence limit
$\hbar\rightarrow0$ may be investigated. A subset $J \subseteq
\dR$, containing $0$ as an accumulation point of
$J_0:=J\!\setminus\!\{0\}$ would be  suf\/f\/icient. But, for
simplicity, in the present work we choose mostly $J=\dR$.

For each $\hbar\in J_0$ let us be given a quantization map
$Q_\hbar:\cP\rightarrow\cA^\hbar$, which tells how the quantum
observables are related to given classical ones. $Q_\hbar$ is
supposed  linear and $*$-preserving, that is
$Q_\hbar(A^*)=Q_\hbar(A)^*$ for all $A\in\cP$. Clearly, $Q_\hbar$
cannot respect the commutative products of $\cP$.

\begin{definition}[Strict deformation quantization]
\label{WQ:defi:9-5:SQ} Let $J\subseteq\dR$ be as specif\/ied
above. A~\emph{strict quantization} $(Q_\hbar)_{\hbar\in J}$ of
the Poisson algebra $(\cP,\{\cdot,\cdot\})$ consists for each
value $\hbar\in J$ of a linear, $*$-preserving map
$Q_\hbar:\cP\rightarrow\cA^\hbar$ ($\cA^\hbar$ being a
$C^*$-algebra), such that $Q_0$ is the identical embedding and
such that for all $A,B\in\cP$ the following conditions are
fulf\/illed:
\begin{enumerate}\itemsep=0pt
\renewcommand{\labelenumi}{(\alph{enumi})}
\renewcommand{\theenumi}{(\alph{enumi})}
\item \label{WQ:defi:9-5:SQ-a} \textbf{[Dirac's condition]} The
$\hbar$-scaled commutator~\eqref{WQ:eq:9-5:hbar-scaled-commut}
approaches the Poisson bracket as $0\neq\hbar\rightarrow0$, that
is, $\lim\limits_{\hbar\rightarrow0} \norm{
\{Q_\hbar(A),Q_\hbar(B)\}_\hbar  -  Q_\hbar(\{A,B\})
                      }_\hbar   =   0$.
\item \label{WQ:defi:9-5:SQ-b} \textbf{[von Neumann's condition]}
As $\hbar\rightarrow0$ one has the asymptotic homomorphism
property $\lim\limits_{\hbar\rightarrow0}
\norm{Q_\hbar(A)Q_\hbar(B)-Q_\hbar(A\cdot_{\!0}B)}_\hbar=0$. \item
\label{WQ:defi:9-5:SQ-c} \textbf{[Rief\/fel's condition]} $J\ni
\hbar\mapsto\norm{Q_\hbar(A)}_\hbar$ is continuous.
\end{enumerate}

The strict quantization $(Q_\hbar)_{\hbar\in J}$ is called a
\emph{strict deformation quantization}, if each $Q_\hbar$ is
injective and if its image $Q_\hbar(\cP)$ is a sub-$*$-algebra of
$\cA^\hbar$.
\end{definition}

For a strict deformation quantization one may def\/ine on $\cP$
the deformed product
\begin{gather}
\label{WQ:eq:9-5:DeformedProduct} A\cdot_{\!\hbar}
B:=Q_\hbar^{-1}(Q_\hbar(A)Q_\hbar(B)), \qquad \forall \, A,B\in
\cP,
\end{gather}
ensuring $(\cP,\cdot_{\!\hbar})$ to be a non-commutative
$*$-algebra, $*$-isomorphic to $Q_\hbar(\cP)$.

In Subsection~\ref{WQ-s5.22xx} we deal with a Banach-$*$-algebra
version of strict deformation quantization, where the above
$C^*$-algebras $\cA^\hbar$ are replaced by Banach-$*$-algebras. We
distinguish the two notions of a strict deformation quantization
by calling them to be of $C^*$-type resp.~of Banach-$*$-type.

\subsection{Weyl algebra}
\label{WQ-s2.2}

Let us have  a \emph{pre-symplectic space} $(E,\sigma)$. Recall by
the way that $\sigma$ is symplectic, if its null space
\begin{gather}
\label{WQ:eq:9-5:nullspace-sigma} \ker_\sigma:=\{f\in E\mid
\sigma(f,g)=0\;\forall \, g\in E\}
\end{gather}
is trivial. The standard example of a symplectic space $E$ is a
complex pre-Hilbert space, regarded as a real vector space, where
$\sigma$ is the imaginary part of its complex inner product
$\skal{\cdot}{\cdot}$.

Let us f\/ix an arbitrary value $\hbar\in\dR$. For $E$, considered
as a topological vector group with respect to the discrete
topology, we def\/ine the multiplier
\begin{gather}
\label{WQB:eq:multiplier-hbar} E\times E\ni(f,g)\longmapsto
\ex{-\tfrac{i}{2}\hbar\sigma(f,g)}.
\end{gather}
In order to construct for $E$ the twisted group Banach-$*$-
resp.~$C^*$-algebra we start from an abstract $*$-algebra, given
as the formal linear hull
\begin{gather}
\label{WQB:eq:9-5:Delta=LH}
\Delta(E,\hbar\sigma):=\mathrm{LH}\{W^\hbar(f)\mid f\in E\}
\end{gather}
of \emph{linearly independent} symbols $W^\hbar(f)$, $f\in E$,
called Weyl elements. Equipped with the Weyl relations
\begin{gather}
\label{WQB:eq:9-5:Weyl-relations} W^\hbar(f)W^\hbar(g)=
\ex{-\tfrac{i}{2}\hbar\sigma(f,g)} W^\hbar(f+g), \qquad
W^\hbar(f)^*=W^\hbar(-f),\quad \forall \, f,g\in E,
\end{gather}
$\Delta(E,\hbar\sigma)$ indeed becomes a $*$-algebra. Its identity
is given by $\1^\hbar:=W^\hbar(0)$, and every Weyl element
$W^\hbar(f)$ is unitary ($W^\hbar(f)^*=W^\hbar(f)^{-1}$).

The completion $\ovl{\Delta(E,\hbar\sigma)}^1$ with respect to the
norm
\begin{gather*}
\left\| \sum\limits_{k=1}^n z_k W^\hbar(f_k)   \right\|_1:=
\sum\limits_{k=1}^n \betr{z_k}, \qquad \text{$n\in\dN$,
$z_k\in\dC$, dif\/ferent $f_k$'s from $E$},
\end{gather*}
is just the twisted group Banach-$*$-algebra of $E$. The Weyl
algebra $\cW(E,\hbar\sigma)$ over the pre-symplectic space is by
def\/inition the twisted group $C^*$-algebra of $E$, i.e., the
enveloping $C^*$-algebra of the Banach-$*$-algebra
$\ovl{\Delta(E,\hbar\sigma)}^1$. (For twisted group algebras we
defer the reader to the citations in Subsection~\ref{WQ-s5.1}, and
for the Weyl algebra with degenerate $\sigma$ see
\cite{MSTV73,BinzHonRie99b}, and references therein.) The
$C^*$-norm on $\cW(E,\hbar\sigma)$ is denoted by
$\norm{\cdot}_\hbar$. It varies with dif\/ferent values of
$\hbar\in\dR$, in contrast to the $\hbar$-independent Banach norm
$\norm{\cdot}_1$. $\ovl{\Delta(E,\hbar\sigma)}^1$ is a proper, but
$\norm{\cdot}_\hbar$-dense sub-$*$-algebra of
$\cW(E,\hbar\sigma)$, where (see the def\/initions at the
beginning of this Section)
\begin{gather*}
\norm{A}_\hbar\leq\norm{A}_1, \qquad\forall \,
A\in\ovl{\Delta(E,\hbar\sigma)}^1.
\end{gather*}
By construction of the enveloping $C^*$-algebra, the states and
representations of $\cW(E,\hbar\sigma)$ and
$\ovl{\Delta(E,\hbar\sigma)}^1$ are in 1:1-correspondence, given
by continuous extension resp.~restriction. The $C^*$-Weyl algebra
$\cW(E,\hbar\sigma)$ is simple, if and only if $\sigma$ is
non-degenerate and $\hbar\neq0$.

If $\pi_\hbar$ is any projective unitary representation of the
additive group $E$ with respect to the
multiplier~\eqref{WQB:eq:multiplier-hbar}, then there exists a
unique non-degenerate representation $\Pi_\hbar$ of
$\cW(E,\hbar\sigma)$ (and by restriction of
$\ovl{\Delta(E,\hbar\sigma)}^1$) such that
\begin{gather}
\label{WQB:eq:pi-Pi}
\pi_\hbar(f)=\Pi_\hbar(W^\hbar(f)),\qquad\forall f\in E,
\end{gather}
and conversely. Whenever we use in the sequel both notions
$\pi_\hbar$ and $\Pi_\hbar$, they are connected as described here.
For $f\neq g$ we have $\norm{W^\hbar(f)-W^\hbar(g)}_\hbar=2$,
which implies discontinuity of $f \mapsto W^\hbar(f)$ in the norm.
But for certain projective group representations $\pi_\hbar$, the
mapping $E\ni f\mapsto\pi_\hbar(f)$ may be strongly continuous
with respect to some topology $\tau$ on $E$. This point is
essential for the present investigation and considered in more
detail in Subsection~\ref{WQ-s4.1} below.

\subsection[$*$-isomorphisms for the quantum Weyl algebras]{$\boldsymbol{*}$-isomorphisms for the quantum Weyl algebras}
\label{WQ-s4.3}

For $\hbar=0$ the multiplier~\eqref{WQB:eq:multiplier-hbar}
becomes trivial, which leads to the commutative $C^*$-Weyl algebra
$\cW(E,0)$. For the quantum cases $\hbar\neq0$ all Weyl algebras
$\cW(E,\hbar\sigma)$ are $*$-iso\-mor\-phic, a fact which we need
in Section~\ref{WQ-s6} to make representations $\Pi_\hbar$ of
$\cW(E,\hbar\sigma)$, with dif\/ferent $\hbar$, compatible with
each other. 
\begin{lemma}
\label{WQ:9-5:lemm:beta-hbar-*iso} For $\hbar\neq0$ let
$T_\hbar:E\rightarrow E$ be an $\dR$-linear bijection such that
$\sigma(T_\hbar f,T_\hbar g)=\hbar\sigma(f,g)$ for all $f,g\in E$.
Then there exists a unique $*$-isomorphism $\beta_\hbar$ from
$\cW(E,\hbar\sigma)$ onto $\cW(E,\sigma)$ (where $\hbar=1$) such
that $\beta_\hbar(W^\hbar(f))=W^1(T_\hbar f)$ for all $f\in E$.

If in addition $T_\hbar$ is a homeomorphism with respect to
$\tau$, then the dual mapping $\beta_\hbar^*$ is an affine
bijection from $\cF_{\hbar=1}^\tau$ onto $\cF_\hbar^\tau$.

(The convex folia $\cF_\hbar^\tau$ of $\tau$-continuous states are
introduced in Subsection~{\rm \ref{WQ-s4.1}}.)
\end{lemma}

\begin{proof}
In virtue of the linear independence of the $W^\hbar(f)$, $f\in
E$, $\beta_\hbar$ is a well-def\/ined $*$-isomorphism from
$\Delta(E,\hbar\sigma)$ onto $\Delta(E,\sigma)$. Extend
norm-continuously using a result in \cite{BinzHonRie99b}.
\end{proof}

A f\/irst idea to construct a $T_\hbar$ would be
\begin{gather}
\label{WQB4:eq:Thbar-1} T_\hbar f:=\sqrt{\betr{\hbar}} f,
\qquad\forall \, f\in E.
\end{gather}
But here it follows for $\hbar<0$ that $\beta_\hbar$ is a
$*$-anti-isomorphism (antilinear $*$-isomorphism), and thus either
the initial or the f\/inal folium has to consist of antilinear
states.

Let us try a second example. Suppose the existence of a
conjugation $C$ on $(E,\sigma)$, that is, $C:E\rightarrow E$ is an
$\dR$-linear mapping satisfying $C^2f=f$ and
$\sigma(Cf,g)=-\sigma(f,Cg)$ for all $f,g\in E$. The eigenspaces
$E_+$ and $E_-$ of $C$ are given by $E_\pm:= \{f\in E\mid Cf=\pm
f\}=\{g\pm Cg\mid g\in E\}$, and the associated spectral
projections $P_\pm$ by $P_\pm g=\frac12(g\pm Cg)$ for $g\in E$.
Clearly $E_+\bigcap E_-=\{0\}$ and every $f\in E$ decomposes
uniquely as $f=P_+f + P_-f$ (in analogy to a polarization of
a~phase space).

One f\/inds $\sigma(P_+f,P_+g)=0=\sigma(P_-f,P_-g)$, thus
$\sigma(f,g)=\sigma(P_+f,P_-g)+\sigma(P_-f,P_+g)$ for all $f,g\in
E$. For $\hbar\neq0$ we def\/ine an $\dR$-linear bijection
\begin{gather}
\label{WQB4:eq:Thbar-2} T_\hbar f:= \theta_+(\hbar) P_+f +
\theta_-(\hbar) P_-f, \qquad\forall\, f\in E,
\end{gather}
where the real values $\theta_+(\hbar)$ and $\theta_-(\hbar)$ have
to satisfy $\theta_+(\hbar)\theta_-(\hbar)=\hbar$. Then
$\sigma(T_\hbar f,T_\hbar g)=\hbar\sigma(f,g)$, and thus
Lemma~\ref{WQ:9-5:lemm:beta-hbar-*iso} is applicable. If $C$ is
$\tau$-continuous, then $P_\pm$ are so, and thus it follows that
$T_\hbar$ is a homeomorphism of $E$.

One may assume in addition a complex structure $j$ on $(E,\sigma)$
($j$ is an $\dR$-linear mapping on $E$ satisfying $j^2f=-f$,
$\sigma(jf,g)=-\sigma(f,jg)$, and $\sigma(f,jf)\geq0$, for all
$f,g\in E$), such that $C$ is just a complex conjugation for $j$,
i.e., $Cj=-jC$. Then $E$ becomes a complex vector space with the
multiplication $z f=\Re(z)f+j\Im(z)f$, $f\in E$, $z\in\dC$,
carrying the associated complex semi-inner product
$\skal{f}{g}_j:= \sigma(f,jg) + i\sigma(f,g)$, for $f,g\in E$.
Since $E_-= j E_+$ resp.~$P_-j=jP_+$, we have $E=E_+ + jE_+$, that
is the complexif\/ication of $E_+$. Now,
$T_\hbar(u+jv)=\theta_+(\hbar)u+j\theta_-(\hbar)v$ for all $u,v\in
E_+$. Clearly, $C$ and $j$ are highly non-unique. A f\/inite
dimensional example, f\/itting to Subsection~\ref{WQ-s1mot}, is
$T_\hbar(u+iv):=\hbar u+iv$ for all $u,v\in\dR^d$.

\subsection[Strict deformation quantization for $C^*$-Weyl algebras]{Strict deformation quantization for $\boldsymbol{C^*}$-Weyl algebras}
\label{WQ-s2.3}

For the classical case $\hbar=0$ it turns out that the abstract
commutative $*$-algebra $\Delta(E,0)$ becomes a Poisson algebra
$(\Delta(E,0),\{\cdot,\cdot\}_0)$, where the Poisson bracket
$\{\cdot,\cdot\}_0$ is given by the bilinear extension of the
algebraic relations
\begin{gather}
\label{WQ:eq:9-5:Poisson-bracket-W0(f)}
\{W^0(f),W^0(g)\}_0=\sigma(f,g)W^0(f+g), \qquad\forall \, f,g\in
E.
\end{gather}

As indicated already in Section~\ref{WQ-s1} for each $\hbar\in\dR$
the Weyl quantization map
$Q_\hbar:\Delta(E,0)\rightarrow\cW(E,\hbar\sigma)$ is def\/ined by
the linear extension of
\begin{gather}
\label{WQ:eq:9-5:Qhbar} Q_\hbar(W^0(f)):=W^\hbar(f),\qquad\forall
\, f\in E,
\end{gather}
(which is well def\/ined since the Weyl elements $W^\hbar(f)$,
$f\in E$, are linearly independent for every $\hbar\in\dR$).
Obviously, $Q_\hbar$ is a linear, $*$-preserving
$\norm{\cdot}_1$-$\norm{\cdot}_1$-isometry from $\Delta(E,0)$ onto
$\Delta(E,\hbar\sigma)$. Using the $C^*$-norms, one recognizes as
an intermediate result in the course of the present investigation:
\begin{theorem}[\cite{BinzHonRie03b}]
\label{WQ:theo:9-5:SDQ-Weylalgebra} The family of mappings
$(Q_\hbar)_{\hbar\in \dR}$ constitutes a strict deformation
quantization (of $C^*$-type) of the Poisson algebra
$(\Delta(E,0),\{\cdot,\cdot\}_0)$.
\end{theorem}

\subsection[Geometry and algebraization of a classical field theory]{Geometry and algebraization of a classical f\/ield theory}
\label{WQ-s2.4}

Suppose a locally convex topology $\tau$ on the real vector space
$E$, the $\tau$-topological dual of which be denoted by $E_\tau'$.
On $E_\tau'$ we choose the $\sigma(E_\tau',E)$-topology and denote
by $(\operatorname{C}_b(E_\tau'),\cdot_{\!0})$ the commutative
$C^*$-algebra of all bounded, $\sigma(E_\tau',E)$-continuous
functions $A:E_\tau\rightarrow\dC$.

For $f\in E$ the periodic functions
\begin{gather}
\label{WQ:eq:9-5:xi(f)} W_c(f):E_\tau'\rightarrow\dC,\qquad
F\mapsto\ex{iF(f)}=W_c(f)[F],
\end{gather}
realize the commutative Weyl relations. With the commutative
$*$-algebraic operations from
equation~\eqref{WQ:eq:9-5:*-algebraic-operations-F} the
trigonometric polynomials
\begin{gather}
\label{WQ:eq:9-5:Xi=LH-xi(f)}
\Delta(E_\tau'):=\mathrm{LH}\{W_c(f)\mid f\in E\}
\end{gather}
constitute a sub-$*$-algebra of
$(\operatorname{C}_b(E_\tau'),\cdot_{\!0})$.

The $\norm{\cdot}_0$-closure of $\Delta(E_\tau')$ within the
$C^*$-algebra $(\operatorname{C}_b(E_\tau'),\cdot_{\!0})$ gives
the proper sub-$C^*$-algebra $\operatorname{AP}(E_\tau')$
consisting of the almost periodic, $\sigma(E_\tau',E)$-continuous
functions on $E_\tau'$, \cite[18.2 and 33.26]{HewittRoss}. The
following result from \cite{BinzHonRie99b} ensures the
independence of $\operatorname{AP}(E_\tau')$, as a $C^*$-algebra,
from the chosen locally convex topology $\tau$. 
\begin{proposition}
\label{WQ:prop:9-5:hbar=0-*iso} There exists a unique
$*$-isomorphism between the commutative Weyl algebra $\cW(E,0)$
and $(\operatorname{AP}(E_\tau'),\cdot_{\!0})$, which identifies
the Weyl element $W^0(f)$ with the periodic function $W_c(f)$ for
every $f\in E$. In this sense,
$\cW(E,0)\cong(\operatorname{AP}(E_\tau'),\cdot_{\!0})$, and
$\Delta(E,0)\cong\Delta(E_\tau')$.
\end{proposition}
For discussing geometric aspects we denote henceforth the phase
space $E_\tau'$, considered (in a~loose sense, see below) as
dif\/ferentiable manifold with respect to the
$\sigma(E_\tau',E)$-topology, by the symbol $\mathsf{P}$.

One may use, as in \cite{BinzHonRie03b}, $T_F\mathsf{P}:=E_\tau'$
as tangent space at each phase space point $F\in \mathsf{P}$.
Hence, the cotangent space is given by
$T_F^*\mathsf{P}:=(E_\tau')'=E$, and its elements are identif\/ied
with the test functions $f$. The total dif\/ferential $dA\in
T^*\mathsf{P}$ of $A:\mathsf{P}\rightarrow\dR$ is def\/ined at
each $F\in \mathsf{P}$ by $d_F A(G):=\frac{d A[c(t)]}{dt}|_{t=0}$
where one dif\/ferentiates along all curves $t\mapsto
c(t)\in\mathsf{P}$ with $c(0)=F$ and $\frac{d c}{dt}|_{t=0}=G\in
T_F\mathsf{P}$ (e.g. along the linear curve $t\mapsto F+tG$ for
any $G\in E_\tau'$). For a $\dC$-valued function $A$ on
$\mathsf{P}$ we put $d_FA:=d_F A_1+id_F A_2$ with its real and
imaginary parts, $A_1$ resp.~$A_2$, which is an element of the
complexif\/ied cotangent space
$T_F^*\mathsf{P}+iT_F^*\mathsf{P}=:{}^\dC\!T_F^*\mathsf{P}$.

From now on we suppose a pre-symplectic form $\sigma$ on $E$, not
necessarily $\tau$-continuous, which may be compared with a
constant bivector f\/ield on the complexif\/ied cotangent bundle.
With this given, a Poisson bracket $\{\cdot,\cdot\}$ may be
def\/ined
\begin{gather}
\{A,B\}[F]:=
-\sigma(d_F A_1,d_F B_1)-i\sigma(d_F A_1,d_F B_2)\nonumber\\
\phantom{\{A,B\}[F]:=}{} -i\sigma(d_F A_2,d_F B_1)+\sigma(d_F
A_2,d_F B_2),\label{eq:9-5:Poisson-bracket-sigma}
\end{gather}
a construction familiar from (f\/inite dimensional) classical
Hamiltonian mechanics, e.g.~\cite{AbrahamMarsden78,
Arnold85,LibMas87,BinzSF88,MarsdenRatiu94}. The
dif\/ferentiability of $A$ and $B$ does not ensure the
dif\/ferentiability of $\mathsf{P}\ni F\mapsto\{A,B\}[F]\in\dC$.
Consequently, in order to obtain a Poisson algebra we have need
for a sub-$*$-algebra $\cP$ of
$(\operatorname{C}_b(\mathsf{P}),\cdot_{\!0})$ consisting of
dif\/ferentiable functions, such that $\{A,B\}\in\cP$ whenever
$A,B\in\cP$.

For the periodic function $W_c(f)$ from
equation~\eqref{WQ:eq:9-5:xi(f)} we calculate
\begin{gather}
\label{WQ:eq:9-5:totdiff-xi(f)-o} d_F
W_c(f)=i\ex{iF(f)}f=iW_c(f)[F]\, f\in {}^\dC\!T_F^*\mathsf{P}
,\qquad \forall \, F\in \mathsf{P}.
\end{gather}
Insertion into the Poisson bracket
\eqref{eq:9-5:Poisson-bracket-sigma} leads to
\begin{gather}
\label{eq:9-5:Poisson-bracket-xi(f)} \{W_c(f),W_c(g)\}=\sigma(f,g)
W_c(f+g), \qquad\forall \, f,g\in E.
\end{gather}
Since $\{W_c(f),W_c(g)\}\in\Delta(E_\tau')$ for every $f,g\in E$,
the trigonometric polynomials
$\Delta(E_\tau')\equiv\Delta(\mathsf{P})$ from
equation~\eqref{WQ:eq:9-5:Xi=LH-xi(f)}, equipped with the Poisson
bracket $\{\cdot,\cdot\}$ from
equation~\eqref{eq:9-5:Poisson-bracket-sigma} constitute a~Poisson
algebra.

By equations~\eqref{WQ:eq:9-5:Poisson-bracket-W0(f)}
and~\eqref{eq:9-5:Poisson-bracket-xi(f)} the following consequence
is immediate.

\begin{corollary}[Poisson isomorphism]
\label{WQ:9-5:coro:Poisson*iso-hbar=0} The $*$-isomorphism in
Proposition~{\rm \ref{WQ:prop:9-5:hbar=0-*iso}} leads by
restriction to a Poisson isomorphism from
$(\Delta(E,0),\{\cdot,\cdot\}_0)$ onto
$(\Delta(E_\tau'),\cdot_{\!0},\{\cdot,\cdot\})$.
\end{corollary}

The Corollary provides a belated motivation for the algebraically
introduced Poisson bracket in
$(\Delta(E,0),\{\cdot,\cdot\}_0)$.The independence of $\tau$ is
essential. Even on a f\/lat inf\/inite dimensional manifold there
is no standard form of a dif\/ferentiable structure, and various
additional requirements are imposed on $\tau$ in this context. Our
foregoing reasoning demonstrates that any notion of
dif\/ferentiability should lead to the same dif\/ferentials of the
classical Weyl elements and thus to the Poisson brackets
\eqref{eq:9-5:Poisson-bracket-xi(f)}.

\begin{summary}[Functorial quantization for $\boldsymbol{C^*}$-Weyl algebras]
\label{WQB:summ:functor} Let be given the pre-symplectic space
$(E, \sigma)$ of any dimension with non-trivial $\sigma$, defining
the Weyl relations for the abstract Weyl elements $W^\hbar(f)$, $f
\in E$. Associated with this input is a complex $*$-algebra
 $\Delta(E,\hbar\sigma)$, linearly spanned by the $W^\hbar(f)$,
for any $\hbar \in \dR$. For $\hbar = 0$ (classical case) the
latter admits an algebraic Poisson bracket, which conforms with
that naturally induced by a differentiable structure on the
$\tau$-dependent, topological dual $E_\tau'$.

For $\hbar \neq 0$ (quantum regime) the non-commutative Weyl
relations define scaled commutators in $\Delta(E,\hbar\sigma)$.
$\Delta(E,\hbar\sigma)$ is easily equipped with an algebra norm,
the closure with which produces the Banach-$*$-algebra
$\ovl{\Delta(E,\hbar\sigma)}^1$. The enveloping $C^*$-algebra
defines the $C^*$-Weyl algebra $\cW(E,\hbar\sigma)$ in this
general frame.

As recapitulated in Theorem~{\rm
\ref{WQ:theo:9-5:SDQ-Weylalgebra}} the family of mappings
$(Q_\hbar)_{\hbar\in\dR}$, which linearly connect the classical
Weyl elements with the quantized, leads to a strict deformation
quantization.

Thus the whole quantization scheme depends functorially on $(E,
\sigma)$.
\end{summary}

In the Sections~\ref{WQ-s5} and~\ref{WQ-s6} we generalize this
functorial concept of strict deformation quantization to much
larger Poisson algebras. There, however, one needs beside
$(E,\sigma)$, a suitable family of representations $\Pi_\hbar$ of
$\cW(E,\hbar\sigma)$ for $\hbar\neq0$ and  a semi-norm $\varsigma$,
satisfying~\eqref{WQ:eq:9-5:semi-norm-sigma}, as the only
additional ingredients.

\section[Strict deformation quantization via
            Banach-$*$-algebras of measures]{Strict deformation quantization\\ via
            Banach-$\boldsymbol{*}$-algebras of measures}
\label{WQ-s5}

From now on we suppose $(E,\sigma)$ to be a pre-symplectic space
of \emph{infinite} dimensions.

Let $I$ index the set of all f\/inite dimensional subspaces
$E_\alpha$ of $E$.
The inductive limit topolo\-gy~$\tau_{\text{il}}$~\cite{Schaefer66,Conway85} 
over the directed set $\{E_\alpha\mid \alpha\in I\}$ is the
f\/inest locally convex topology on $E$. Observe that we have
chosen the f\/inest directed set, given by all f\/inite
dimensional subspaces $E_\alpha$ of $E$, but for the construction
of $\tau_{\text{il}}$ it would be suf\/f\/icient to use any
absorbing directed subset of f\/inite dimensional subspaces.

Especially we are interested in the sub-index set $I_\sigma$
consisting of those indexes $\alpha\in I$ for which the
restriction of the pre-symplectic form $\sigma$ is non-degenerate
on $E_\alpha$. If $\sigma$ is non-degenerate on all of $E$, we
perform the inductive limit along $I_\sigma$, keeping in mind
$E=\bigcup_{\alpha\in I_\sigma}E_\alpha$,
(\cite[\S~4]{Kastler65}), which leads us also to the f\/inest
locally convex topology $\tau_{\text{il}}$ on $E$
(\cite[\S~IV.5]{Conway85}). 

\subsection{Inductive limit of regular Borel measures}
\label{WQ-s5.0}

 Let us denote by $\mathsf{B}^\tau(E)$ the Borel subsets
of $E$ with respect to the locally convex topology~$\tau$ on~$E$,
where for $\tau=\tau_{\text{il}}$ we simply use ``${\text{il}}$''
as index. By $\mathsf{B}(E_\alpha)$ we mean the Borel subsets of
the f\/inite dimensional $E_\alpha$ equipped with its vector space
topology.

\begin{lemma}
\label{WQ:lemm:Borel-sets} If $\Lambda\in\mathsf{B}(E_\alpha)$,
then $\Lambda\in\mathsf{B}^\tau(E)$. If
$\Lambda\in\mathsf{B}^\tau(E)$, then $\Lambda\bigcap E_\alpha\in
\mathsf{B}(E_\alpha)$.

The partial ordering of being ``finer than'' for the locally
convex topologies on $E$ carries over to the partial ordering on
the Borel sigma algebras given by the inclusion relation.
\end{lemma}
\begin{proof}
First note that the relative topology of $\tau$ on $E_\alpha$ is
just its unique natural vector space topology. That is, if
$\Lambda$ is an open subset of $E_\alpha$, then there exists a
$\Lambda'\in\tau$ with $\Lambda'\bigcap E_\alpha=\Lambda$.
$E_\alpha$~is a closed subset of $E$, since it contains each of
its limit points, and consequently,
$E_\alpha\in\mathsf{B}^\tau(E)$. The rest is immediate by the
construction of the Borel sigma algebras.
\end{proof}

Let $M^\tau(E)$ denote the Banach space of the f\/inite complex
Borel measures on $E$ with respect to the locally convex $\tau$ on
$E$ (which are not necessarily regular as for the locally compact
$E_\alpha$). The norm on $M^\tau(E)$ is given by the total
variation $\norm{\mu}_1:=|\mu|(E)$, where $\betr{\mu}\in
M^\tau(E)$ denotes the variation measure of $\mu\in M^\tau(E)$.
(For later proofs recall that $|\mu|(\Lambda)$ is the supremum of
the numbers $\textstyle\sum\limits_{k=1}^n \betr{\mu(\Lambda_k)}$,
where $\{\Lambda_k\mid k=1,\ldots,n\}$ ranges over all f\/inite
partitions of $\Lambda\in\mathsf{B}^\tau(E)$ into
$\mathsf{B}^\tau(E)$-measurable sets
 \cite[Section~4.1]{Cohn80}.)
A Borel measure $\mu$ on $E$ is called positive -- written as
$\mu\geq0$ --, if $\mu(\Lambda)\geq0$ for each $\Lambda\in
\mathsf{B}^\tau(E)$. (Sets of positive measures will be indicated
by the upper index ``$+$''.)

The f\/inite (regular) complex Borel measures on the f\/inite
dimensional $E_\alpha$ are denoted by $\cM(E_\alpha)$. According
to Lemma~\ref{WQ:lemm:Borel-sets} we may identify each $\mu\in
\cM(E_\alpha)$ with a measure from~$M^\tau(E)$ concentrated
on~$E_\alpha$. The total variation norm $\norm{\cdot}_1$
calculated on $E_\alpha$ coincides with that on $E$, and
subsequently we understand $\cM(E_\alpha)$ as a
$\norm{\cdot}_1$-closed subspace of $M^\tau(E)$. Since
$\cM(E_\alpha)\subseteq \cM(E_\beta)$ for $\alpha\leq\beta$, these
measure spaces constitute an inductive system.

Our inf\/inite dimensional $E$ is far from being locally compact
(what each $E_\alpha$ is). Since each $\mu\in \cM(E_\alpha)$ is a
regular Borel measure on $E_\alpha$, we may consider the
$\norm{\cdot}_1$-closure
\begin{gather}
\label{WQB:eq:reg-M(E)}
\cM(E):=\ovl{\textstyle\bigcup\limits_{\alpha\in I}
\cM(E_\alpha)}^{\norm{\cdot}_1}
\end{gather}
as a closed subspace of $M^\tau(E)$. The measures in $\cM(E)$ are
regular (what we indicate by the italic $\cM$) for each locally
convex $\tau$, a universality property.

A further reason for considering $\cM(E)$ instead of $M^\tau(E)$
lies in the applicability of Fubini's theorem in introducing
associative products (see next Subsection). Note that
\begin{gather*}
\cM(E)^+=\cM(E){\textstyle\bigcap} M^\tau(E)^+ =
\ovl{\textstyle\bigcup\limits_{\alpha\in I}
\cM(E_\alpha)^+}^{\norm{\cdot}_1}.
\end{gather*}
Especially it holds for $\mu\in\cM(E)$, with Jordan decomposition
$\mu=\mu_1-\mu_2+i(\mu_3-\mu_4)$, that the $\mu_k$ as well as
$\betr{\mu}$ are in $\cM(E)^+$.

Subsequently we use the following two facts without mentioning:
$\bigl| \int_E d\mu\, a \bigr| \leq \int_E d|\mu|\,|a|$ for every
bounded Borel measurable function $a:E\rightarrow\dC$. If
$\mu\in\cM(E)$ and $a:E\rightarrow\dC$ is Borel measurable with
$\int_E d|\mu|\,|a| <\infty$, then $\mu_a\in\cM(E)$, where
$d\mu_a(f):=a(f)d\mu(f)$.

For each $\alpha\in I$ the measure Banach space $\cM(E_\alpha)$
decomposes uniquely according to
\begin{gather*}
\cM(E_\alpha)=\cM(E_\alpha)_d\oplus
\underbrace{\cM(E_\alpha)_s\oplus
\cM(E_\alpha)_a}_{\mbox{$=:\cM(E_\alpha)_c$}}
\end{gather*}
into three $\norm{\cdot}_1$-closed subspaces: the discrete
measures $\cM(E_\alpha)_d$, the singularly continuous measures
$\cM(E_\alpha)_s$, and the absolutely continuous measures
$\cM(E_\alpha)_a$ with respect to the Haar measure on $E_\alpha$.
$\cM(E_\alpha)_c$ are the continuous measures on $E_\alpha$. For
the absolutely continuous measures one has
$\cM(E_\alpha)_a\not\subseteq \cM(E_\beta)_a$ for $\alpha<\beta$,
rather $\cM(E_\alpha)_a\subseteq \cM(E_\beta)_s$, and so they do
not constitute an inductive system. Inductive systems over $I$ are
obtained, however, for the discrete resp.~continuous measures,
since $\cM(E_\alpha)_d\subseteq \cM(E_\beta)_d$
resp.~$\cM(E_\alpha)_c\subseteq \cM(E_\beta)_c$ for $\alpha\leq
\beta$. So we def\/ine the $\norm{\cdot}_1$-closed subspaces
\begin{gather}
\label{WQB:eq:reg-Md(E)}
\cM(E)_d:=\ovl{\textstyle\bigcup\limits_{\alpha\in I}
\cM(E_\alpha)_d}^{\norm{\cdot}_1}, \qquad
\cM(E)_c:=\ovl{\textstyle\bigcup\limits_{\alpha\in I}
\cM(E_\alpha)_c}^{\norm{\cdot}_1}
\end{gather}
of $\cM(E)$. We denote by $\delta(f)$ the point measure at $f\in
E$, which for each $\Lambda\in\mathsf{B}^\tau(E)$ gives
$\delta(f)(\Lambda)=1$, if $f\in \Lambda$, and
$\delta(f)(\Lambda)=0$, if $f\not\in\Lambda$. Each element $\mu$
of $\cM(E)_d$ is of the form $\mu=\sum\limits_{k=1}^\infty z_k
\delta(f_k)$ with $\norm{\mu}_1=\sum\limits_{k=1}^\infty
\betr{z_k}<\infty$ and with dif\/ferent $f_k$'s from $E$, where
$\betr{\mu}=\sum\limits_{k=1}^\infty \betr{z_k}\delta(f_k)$ is the
associated variation measure.

\subsection[Non-commutative Banach-$*$-algebras of measures]{Non-commutative Banach-$\boldsymbol{*}$-algebras of measures}
\label{WQ-s5.1}

Suppose $\hbar\in\dR$. It is well known that for each $\alpha\in
I$ the measure space $\cM(E_\alpha)$ becomes a Banach-$*$-algebra
with respect to the  $*$-operation $\mu\mapsto\mu^*$ def\/ined by
\begin{gather}
\label{WQ:eq:9-5:*-involution}
\mu^*(\Lambda):=\ovl{\mu(-\Lambda)}, \qquad\forall
\,\Lambda\in\mathsf{B}(E_\alpha),
\end{gather}
and the associative product $\star_{\!\hbar}$ given as the twisted
resp.~deformed convolution
\begin{gather}
\label{WQ:eq:9-5:deformed-convolution}
\mu\star_{\!\hbar}\nu(\Lambda):= \int_{E_\alpha} d\mu(f)
\int_{E_\alpha} d\nu(g)
\ex{-\textstyle\frac{i}{2}\hbar\sigma(f,g)}\Lambda(f+g),
\qquad\forall \,\Lambda\in\mathsf{B}(E_\alpha).
\end{gather}
$E_\alpha\ni h\mapsto\Lambda(h)$ denotes the characteristic
function of $\Lambda$, (with $\Lambda(h)=1$ for $h\in \Lambda$,
and $\Lambda(h)=0$ elsewhere). The above construction is found for
the commutative case $\hbar=0$ in textbooks on measure theory and
harmonic analysis, whereas for (twisted) group algebras we refer
e.g. to \cite{EdwardsLewis69, HewittRoss, BusbySmith70, Dixmier77,
Pedersen79, PackerRaeburn89, Grundling97}. The relations
$\norm{\mu^*}_1=\norm{\mu}_1$ and
$\norm{\mu\star_{\!\hbar}\nu}_1\leq\norm{\mu}_1\norm{\nu}_1$ of
the $*$-algebraic operations extend $\norm{\cdot}_1$-continuously
to the inductive Banach space $\cM(E)$ from
equation~\eqref{WQB:eq:reg-M(E)} making it to a
Banach-$*$-algebra, which we denote by
\begin{gather*}
(\cM(E),\star_{\!\hbar})
\end{gather*}
(cf.~Theorem~\ref{WQ:9-5:theo:Banach-hbar-algebra} below). Its
identity is realized by the point measure at zero $\delta(0)$. The
Banach-$*$-algebra $(\cM(E),\star_{\!\hbar})$ is commutative, if
and only if $\hbar=0$, provided that  $\sigma\neq0$.

Fubini's theorem plays the essential role in proving that the
product in equation~\eqref{WQ:eq:9-5:deformed-convolution} is
associative, resp.~commutative for the case $\hbar=0$. Fubini's
theorem, however, works well for regular Borel measures on locally
compact spaces. Since our test function space $E$ has inf\/inite
dimension, $\star_{\!\hbar}$ may not be associative on
$M^\tau(E)$. That is why we restrict ourselves to the regular
measures $\cM(E)$. In \cite{Kastler65} the Banach-$*$-algebra
$(\cM(E),\star_{\!\hbar})$ is performed as a purely inductive
limit not taking into account that in the completion one gains
further measures out of~$M^\tau(E)$. In contradistinction to the
product $\star_{\!\hbar}$, the $*$-involution is well def\/ined
also on $M^\tau(E)$ by equation~\eqref{WQ:eq:9-5:*-involution}.

\begin{theorem}
\label{WQ:9-5:theo:Banach-hbar-algebra} The mapping $\dR\times
\cM(E)\times \cM(E)\rightarrow \cM(E)$, $(\hbar,\mu,\nu)\mapsto
\mu\star_{\!\hbar}\nu$ is jointly continuous (i.e., continuous
with respect to the product topology on $\dR\times \cM(E)\times
\cM(E)$, where on $\cM(E)$ we have the topology arising from the
total variation norm $\norm{\cdot}_1$).
\end{theorem}

\begin{proof}
For each $\mu,\nu\in \cM(E)$ and all $\hbar,\lambda\in\dR$ we
obtain
\begin{gather*}
\betr{(\mu\star_{\!\hbar}\nu - \mu\star_{\!\lambda}\nu)(\Lambda)}   \\
\qquad ={} \betr{ \int_E d\mu(f) \int_E d\nu(g) \klr{
\ex{-\textstyle\frac{i}{2}\hbar\sigma(f,g)} -
\ex{-\textstyle\frac{i}{2}\lambda\sigma(f,g)}
    }  \Lambda(f+g)
     }   \\
\qquad{}\leq \int_E d|\mu|(f) \int_E d|\nu|(g) \betr{
\ex{-\textstyle\frac{i}{2}\hbar\sigma(f,g)} -
\ex{-\textstyle\frac{i}{2}\lambda\sigma(f,g)}
    }  \Lambda(f+g),
\end{gather*}
from which we conclude
\begin{gather*}
|\mu\star_{\!\hbar}\nu - \mu\star_{\!\lambda}\nu|(\Lambda) \leq
\int_E\! d|\mu|(f) \int_E\! d|\nu|(g) \betr{
\ex{-\tfrac{i}{2}\hbar\sigma(f,g)} -
\ex{-\tfrac{i}{2}\lambda\sigma(f,g)}
    }  \Lambda(f+g)
\end{gather*}
for all $\Lambda\in\mathsf{B}^\tau(E)$. Taking into account
$\norm{\mu}_1=|\mu|(E)$ for $\mu\in \cM(E)$ yields that
\begin{gather*}
\norm{\mu\star_{\!\hbar}\nu - \mu\star_{\!\lambda}\nu}_1 \leq
\int_E d|\mu|(f) \int_E d|\nu|(g) \betr{
\ex{-\tfrac{i}{2}\hbar\sigma(f,g)} -
\ex{-\tfrac{i}{2}\lambda\sigma(f,g)}
    }
\;\stackrel{\hbar\rightarrow\lambda}{\longrightarrow}\;0.
\end{gather*}
Now the result follows from standard $\norm{\cdot}_1$-estimations.
\end{proof}

The $*$-algebraic structure (consisting of the $*$-operation and
the product $\star_{\!\hbar}$) allows further concept of
positivity. For a measure $\mu\in\cM(E)^+$ its positivity is for
no $\hbar\in\dR$ connected with the $*$-algebraic positivity.
Recall that the latter means $\mu=\nu^*\star_{\!\hbar}\nu$ for
some $\nu\in \cM(E)$. The positive measures $\cM(E)^+$, however,
constitute a closed convex cone in $\cM(E)$ satisfying
$\cM(E)^+\bigcap (-\cM(E)^+)=\{0\}$ (pointedness) and
$\cM(E)^+\star_{\!0}\cM(E)^+\subseteq \cM(E)^+$.

For each $\alpha\in I$ we have that $\cM(E_\alpha)_d$ is a
sub-Banach-$*$-algebra, whereas $\cM(E_\alpha)_a$ and
$\cM(E_\alpha)_c$ constitute closed (two-sided) $*$-ideals of the
Banach-$*$-algebra $(\cM(E_\alpha),\star_{\!\hbar})$, (e.g.
\cite{HewittRoss}). Thus also in the inductive limit the discrete
measures $\cM(E)_d$ constitute a sub-Banach-$*$-algebra and the
continuous measures $\cM(E)_c$ a closed (two-sided) $*$-ideal of
the Banach-$*$-algebra $(\cM(E),\star_{\!\hbar})$. Especially
$\cM(E)_{df}$, the subspace of measures in $\cM(E)_d$ with
f\/inite support (for which $\sum_k z_k\delta(f_k)$ ranges over a
f\/inite number of terms), is a $\norm{\cdot}_1$-dense
sub-$*$-algebra of~$(\cM(E)_d,\star_{\!\hbar})$.

The following result is immediate with the construction in
Subsection~\ref{WQ-s2.2}. 
\begin{proposition}[Measure realization of the Weyl algebra]
\label{WQB:9-5:obse:Banach-hbar-discrete} Let $\hbar\in\dR$, and
let $(E, \sigma)$ be an arbitrary pre-symplectic space. If we
associate for each $f\in E$ the abstract Weyl element~$W^\hbar(f)$
with the point measure~$\delta(f)$, the Weyl relations
\begin{gather*}
W^\hbar(f)W^\hbar(g)= \ex{-\textstyle\frac{i}{2}\hbar\sigma(f,g)}
W^\hbar(f+g), \qquad W^\hbar(f)^*=W^\hbar(-f),\qquad \forall \,
f,g\in E,
\end{gather*}
from equation~\eqref{WQB:eq:9-5:Weyl-relations} are transformed
into
\begin{gather}
\label{WQ:eq:9-5:hbar-Weyl-rel-delta(f)}
\delta(f)\star_{\!\hbar}\delta(g)
=\ex{-\tfrac{i}{2}\hbar\sigma(f,g)}\delta(f+g), \qquad
\delta(f)^*=\delta(-f), \qquad \forall \, f,g\in E.
\end{gather}
Moreover, the $*$-algebra $\Delta(E,\hbar\sigma)$ is bijectively
transformed onto the $*$-algebra $(\cM(E)_{df},\star_{\!\hbar})$.
Closure in the $\norm{\cdot}_1$-norm leads to the
Banach-$*$-algebra
$\ovl{\Delta(E,\hbar\sigma)}^1=(\cM(E)_d,\star_{\!\hbar})$, for
which the enveloping $C^*$-algebra is $*$-isomorphic to
$\ovl{\Delta(E,\hbar\sigma)}^{\norm{\cdot}}=\cW(E,\hbar\sigma)$,
the closure in the $C^*$-norm.

Thus we may identify the boson field algebra $\cW(E,\hbar\sigma)$
with the enveloping $C^*$-algebra of the discrete measure algebra
$(\cM(E)_d,\star_{\!\hbar})$.
\end{proposition}

\subsection{Poisson algebras of measures}
\label{WQ-s5.2}

Since $\tau_{\text{il}}$ is the f\/inest locally convex topology
on $E$, every semi-norm $\kappa$ on $E$ is
$\tau_{\text{il}}$-continuous. Thus from $\cM(E)\subseteq
M^{\text{il}}(E)$ it follows that the integral $\int_E \kappa(f)^m
d\betr{\mu}(f)$ is well def\/ined for all $\mu\in\cM(E)$ with
respect to the Borel sigma algebra $\mathsf{B}^{\text{il}}(E)$,
but possibly leads to the value $\infty$. If $\kappa$ is a
$\tau$-continuous semi-norm with respect to any other locally
convex topology $\tau$ on $E$, then the integral $\int_E
\kappa(f)^m d\betr{\mu}(f)$ may be understood also in terms of the
Borel sigma algebra~$\mathsf{B}^\tau(E)$.

Let $n\in\dN\bigcup\{0,\infty\}$. We def\/ine for each semi-norm
$\kappa$ on $E$ the space
\begin{gather*}
\cM_\kappa^n(E):= \{\mu\in \cM(E)  \mid \text{
$\textstyle\int_E \kappa(f)^m d|\mu|(f)<\infty$
                                      for all $0\leq m\leq n$    }
\}.
\end{gather*}
We call $\mu^m_\kappa\in \cM(E)$ with
$d\mu^m_\kappa(f):=\kappa(f)^m d\mu(f)$ the \emph{$m$-th moment
measure} for $\mu\in \cM_\kappa^n(E)$ with respect to the
semi-norm $\kappa$. Note that $\mu\in \cM_\kappa^n(E)$ implies
$|\mu|\in \cM_\kappa^n(E)$, and $|\mu^m_\kappa|=|\mu|^m_\kappa$ by
def\/inition. For f\/inite $n\in\dN$ it is easily shown that
$\cM_\kappa^n(E)$ is a Banach space with respect to the norm
\begin{gather*}
\norm{\mu}_\kappa^n:=\norm{\mu}_1+\sum_{m=1}^n\norm{\mu_\kappa^m}_1.
\end{gather*}
$\cM_\kappa^\infty(E)=\bigcap_n \cM_\kappa^n(E)$ turns out to be a
Fr\'{e}chet space with respect to the metrizable locally convex
topology $\upsilon_\kappa$ arising from the increasing system of
norms $\norm{\cdot}_\kappa^n$, $n\in\dN$. For $n=0$ we re-obtain
$\cM_\kappa^0(E)=\cM(E)$ and
$\norm{\cdot}_\kappa^0=\norm{\cdot}_1$. Obviously,
$\bigcup_{\alpha\in I}\cM_\kappa^n(E_\alpha)$ is dense in
$\cM_\kappa^n(E)$ with respect to $\norm{\cdot}_\kappa^n$, for
$n<\infty$, resp.~to $\upsilon_\kappa$, for $n=\infty$. 
\begin{lemma}
\label{WQ:9-5:lemm:sub-*-algebra-M(E)} For each
$n\in\dN\bigcup\{\infty\}$, each $\hbar\in\dR$, and each semi-norm
$\kappa$ on the pre-symplectic space $E$ the previously introduced
linear space of complex measures $\cM_\kappa^n(E)$ is a
sub-$*$-algebra of the Banach-$*$-algebra
$(\cM(E),\star_{\!\hbar})$. The following assertions are valid:
\begin{enumerate}\itemsep=0pt
\renewcommand{\labelenumi}{(\alph{enumi})}
\renewcommand{\theenumi}{(\alph{enumi})}
\item \label{WQ:9-5:lemm:sub-*-algebra-M(E)-a} For each $n\in\dN$
$\norm{\mu}_\kappa^n=\norm{\mu^*}_\kappa^n$ and
$\norm{\mu\star_{\!\hbar}\nu}_\kappa^n \leq
c_n\norm{\mu}_\kappa^n\norm{\nu}_\kappa^n$ for all
$\mu,\nu\in\cM_\kappa^n(E)$ with some constant $c_n\geq1$ defined
in equation~\eqref{WQB:eq:lemm:defi:c-n} below. (One has $c_{1}=1$
for $n=1$ but necessarily $c_n>1$ for $n\geq2$, and so
$\cM_\kappa^n(E)$ is not a Banach$*$-algebra with respect to
$\norm{\cdot}_\kappa^n$ but with respect to an equivalent norm.)
\item \label{WQ:9-5:lemm:sub-*-algebra-M(E)-b}
$\cM_\kappa^\infty(E)$ is a Fr\'{e}chet-$*$-algebra with respect
to its Fr\'{e}chet topology $\upsilon_\kappa$ (i.e., the product
is jointly $\upsilon_\kappa$-continuous, and the $*$-operation is
$\upsilon_\kappa$-continuous).
\end{enumerate}
\end{lemma}
\begin{proof}
$|\mu|^*(\Lambda)=|\mu^*|(\Lambda)=|\mu|(-\Lambda)$ for all
$\Lambda\in\mathsf{B}^{\text{il}}(E)$ together with the semi-norm
property $\kappa(f)=\kappa(-f)$ imply $\cM_\kappa^n(E)$ to be
invariant under the $*$-operation $\mu\mapsto\mu^*$ and
$\norm{\mu}_\kappa^n=\norm{\mu^*}_\kappa^n$.

Let $\mu,\nu\in \cM_\kappa^n(E)$. Then
$\mu\star_{\!\hbar}\nu\in\cM(E)$. We show
$\mu\star_{\!\hbar}\nu\in\cM_\kappa^n(E)$. From the semi-norm
property $\kappa(f+g)\leq\kappa(f)+\kappa(g)$ we obtain for all
$m\in\dN$ that
\begin{gather*}
\betr{(\mu\star_{\!\hbar}\nu)^m_\kappa(\Lambda)} = \betr{ \int_E
d\mu(f) \int_E d\nu(g) \ex{-\textstyle\frac{i}{2}\hbar\sigma(f,g)}
\kappa(f+g)^m \Lambda(f+g)
     }    \\
\phantom{\betr{(\mu\star_{\!\hbar}\nu)^m_\kappa(\Lambda)}}{} \leq
\int_E \! d|\mu|(f) \int_E\!  d|\nu|(g) (\kappa(f)+\kappa(g))^m
\Lambda(f+g) = \sum_{k=0}^m\!
\begin{pmatrix} m\\ k \end{pmatrix}
|\mu|_\kappa^{m-k}\star_{\!0}|\nu|_\kappa^{k}(\Lambda).
\end{gather*}
Thus, $|\mu\star_{\!\hbar}\nu|^m_\kappa \leq \sum\limits_{k=0}^m
\left(\begin{smallmatrix} m\\ k \end{smallmatrix}\right)
|\mu|_\kappa^{m-k}\star_{\!0}|\nu|_\kappa^{k}$, resp.\
$\|(\mu\star_{\!\hbar}\nu)^m_\kappa\|_1 \leq \sum\limits_{k=0}^m
\left(\begin{smallmatrix} m\\ k \end{smallmatrix}\right)
\|\mu_\kappa^{m-k}\|_1  \|\nu_\kappa^{k}\|_1 <\infty$ for all
$m\leq n$. Consequently, $\mu\star_{\!\hbar}\nu\in
\cM^n_\kappa(E)$. Moreover,
\begin{gather}
\notag \|\mu\star_{\!\hbar}\nu\|_\kappa^n = \sum_{m=0}^n
\|(\mu\star_{\!\hbar}\nu)^m_\kappa\|_1 \leq \sum_{m=0}^n
\sum_{k=0}^m\begin{pmatrix} m\\ k \end{pmatrix}
\|\mu_\kappa^{m-k}\|_1  \|\nu_\kappa^{k}\|_1
\\ \label{WQB:eq:lemm:defi:c-n}
\phantom{\|\mu\star_{\!\hbar}\nu\|_\kappa^n}{} \leq \underbrace{
\sup\{ {  \textstyle \left(\begin{smallmatrix} n\\ k
\end{smallmatrix}\right)  } \mid k=0,1,\ldots,n
    \}
           }_{\mbox{$=:c_n$}}
\norm{\mu}_\kappa^n\norm{\nu}_\kappa^n.
\end{gather}
Hence part~\ref{WQ:9-5:lemm:sub-*-algebra-M(E)-a} is proved, which
also leads to proving of the
part~\ref{WQ:9-5:lemm:sub-*-algebra-M(E)-b}.
\end{proof}

For each f\/ixed $n\in\dN\bigcup\{\infty\}$ the moment measure
spaces $\cM_\kappa^n(E)$ are in inverse-order-preserving
correspondence with the semi-norms $\kappa$ on $E$: If the
semi-norm $\eta$ is stronger than~$\kappa$~-- i.e.,
$0\leq\kappa\leq c\eta$ for some $c\geq0$ --, then $\cM(E)
\supseteq \cM_\kappa^n(E) \supseteq \cM_\eta^n(E)$, where $\cM(E)$
corresponds to the trivial semi-norm. If two semi-norms $\kappa$
and $\eta$ are equivalent -- $c_1\eta\leq\kappa\leq c_2\eta$ with
some constant $0<c_1\leq c_2$ --, then $\cM_\kappa^n(E)=
\cM_\eta^n(E)$. For a collection of semi-norms $\kappa_\beta$,
$\beta\in B$, the intersection $\bigcap_{\beta\in B}
\cM_{\kappa_\beta}^n(E)$ gives a further sub-$*$-algebra of
$(\cM(E),\star_{\!\hbar})$.

Analogously to equation~\eqref{WQ:eq:9-5:hbar-scaled-commut} we
introduce for each $\hbar\neq0$ the scaled commutator
\begin{gather}
\label{WQ:eq:9-5:hbar-scaled-commut-M(E)} \{\mu,\nu\}_\hbar :=
\tfrac{i}{\hbar}
\klr{\mu\star_{\!\hbar}\nu-\nu\star_{\!\hbar}\mu}, \qquad\forall
\,\mu,\nu\in \cM(E),\qquad\hbar\neq0.
\end{gather}
For $\hbar=0$, instead of a commutator, we construct a Poisson
bracket $\{\cdot,\cdot\}_0$. The starting point for this
construction is a semi-norm $\varsigma$ on $E$ satisfying
\begin{gather}
\label{WQ:eq:9-5:semi-norm-sigma} \betr{\sigma(f,g)}\leq
c\:\varsigma(f)\,\varsigma(g), \qquad\forall\, f,g\in E,
\end{gather}
for some constant $c>0$. Whenever we write $\varsigma$ below, we
assume this inequality. Since each semi-norm is continuous with
respect to the f\/inest locally convex topology
$\tau_{\text{il}}$, we conclude that a semi-norm $\varsigma$
satisfying equation~\eqref{WQ:eq:9-5:semi-norm-sigma} exists, if
and only if the pre-symplectic form $\sigma$ is jointly continuous
with respect to some locally convex topology~$\tau$ on~$E$.
Clearly, $\varsigma$ has to be a norm on~$E$, if $\sigma$ is
non-degenerate on~$E$.

\begin{theorem}
\label{WQ:9-5:theo:PoissonBracket-M(E)} For every $\mu,\nu\in
\cM_\varsigma^1(E)$ the expression
\begin{gather}
\label{WQ:9-5:eq:PoissonBracket-M(E)} \{\mu,\nu\}_0(\Lambda):=
\int_E d\mu(f)  \int_E d\nu(g) \:\sigma(f,g)\Lambda(f+g),
\qquad\forall \, \Lambda\in\mathsf{B}^{\text{\rm il}}(E),
\end{gather}
gives a well-defined measure $\{\mu,\nu\}_0\in \cM(E)$.
Furthermore, if $\mu,\nu\in \cM_\varsigma^n(E)$ for some finite
$n\in\dN$, then $\{\mu,\nu\}_0\in \cM_\varsigma^{n-1}(E)$, and
\begin{gather*}
\norm{  \{\mu,\nu\}_0  }_\varsigma^{n-1} \leq c\, c_{n-1}
\norm{\mu}_\varsigma^n\norm{\nu}_\varsigma^n , \qquad\forall \,
\mu,\nu\in \cM_\varsigma^n(E).
\end{gather*}

With the scaled commutators $\{\cdot,\cdot\}_\hbar$ from
equation~\eqref{WQ:eq:9-5:hbar-scaled-commut-M(E)} for
$\hbar\neq0$, and the bracket $\{\cdot,\cdot\}_0$ for $\hbar=0$,
the mapping
\begin{gather}
\label{WQ:9-5:eq:PoissonBracket-M(E)-2} \dR\times
\cM^{n}_\varsigma(E)\times \cM^{n}_\varsigma(E)\rightarrow
\cM^{n-1}_\varsigma(E),\qquad (\hbar,\mu,\nu)\mapsto
\{\mu,\nu\}_\hbar
\end{gather}
is jointly continuous for every $n\in \dN\bigcup\{\infty\}$, that
means, continuous with respect to the product topology on
$\dR\times \cM^n_\varsigma(E)\times \cM^n_\varsigma(E)$ arising
from the norm $\norm{\cdot}^n_\varsigma$ and the norm
$\norm{\cdot}^{n-1}_\varsigma$ on $\cM^{n-1}_\varsigma(E)$ for
$n<\infty$, resp.~the Fr\'{e}chet topology $\upsilon_\kappa$ for
$n=\infty$.

Furthermore, the mapping
\begin{gather*}
\{\cdot,\cdot\}_0:\cM_\varsigma^\infty(E)\times
\cM_\varsigma^\infty(E)\rightarrow \cM_\varsigma^\infty(E), \quad
(\mu,\nu)\mapsto \{\mu,\nu\}_0
\end{gather*}
defines a jointly continuous Poisson bracket $\{\cdot,\cdot\}_0$
with respect to the Fr\'{e}chet topology $\upsilon_\varsigma$,
which makes
$(\cM_\varsigma^\infty(E),\star_{\!0},\{\cdot,\cdot\}_0)$ to a
Poisson algebra. Suppose $\cP$ to be one of the following cases:
\begin{enumerate}\itemsep=0pt
\renewcommand{\labelenumi}{(\Roman{enumi})}
\renewcommand{\theenumi}{(\Roman{enumi})}
\item \label{WQ:9-5:theo:PoissonBracket-M(E)-a}
$\cP=\cM_\kappa^\infty(E)$ with respect to some semi-norm $\kappa$
on $E$ stronger than, or equivalent to $\varsigma$. \item
\label{WQ:9-5:theo:PoissonBracket-M(E)-b}
$\cP=\bigcap\limits_{\beta\in B}\cM_{\kappa_\beta}^\infty(E)$ for
a collection of semi-norms $\kappa_\beta$, $\beta\in B$,
satisfying $\varsigma\leq\sum\limits_{\beta\in \Gamma}
c_\beta\kappa_\beta$ for some finite subindex set $\Gamma\subseteq
B$ and some constants $c_\beta>0$. \item
\label{WQ:9-5:theo:PoissonBracket-M(E)-c} $\cP$ is the
intersection of $\cM(E)_d=\ovl{\Delta(E,0)}^1$ or of $\cM(E)_c$
with one of the above cases. One may also take intersections with
the absolutely continuous complex measures $\cM(E_\alpha)_a$ on a
finite dimensional subspace $E_\alpha$ of $E$. \item
\label{WQ:9-5:theo:PoissonBracket-M(E)-d}
$\cP=\cM(E)_{df}=\Delta(E,0)$.
\end{enumerate}
Then $\cP\subseteq \cM_\varsigma^\infty(E)$, and
$(\cP,\star_{\!0},\{\cdot,\cdot\}_0)$ constitutes a Poisson
algebra, which in addition is invariant under the product
$\star_{\!\hbar}$ for each $\hbar\in\dR$.
\end{theorem}

\begin{proof}
Estimation~\eqref{WQ:eq:9-5:semi-norm-sigma} ensures that for all
$\Lambda\in\mathsf{B}^{\text{il}}(E)$ in
equation~\eqref{WQ:9-5:eq:PoissonBracket-M(E)} we have that
\begin{gather*}
\betr{(\{\mu,\nu\}_0)^m_\varsigma(\Lambda)} \leq \int_E d|\mu|(f)
\int_E d|\nu|(g) \varsigma(f+g)^m
\betr{\sigma(f,g)}\Lambda(f+g)   \\
\phantom{\betr{(\{\mu,\nu\}_0)^m_\varsigma(\Lambda)}}{} \leq c
\int_E d|\mu|(f)  \int_E d|\nu|(g)
\bigl(\varsigma(f)+\varsigma(g)\bigr)^m
\varsigma(f)\,\varsigma(g)\,\Lambda(f+g)
\\
\phantom{\betr{(\{\mu,\nu\}_0)^m_\varsigma(\Lambda)}}{}\leq c
\sum_{k=0}^m\begin{pmatrix} m\\ k \end{pmatrix} \int_E
d|\mu|_\varsigma^{m-k+1}(f)  \int_E d|\nu|_\varsigma^{k+1}(g)
\Lambda(f+g),
\end{gather*}
implying $\|(\{\mu,\nu\}_0)_\varsigma^m\|_1 \leq c
\sum\limits_{k=0}^m\left(\begin{smallmatrix} m\\ k
\end{smallmatrix}\right) \|\mu_\varsigma^{m-k+1}\|_1
\|\nu_\varsigma^{k+1}\|_1$. Thus
\begin{gather*}
\norm{  \{\mu,\nu\}_0  }_\varsigma^{n-1} = \sum_{m=0}^{n-1}
\|(\{\mu,\nu\}_0)_\varsigma^m\|_1 \leq c \sum_{m=0}^{n-1}
\sum_{k=0}^m \begin{pmatrix} m\\ k \end{pmatrix}
\|\mu_\varsigma^{m-k+1} \|_1\|\nu_\varsigma^{k+1}\|_1
\\
\phantom{\norm{  \{\mu,\nu\}_0  }_\varsigma^{n-1}}{} \leq c
\underbrace{ \sup\{ {  \textstyle \left(\begin{smallmatrix} n-1\\
k \end{smallmatrix}\right)  } \mid k=0,1,\ldots,n-1
    \}
           }_{\mbox{$=c_{n-1}$\ \
                     (def\/ined in equation~\eqref{WQB:eq:lemm:defi:c-n})}}
\norm{\mu}_\varsigma^n \norm{\nu}_\varsigma^n.
\end{gather*}
Thus $\{\mu,\nu\}_0\in \cM_\varsigma^{n-1}(E)$, whenever
$\mu,\nu\in \cM_\varsigma^{n}(E)$.

We now turn to the joint continuity
in~\eqref{WQ:9-5:eq:PoissonBracket-M(E)-2}. We demonstrate here
only the case $n=1$, for $n\geq2$ one may proceed similarly as in
the previous argumentations. Let $\mu,\nu\in \cM_\varsigma^1(E)$.
Then
\begin{gather*}
\betr{  \klr{  \{\mu,\nu\}_\hbar  -  \{\mu,\nu\}_0  }\!(\Lambda)
} \leq
\betr{  \{\mu,\nu\}_\hbar  -  \{\mu,\nu\}_0  }\!(\Lambda)   \\
\qquad{} \leq \int_E d|\mu|(f) \int_E d|\nu|(g) \betr{
\tfrac{i}{\hbar} \klr{\ex{-\tfrac{i}{2}\hbar\sigma(f,g)}-
     \ex{\tfrac{i}{2}\hbar\sigma(f,g)}
     }       - \sigma(f,g)
     } \Lambda(f+g).
\end{gather*}
Inserting $\Lambda=E$ f\/inally gives
$\lim\limits_{\hbar\rightarrow0}\norm{\{\mu,\nu\}_\hbar-\{\mu,\nu\}_0}_1=0$
by use of the dif\/ferential quotient limit
$\lim\limits_{\hbar\rightarrow0}\textstyle\frac{i}{\hbar}
\klr{\ex{-\tfrac{i}{2}\hbar\sigma(f,g)}-
     \ex{\tfrac{i}{2}\hbar\sigma(f,g)}
     }
=\sigma(f,g)$. Note that by the mean value theorem of
dif\/ferential calculus with one real variable we obtain a bound
by
\begin{gather*}
\betr{ \frac{\ex{\pm\tfrac{i}{2}\hbar\sigma(f,g)}-1}{\hbar}
      }
\leq \tfrac12\betr{\sigma(f,g)} \leq
\frac{c}{2}\:\varsigma(f)\,\varsigma(g) ,\qquad\forall\, 0\neq
\hbar\in\dR,\qquad\forall \, f,g\in E,
\end{gather*}
which allows application of Lebesgue's dominated convergence
theorem. Consequently,
\begin{gather*}
\norm{\{\mu,\nu\}_\hbar - \{\mu',\nu'\}_0}_1 \leq
\norm{\{\mu,\nu\}_\hbar - \{\mu,\nu\}_0}_1
+\norm{\{\mu-\mu',\nu\}_0}_1
+\norm{\{\mu',\nu-\nu'\}_0}_1 \\
\phantom{\norm{\{\mu,\nu\}_\hbar - \{\mu',\nu'\}_0}_1}{} \leq
\norm{\{\mu,\nu\}_\hbar - \{\mu,\nu\}_0}_1
+\norm{(\mu-\mu')^1_\varsigma}_1 \norm{\nu^1_\varsigma}_1
+\bigl\|{\mu'}^1_\varsigma\bigr\|_1
\norm{(\nu-\nu')^1_\varsigma}_1,
\end{gather*}
implying the joint continuity at $\hbar=0$. Now the rest is easily
shown.
\end{proof}

Of course, one obtains a larger Poisson algebra than
$\cM^\infty_\varsigma(E)$, if one f\/inds a semi-norm
$\varsigma'$, which is smaller than $\varsigma$ but also
satisf\/ies the estimate~\eqref{WQ:eq:9-5:semi-norm-sigma}.

\subsection[Strict deformation quantization of Banach-$*$-type]{Strict deformation quantization of Banach-$\boldsymbol{*}$-type}
\label{WQ-s5.22xx}

With our previous results we arrive at a version of strict
deformation quantization, which is formulated in terms of our
measure Banach-$*$-algebras $(\cM(E),\star_{\!\hbar})$, instead of
$C^*$-algebras as in Def\/inition~\ref{WQ:defi:9-5:SQ}. In some
sense, the basic idea of deformation quantization is realized in
this version more exactly than in the $C^*$-algebraic manner: One
has the same mathematical quantities for observables in the
classical and quantum regime. One has even the same topology in
both cases, given by the $\norm{\cdot}_1$-norm, what makes the
classical limit easier. The classical product is now the usual
convolution and the quantum product is the twisted convolution.

\begin{theorem}[Functorial Banach-$*$-strict deformation quantization]
\label{WQ:9-5:coro:Neumann-Dirac-M(E)}
\mbox{ }\\
Theorem~{\rm \ref{WQ:9-5:theo:Banach-hbar-algebra}} implies
\textbf{von Neumann's condition}
\begin{gather*}
\lim\limits_{\hbar\rightarrow 0}
\norm{\mu\star_{\!\hbar}\nu-\mu\star_{\!0}\nu}_1=0, \qquad\forall
\, \mu,\nu\in \cM(E).
\end{gather*}
From Theorem~{\rm \ref{WQ:9-5:theo:PoissonBracket-M(E)}} follows
\textbf{Dirac's condition}
\begin{gather*}
\lim\limits_{\hbar\rightarrow0} \norm{  \{\mu,\nu\}_\hbar  -
\{\mu,\nu\}_0  }_1   =   0, \qquad\forall \,\mu,\nu\in
\cM_\varsigma^1(E).
\end{gather*}
\textbf{Rieffel's condition} is trivially fulfilled, since on all
the Banach-$*$-algebras $(\cM(E),\star_{\!\hbar})$, $\hbar\in\dR$,
we have the same norm, namely the total variation norm
$\norm{\cdot}_1$.

Now select an arbitrary Poisson algebra
$(\cP,\star_{\!0},\{\cdot,\cdot\}_0)$ from Theorem~{\rm
\ref{WQ:9-5:theo:PoissonBracket-M(E)}} and define for each
$\hbar\in\dR$ the identical (thus injective) embedding
\begin{gather*}
Q_\hbar^B:\cP\longrightarrow(\cM(E),\star_{\!\hbar}),
\qquad\mu\longmapsto\mu,
\end{gather*}
as quantization map. Then the image $Q_\hbar^B(\cP)$ is a
sub-$*$-algebra of $(\cM(E),\star_{\!\hbar})$ for every
$\hbar\in\dR$.

Summarizing it follows that $(Q_\hbar^B)_{\hbar\in \dR}$
constitutes a Banach-$*$-algebra version of strict deformation
quantization, which depends functorially on the pre-symplectic
space $(E,\sigma)$, provided the semi-norms are chosen, which
characterize the enlarged measure  Poisson algebras.
\end{theorem}

\section{Poisson algebras of phase space functions}
\label{WQ-s5xt}

By means of Fourier transformation we are going to realize the
measure Poisson algebras $(\cP,\star_{\!0},\{\cdot,\cdot\}_0)$
from Subsection~\ref{WQ-s5.2} (see especially
Theorem~\ref{WQ:9-5:theo:PoissonBracket-M(E)}) in terms of
functions on $E_\tau'$, the $\tau$-dual of $E$. For
$\tau=\tau_{\text{il}}$ it coincides with the space $E'$ of all
$\dR$-linear functionals on $E$.

\subsection{Fourier transformation of the measure algebras}
\label{WQ-s5.3}

\begin{lemma}
\label{WQ:lemm:tau-dual} The partial ordering of being ``finer
than'' of the locally convex topologies on $E$ carries over to the
partial ordering of the dual spaces given by inclusion.
Furthermore, if $\rho\leq\tau$, then $E_\rho'$ is dense in
$E_\tau'$ with respect to the $\sigma(E_\tau',E)$-topology.
\end{lemma}

\begin{proof}
The $\sigma(E_\tau',E)$-dual of $E_\tau'$ is $E$ itself, and thus
consists of the functionals $F\mapsto F(f)$, $f\in E$. Since
$\rho$ is Hausdorf\/f, it holds that $F(g)=0$ for all $F\in
E_\rho'$ implies $g=0$. So the ``annihilator'' of $E_\rho'$
vanishes, and thus $E_\rho'$ is dense in $E_\tau'$ according to
the Hahn--Banach
theorem \cite[Corollary~IV.3.14]{Conway85}. 
\end{proof}
For every locally convex $\tau$ the Fourier transformation $\dF$,
acting on all f\/inite Borel measures $M^\tau(E)$, is def\/ined
\begin{gather}
\label{WQ:eq:9-5:FourierTransForm} \dF\mu[F]\equiv\wh{\mu}[F]:=
\int_E d\mu(f)\,\ex{iF(f)},\qquad F\in E_\tau',\qquad \mu\in
M^\tau(E).
\end{gather}
In terms of the Weyl functions $W_c(f)[F]=\ex{iF(f)}$, $f \in E$,
the Fourier transformation writes
\begin{gather*}
\wh{\mu}=\int_E d\mu(f)W_c(f), \qquad \mu\in M^\tau(E),
\end{gather*}
and may now be read as an integral over phase space functions.

For $\alpha\in I$ a measure $\mu\in\cM(E_\alpha)$ is concentrated
on $E_\alpha$, and hence its Fourier transform satisf\/ies
$\wh{\mu}[F]=\wh{\mu}[G]$ for all $F,G\in E_\tau'$ with
$F(f)=G(f)$ $\forall f\in E_\alpha$. This is just the
identif\/ication of the dual $E_\alpha'$ of $E_\alpha$ with
classes in $E_\tau'$. In this sense we identify subsequently
$E_\alpha'$ with a part of $E_\tau'$.

We denote the Fourier image of a set of measures by a hat,
indicating by $(E_\tau')$ the independent variables. Thus we write
e.g. $\wh{M}(E_\tau'):=\dF M^\tau(E)$. We know
$\wh{M}(E_\tau')\subseteq \operatorname{C}_b(E_\tau')$. For the
point measures one has $\wh{\delta(f)}=W_c(f)$. Thus we obtain the
following subspace structure for some Fourier transformed measure
spaces
\begin{gather*}
\wh{\cM}(E_\tau')_{df} \subset \wh{\cM}(E_\tau')_{d} \subset
\operatorname{AP}(E_\tau') \subset\operatorname{C}_b(E_\tau'),
\end{gather*}
where the $*$-algebra $\wh{\cM}(E_\tau')_{df}=\Delta(E_\tau')$ of
trigonometric functions and the $C^*$-algebra of all almost
periodic functions $\operatorname{AP}(E_\tau')$ are introduced in
Subsection~\ref{WQ-s2.4}. They are independent of $\tau$ as
algebras. In the set of special almost periodic functions
$\wh{\cM}(E_\tau')_{d}$, one can consider the
$\norm{\cdot}_1$-norm, making it $*$-isomorphic to the
$\tau$-independent Banach-$*$-algebra $\ovl{\Delta(E, \hbar
\sigma)}^1$ of Subsection~\ref{WQ-s2.2}.

Since the Fourier transforms are $\sigma(E_\tau',E)$-continuous
functions, we conclude from the above Lemma that the sup-norm may
be evaluated on smaller dual spaces,
\begin{gather}
\label{WQ:eq:9-5:sup-norm-2} \|\wh{\mu}\|_0 =
\sup\{|\wh{\mu}[F]|\mid F\in E_\tau'\} = \sup\{|\wh{\mu}[F]|\mid
F\in E_\rho'\}, \qquad\forall\, \mu\in M^\tau(E),
\end{gather}
where $\rho$ is any locally convex topology on $E$ with
$\rho\leq\tau$.

Because of the mentioned universality of $\cM(E)$ we have
\begin{gather}
\label{extchain} \wh{\cM}(E_\tau')_{d} \subset
\dF\cM(E)=\wh{\cM}(E_\tau')\subset \operatorname{C}_b(E_\tau'),
\end{gather}
for every locally convex topology $\tau$ on $E$. This means that
the extension of the considered phase space functions from
$\wh{\cM}(E_\tau')_{d}$ to $\wh{\cM}(E_\tau')$ depends
functorially on $E$. A concrete topology~$\tau$ comes  into play
only if the measures $\mu\in M^\tau(E)$ are considered which are
not in $\cM(E)$, respectively, which possess Fourier transforms
$\wh{\mu}\in\wh{M}(E_\tau')$ not contained in $\wh{\cM}(E_\tau')$.

If we use in $\wh{M}(E_\tau')$ the $*$-operation of pointwise
complex conjugation inherited from $\operatorname{C}_b(E_\tau')$
(see equation~\eqref{WQ:eq:9-5:*-algebraic-operations-F}), then
the Fourier transformation $\dF$ is $*$-preserving, that is,
$\wh{\mu^*\:}\!\!={\wh{\mu}}^*$ for all $\mu\in M^\tau(E)$. The
Fourier transformation of the deformed resp.~twisted convolution
$\star_{\!\hbar}$ from
equation~\eqref{WQ:eq:9-5:deformed-convolution} leads to the
\emph{deformed product} $\cdot_{\!\hbar}$ on $\wh{\cM}(E_\tau')$,
\begin{gather}
\label{WQ:eq:9-5:deformedproducts}
\wh{\mu}\cdot_{\!\hbar}\wh{\nu}:=\wh{\mu\star_{\!\hbar}\nu},
\qquad\forall \, \mu,\nu\in \cM(E),\qquad \hbar\in\dR.
\end{gather}
This agrees with the deformed product from
equation~\eqref{WQ:eq:9-5:DeformedProduct} arising from a strict
deformation quantization (cf.~also Subsection~\ref{WQ-s1mot}). By
construction it follows for each $\hbar\in\dR$ that $\dF$ is a $*$-isomorphism from the Banach-$*$-algebra
$(\cM(E),\star_{\!\hbar})$ onto the $*$-algebra
$(\wh{\cM}(E_\tau'),\cdot_{\!\hbar})$ consisting of functions.

The Weyl relations \eqref{WQ:eq:9-5:hbar-Weyl-rel-delta(f)} for
the point measures $\delta(f)\equiv W^\hbar(f)$ lead by Fourier
transformation to the deformed $\cdot$-product for the periodic
Weyl functions $W_c(f)$
\begin{gather}
\label{WQ:eq:9-5:hbar-Weyl-rel-xi(f)} W_c(f)\cdot_{\!\hbar}W_c(g)
=\ex{-\tfrac{i}{2}\hbar\sigma(f,g)}W_c(f+g), \qquad
W_c(f)^*=W_c(-f), \qquad \forall \, f,g\in E.\!\!
\end{gather}

In case of $\hbar=0$ equation~\eqref{WQ:eq:9-5:deformedproducts}
reduces to the commutative pointwise product of functions. We see
that $\dF$ is a non-surjective, but injective $*$-homomorphism
from the commutative Banach-$*$-algebra $(\cM(E),\star_{\!0})$
into the commutative $C^*$-algebra
$(\operatorname{C}_b(E_\tau'),\cdot_{\!0})$. Especially,
$(\wh{\cM}(E)_{df},\cdot_{\!0})$
resp.~$(\wh{\cM}(E)_{d},\cdot_{\!0})$ are $\norm{\cdot}_0$-dense
sub-$*$-algebras of the $C^*$-algebra
$(\operatorname{AP}(E_\tau'),\cdot_{\!0})$ of the almost periodic
functions.

\subsection{Pre-symplectic geometry revisited}
\label{WQ-s5.4}

We assume in the present Subsection a f\/ixed locally convex
topology $\tau$ on $E$, for which a $\tau$-continuous semi-norm
$\varsigma$, satisfying \eqref{WQ:eq:9-5:semi-norm-sigma}, exists.
Thus, in contradistinction to Subsection~\ref{WQ-s2.4}, $\sigma$
is now jointly $\tau$-continuous. We denote again the phase space
manifold $E_\tau'$ by $\mathsf{P}$. We modify, however, the
notions of tangent and cotangent spaces, which is suggested from
our $\tau$-continuous semi-norm $\varsigma$ on $E$. Let $f\mapsto
[f]$ be the quotient map to form the quotient $E/\ker(\varsigma)$
with respect to the kernel $\ker(\varsigma):=\{f\in E\mid
\varsigma(f)=0\}$. Then $\norm{[f]}_\varsigma:=\varsigma(f)$
def\/ines a norm on $E/\ker(\varsigma)$, the completion of which
is denoted by $E_\varsigma$. Because
of~\eqref{WQ:eq:9-5:semi-norm-sigma} the pre-symplectic form
$\sigma$ extends $\norm{\cdot}_\varsigma$-continuously to
$E_\varsigma$ with the estimation
\begin{gather}
\label{WQ:eq:9-5:semi-norm-sigma-2}
\betr{\sigma_\varsigma(\phi,\psi)}\leq c \norm{\phi}_\varsigma
\norm{\psi}_\varsigma, \qquad\forall \,\phi,\psi\in E_\varsigma,
\end{gather}
where we have set $\sigma_\varsigma([f],[g]):=\sigma(f,g)$ for all
$f,g\in E$.

Let $E_\varsigma'$ consist of all $\dR$-linear functionals
$G:E\rightarrow\dR$ for which there exists a constant $k\geq0$
(depending on $G$) with $\betr{G(f)}\leq k \,\varsigma(f)$
$\forall \, f\in E$. By construction, $E_\varsigma'\subseteq
E_\tau'$. Since $\ker(G)\supseteq\ker(\varsigma)$ for all $G\in
E_\varsigma'$ it follows that $E_\varsigma'$ is the topological
dual of $E_\varsigma$ with respect to the norm
$\norm{\cdot}_\varsigma$, identifying $G([f])$ with $G(f)$ for all
$f\in E$.

In order to allow in
\begin{gather}
\label{WQ:eq:9-5:differential-G-direction} d_F A(G):=\left.\frac{d
A[c(t)]}{dt}\right|_{t=0}
\end{gather}
the dif\/ferentiation only in the directions $G$ from
$E_\varsigma'$, we restrict at each $F\in \mathsf{P}$ the tangent
space to
\begin{gather*}
{}_\varsigma\!T_F \mathsf{P}:=E_\varsigma'\;\subseteq E_\tau'.
\end{gather*}
Considering on $E_\varsigma'$ the $\sigma(E_\varsigma',
E_\varsigma)$-topology, we obtain the cotangent space
\begin{gather*}
{}_\varsigma\!T_F^* \mathsf{P}:=(E_\varsigma')'=E_\varsigma
\;\supseteq \{[f]\mid f\in E\}.
\end{gather*}

In generalization of equation~\eqref{eq:9-5:Poisson-bracket-sigma}
we def\/ine a constant Poisson tensor f\/ield $\Sigma$ on the
cotangent bundle ${}_\varsigma\!T^*\mathsf{P}$ by
$\Sigma_F:=-\sigma_\varsigma:E_\varsigma\times
E_\varsigma\rightarrow\dR$ for every $F\in \mathsf{P}$, which we
extend complex bilinearly to the complexif\/ied cotangent bundle
${}_\varsigma^\dC\!T^*\mathsf{P}$. The associated Poisson bracket
at the phase space point $F\in \mathsf{P}$ is canonically
introduced as
\begin{gather}
\label{WQ:eq:9-5:Poisson-bracket-sigma} \{A,B\}[F]
:=\Sigma_F(d_F A,d_F B)   \\
\phantom{\{A,B\}[F]}{} = -\sigma_\varsigma(d_F A_1,d_F
B_1)-i\sigma_\varsigma(d_F A_1,d_F B_2) -i\sigma_\varsigma(d_F
A_2,d_F B_1)+\sigma_\varsigma(d_F A_2,d_F B_2),\nonumber
\end{gather}
where $A$ and $B$ are $\dC$-valued functions on $\mathsf{P}$,
which are dif\/ferentiable in all directions $G\in E_\varsigma'$
with $d_F A,d_F B\in{}_\varsigma^\dC\!T^*_F
\mathsf{P}=E_\varsigma+iE_\varsigma$ for all $F\in \mathsf{P}$.
The latter condition means for a function
$A:\mathsf{P}\rightarrow\dC$ that its total dif\/ferential
$\mathsf{P}\ni F\mapsto d_F A$ is continuous with respect to the
$\sigma(E_\varsigma, E_\varsigma')$-topology. But this weak
continuity does in general not ensure that $\mathsf{P}\ni
F\mapsto\{A,B\}[F]\in\dC$ is continuous. But, in virtue of
\eqref{WQ:eq:9-5:semi-norm-sigma-2}, the Poisson bracket
continuity is here automatically fulf\/illed, if $\mathsf{P}\ni
F\mapsto d_F A$ and $\mathsf{P}\ni F\mapsto d_F B$ are strongly
continuous, i.e., with respect to the norm
$\norm{\cdot}_\varsigma$ (and so in the specif\/ied
weak$*$-topology on $\mathsf{P}$).

For the Weyl functions $W_c(f):\mathsf{P}\rightarrow\dC$ from
equation~\eqref{WQ:eq:9-5:xi(f)} we obtain with
def\/inition~\eqref{WQ:eq:9-5:differential-G-direction} $d_F
W_c(f)(G)=i G(f)W_c(f)[F]$. If we restrict dif\/ferentiation to
the directions from $E_\varsigma'$, then the identif\/ication
$G([f])\equiv G(f)$ leads to
\begin{gather}
\label{WQ:eq:9-5:totdiff-xi(f)} d_F
W_c(f)=i\ex{iF(f)}[f]=iW_c(f)[F]\, [f] \in
{}_\varsigma^\dC\!T_F^*\mathsf{P}=E_\varsigma'+ iE_\varsigma',
\qquad \forall \, F\in \mathsf{P},
\end{gather}
which deviates from equation~\eqref{WQ:eq:9-5:totdiff-xi(f)-o}.
(Note that $E_\varsigma'$ separates points not only on
$E_\varsigma$, but also on $E$, if and only if $\varsigma$ is a
proper norm and not only a semi-norm). Inserting into
equation~\eqref{WQ:eq:9-5:Poisson-bracket-sigma} and noting
$\sigma_\varsigma([f],[g])=\sigma(f,g)$ yields
\begin{gather}
\label{WQ:eq:9-5:Poisson-bracket-xi(f)} \{W_c(f),W_c(g)\}
=\sigma(f,g)W_c(f+g) =\sigma_\varsigma([f],[g])W_c(f+g),
\qquad\forall \, f,g\in E,
\end{gather}
which coincides with the former Poisson bracket formula of
equation~\eqref{eq:9-5:Poisson-bracket-xi(f)}.

If the Poisson bracket $\{\cdot,\cdot\}$
of~\eqref{WQ:eq:9-5:Poisson-bracket-sigma} would commute with the
Fourier integration $\wh{\mu}=\int_E d\mu(f)W_c(f)$, then we would
obtain from~\eqref{WQ:eq:9-5:Poisson-bracket-xi(f)}
\begin{gather}
\{\wh{\mu},\wh{\nu}\}[F] =
\int_E d\mu(f)  \int_E d\nu(g)\: \{W_c(f),W_c(g)\}[F] \nonumber\\
\phantom{\{\wh{\mu},\wh{\nu}\}[F]}{} = \int_E d\mu(f)  \int_E
d\nu(g) \:\sigma(f,g)\ex{iF(f+g)}
=\dF\{\mu,\nu\}_0[F],\label{WQ:eq:9-5:FT-PB0}
\end{gather}
where $\{\cdot,\cdot\}_0$ is the Poisson bracket for the measures
from equation~\eqref{WQ:9-5:eq:PoissonBracket-M(E)}. This suggests
to take as Poisson algebras the Fourier transforms of the spaces
$\cP$ from Theorem~\ref{WQ:9-5:theo:PoissonBracket-M(E)}. 
\begin{theorem}
\label{WQ:9-5:theo:PA-B} Let $n\in\dN\bigcup\{\infty\}$. Then
$\wh{\mu}\in \wh{\cM}_\varsigma^n(E_\tau')$ is $n$-times
continuously differentiable in all the directions of
$E_\varsigma'$, and moreover, for each $F\in \mathsf{P}\equiv
E_\tau'$ its total differential
$d_F\wh{\mu}\in{}_\varsigma^\dC\!T_F^*\mathsf{P}$ is given by the
Bochner integral
\begin{gather}
\label{WQ:9-5:eq:PA-B-1} d_F\wh{\mu}= i \int_E d\mu(f) \ex{iF(f)}
[f] = \int_E d\mu(f)\, d_FW_c(f),
\end{gather}
which converges with respect to the norm $\norm{\cdot}_\varsigma$
on the Banach space
${}_\varsigma^\dC\!T_F^*\mathsf{P}=E_\varsigma+iE_\varsigma$.

The Fourier transformation $\dF$ is a Poisson automorphism
transforming the measure Poisson bracket $\{\cdot,\cdot\}_0$ from
equation~\eqref{WQ:9-5:eq:PoissonBracket-M(E)} into the function
Poisson bracket $\{\cdot,\cdot\}$ from
equation~\eqref{WQ:eq:9-5:Poisson-bracket-sigma}, that is,
\begin{gather}
\label{WQ:9-5:eq:PA-B-Fou} \{\wh{\mu},\wh{\nu}\}=
\dF\{\mu,\nu\}_0, \qquad\forall \, \mu,\nu\in \cM_\varsigma^1(E).
\end{gather}

Let $\wh{\cP}:=\dF\cP$, with the choices for $\cP$ as in
Theorem~{\rm \ref{WQ:9-5:theo:PoissonBracket-M(E)}}. Then
$(\wh{\cP},\cdot_{\!0},\{\cdot,\cdot\})$ constitutes a Poisson
algebra contained in $\wh{\cM}_\varsigma^\infty(E_\tau')$, which
in addition is invariant under all deformed products
$\cdot_{\!\hbar}$, $\hbar\in\dR$. Especially
$(\wh{\cM}_\varsigma^\infty(E_\tau'),\cdot_{\!0},\{\cdot,\cdot\})$
is a Poisson algebra of this type.
\end{theorem}

\begin{proof}
For $\mu\in \cM^n_\varsigma(E)$ and $G\in E_\varsigma'$ let us
def\/ine the measure $\mu_G$ by $d\mu_G(f)=G(f)d\mu(f)$. Since
$\betr{G(f)}=\betr{G([f])}\leq
\norm{G}_\varsigma\norm{[f]}_\varsigma =\norm{G}_\varsigma
\varsigma(f)$ $\forall\, f\in E$, we conclude that
$\betr{\mu_G}\leq\norm{G}_\varsigma |\mu|^1_\varsigma$, and hence
$\mu_G\in \cM^{n-1}_\varsigma(E)$. Consequently, we may
dif\/ferentiate according to
equation~\eqref{WQ:eq:9-5:differential-G-direction} as
\begin{gather*}
d_F\wh{\mu}(G) = \left.\frac{d \wh{\mu}[F+tG]}{dt}\right|_{t=0} =
\left.\frac{d}{dt}\klr{ \int_E d\mu(f)\ex{iF(f)+itG(f)}
}\right|_{t=0}
\\
\phantom{d_F\wh{\mu}(G)}{} = i \int_E d\mu(f)\, G(f) \ex{iF(f)} =
\wh{\mu_G}[F],
\end{gather*}
implying $\mathsf{P}\ni F\mapsto d_F\wh{\mu}(G)$ to be an element
of $\wh{\cM}^{n-1}_\varsigma(\mathsf{P})$, especially being
continuous. Iteration shows that $\wh{\mu}$ is $n$-times
continuously dif\/ferentiable in all the directions of
$E_\varsigma'$. Now $G([f])=G(f)$ yields~\eqref{WQ:9-5:eq:PA-B-1}.
Note, for the Bochner integral it is necessary that the range of
the weak measurable quotient map $\operatorname{supp}(\mu)\ni
f\mapsto [f]$
has to be separable~\cite[p.~350]{Cohn80}, 
which is ensured because $\mu$ is approximable by a sequence of
measures on f\/inite dimensional subspaces of $E$.
Inserting~\eqref{WQ:9-5:eq:PA-B-1} into the Poisson bracket from
equation~\eqref{WQ:eq:9-5:Poisson-bracket-sigma} makes our
consideration in equation~\eqref{WQ:eq:9-5:FT-PB0} rigorous, and
thus yields $\{\wh{\mu},\wh{\nu}\}=\dF\{\mu,\nu\}_0$. The
assertions concerning the Poisson algebras $\wh{\cP}$ are
immediate.
\end{proof}

Especially, with the identif\/ication of
$W^\hbar(f)\equiv\delta(f)$ of
Proposition~\ref{WQB:9-5:obse:Banach-hbar-discrete}, the Poisson
isomorphism in Corollary~\ref{WQ:9-5:coro:Poisson*iso-hbar=0}
agrees just with the Fourier transformation. This case is covered
by the measure Poisson algebra
\ref{WQ:9-5:theo:PoissonBracket-M(E)-d} of
Theorem~\ref{WQ:9-5:theo:PoissonBracket-M(E)}.

\section[Representations, folia, and enveloping $C^*$-algebras]{Representations, folia, and enveloping $\boldsymbol{C^*}$-algebras}
\label{WQ-s4}

The present Section makes representation theory available for the
generalized Weyl quantization in Section~\ref{WQ-s6}. Recall that
we assumed our pre-symplectic space $(E,\sigma)$ inf\/inite
dimensional.

\subsection{Folia and quasi-equivalence classes of representations}
\label{WQ-s4.0}

Let us sketch the connection between folia and representations, as
it is
treated e.g.\ in \cite{HaagKadisonKastler70,Sewell73,Pedersen79, 
KadisonRingrose}; an overview is given in~\cite{Hon96a}.

A folium $\cF$ of a $C^*$-algebra $\cA$ is a specif\/ic face of
the state space $\cS(\cA)$: it is a norm-closed, convex subset of
$\cS(\cA)$, which is invariant under so-called small
perturbations, meaning that $\omega\in \cF$ implies
$\omega_B\in\cF$ for all $B\in\cA$ with
$\dual{\omega}{B^*B}\neq0$. Here $\omega_B$ is the state on $\cA$
given by $\dual{\omega_B}{\cdot}=\dual{\omega}{B^*\cdot B}
\dual{\omega}{B^*B}^{-1}$. The collection of all folia of $\cA$ is
denoted by $\operatorname{fol}(\cA)$.

Two non-degenerate representations $\Pi_1$ and $\Pi_2$ of the
$C^*$-algebra $\cA$ are quasi-equivalent (see e.g.~\cite[Section~2.4.4]{BratteliRobinson1}), if and only if there
exists a $*$-isomorphism $\alpha$ from the von Neumann algebra
$\ovl{\Pi_1(\cA)}^w$ onto the von Neumann algebra
$\ovl{\Pi_2(\cA)}^w$ with $\alpha(\Pi_1(A))=\Pi_2(A)$, $\forall\,
A\in\cA$. The upper index ``$w$'' indicates closure with respect
to the $\sigma$-weak operator topology of the representation
Hilbert space. We denote by $\operatorname{rep}(\cA)$ the set of
all quasi-equivalence classes of representations. So, writing
subsequently $\Pi\in\operatorname{rep}(\cA)$ we denote a
quasi-equivalence class of representations, and $\ovl{\Pi(\cA)}^w$
indicates the associated W$*$-algebra, which is abstracted from
the $*$-isomorphic von Neumann algebras given by the
representations in the quasi-equivalence class.

There is a one-to-one correspondence $\cF\mapsto\Pi_\cF$ from
$\operatorname{fol}(\cA)$ onto $\operatorname{rep}(\cA)$, which
preserves the partial order relations
\begin{gather}
\label{WQB:eq:ordering:F-Pi} \cF_1\subseteq\cF_2\subseteq\cS(\cA)
\quad \mapsto \quad \Pi_{\cF_1}\leq\Pi_{\cF_2}\leq\Pi_u, \qquad
\cF_1, \cF_2 \in \operatorname{fol}(\cA).
\end{gather}
The partial ordering in the family of folia
$\operatorname{fol}(\cA)$ is the set inclusion, and ``$\leq$''
means the partial ordering of being a subrepresentation, up to
quasi-equivalence. The quasi-equivalence class
$\Pi_\cF\in\operatorname{rep}(\cA)$, corresponding to a given $\cF
\in \operatorname{fol}(\cA)$, may be constructively determined by
means of the special representative obtained as the direct sum of
GNS representations over all states in~$\cF$.

$\Pi_u$ denotes the universal representation of the $C^*$-algebra.
For $\Pi_{\cF_1}\leq\Pi_{\cF_2}$ there exists a~central projection
$P\in\ovl{\Pi_{\cF_2}(\cA)}^w$ such that
$\ovl{\Pi_{\cF_1}(\cA)}^w=\ovl{\Pi_{\cF_2}(\cA)}^w P$, and we
sometimes identify~$\Pi_{\cF_1}(A)$  with~$\Pi_{\cF_1}(A)P$,
for~$A\in\cA$.

In the mentioned correspondence, the smallest folium $\cF_\omega$
containing the state $\omega$ corresponds to the equivalence class
of its GNS representation $\Pi_\omega$, that is, $\Pi_\omega$ is a
representative of the class
$\Pi_{\cF_\omega}\in\operatorname{rep}(\cA)$.
$\cF\in\operatorname{fol}(\cA)$ consists just of the
$\Pi_\cF$-normal states of $\cA$, from which results the Banach
space duality $\ovl{\Pi_\cF(\cA)}^w=\mathrm{LH}(\cF)^*$. Generally
we do not distinguish notationally between $\omega\in\cF$ as a
state on $\cA$ and its unique normal extension $\omega$ to the
W$*$-algebra $\ovl{\Pi_\cF(\cA)}^w$ (writing $\dual{\omega}{A}$
for $A\in\cA$ as well as for $A\in \ovl{\Pi_\cF(\cA)}^w$).

If $\cA=C^*(\cB)$ is the enveloping $C^*$-algebra of a
Banach-$*$-algebra $\cB$ (see e.g. \cite{Dixmier77}), then $\cA$
and $\cB$ have the same representations given by continuous
extensions resp.~restrictions, in which case we sometimes use the
notation $\operatorname{fol}(\cB):=\operatorname{fol}(\cA)$ and
$\operatorname{rep}(\cB):=\operatorname{rep}(\cA)$. Throughout the
present work we do not distinguish notationally between
representations of $\cB$ and their unique continuous extension to
the enveloping $C^*$-algebra $\cA$.

\subsection[$\tau$-continuous representations of the Weyl algebra]{$\boldsymbol{\tau}$-continuous representations of the Weyl algebra}
\label{WQ-s4.1}

We denote by $\cT(E,\sigma)$ the set of all topologies $\tau$ on
$E$ such that for each $f\in E$ the maps $E\ni g\mapsto f+g\in E$
and $E\ni g\mapsto\sigma(f,g)\in\dR$ are $\tau$-continuous, what
means just the separate $\tau$-continuity of the addition and of
the antisymmetric form $\sigma$. The set $\cT(E,\sigma)$ of
topologies is directed with respect to the natural partial
ordering ``f\/iner than''. So, each vector space topology on $E$,
for which $\sigma$ is separately continuous, is an element of
$\cT(E,\sigma)$.

Let $\tau\in\cT(E,\sigma)$. A state
$\omega\in\cS(\cW(E,\hbar\sigma))$ is called
\emph{$\tau$-continuous}, if its \emph{characteristic function}
\begin{gather}
\label{WQB4:eq:defi:CF} E\ni
f\longmapsto\dual{\omega}{W^\hbar(f)}=:C_\omega(f)
\end{gather}
is $\tau$-continuous. A representation $\Pi_\hbar$ of
$\cW(E,\hbar\sigma)$ is denoted $\tau$-continuous, if $E\ni
f\mapsto \Pi_\hbar(W^\hbar(f))$ is continuous with respect to the
topology $\tau$ on $E$ and to any weak operator topology on the
image space. Recall that all of the weak operator topologies
(weak, strong, $\sigma$-strong*, \ldots) are equivalent on the
group of unitary operators. Clearly $\tau$-continuity carries over
to the whole quasi-equivalence class
$\Pi_\hbar\in\operatorname{rep}(\cW(E,\hbar\sigma))$, and hence
the mapping (see equation~\eqref{WQB:eq:pi-Pi})
\begin{gather}
\label{WQB:eq:pi-Pi-quasiequi} E\ni f\longmapsto
\Pi_\hbar(W^\hbar(f))=\pi_\hbar(f)\in
\ovl{\Pi_\hbar(\cW(E,\hbar\sigma))}^w
\end{gather}
is continuous with respect to $\tau$ on $E$ and to the
$\sigma$-strong (equivalently $\sigma$-weak) topology on the
W$*$-algebra $\ovl{\Pi_\hbar(\cW(E,\hbar\sigma))}^w$. If for
$\tau$ the addition is jointly continuous, then we may regard~$E$
as a topological group, and $\pi_\hbar$ may be viewed as a
quasi-equivalence class of $\tau$-continuous, projective, unitary
group representations.

Let us say a word about the existence of $\tau$-continuous states,
and, via GNS construction, about the existence of
$\tau$-continuous representations. By
\cite{BinzHonRie99b,BinzHonRie99c} the characteristic
functions~\eqref{WQB4:eq:defi:CF} of the states on
$\cW(E,\hbar\sigma)$ coincide with the convex set
$\cC(E,\hbar\sigma)$ of functions $C:E\rightarrow\dC$, which are
projective positive-def\/inite (i.e.
$\sum\limits_{i,j=1}^n\ovl{z_i}z_j
\ex{{\textstyle\frac{i}{2}}\hbar\sigma(f_i,f_j)}  C(f_j-f_i)\geq0$
for arbitrary $z_j\in\dC$, $f_j\in E$, $n\in\dN$), and are
normalized (i.e.\ $C(0)=1$). $\cC(E,\hbar\sigma)$ is always
non-empty, but a~priori it is not known, if it contains elements
which are $\tau$-continuous. Let us f\/irst consider the case
$\sigma\neq0$ and $\hbar\neq0$. It is well known that a Gaussian
function $E\ni f\mapsto\ex{-\frac14 s(f,f)}$ ($s$ is a positive
symmetric $\dR$-bilinear form on $E$) is an element of
$\cC(E,\hbar\sigma)$, if and only if $\hbar^2\sigma(f,g)^2\leq
s(f,f)\,s(g,g)$ for all $f,g\in E$. So, provided there exists such
an jointly $\tau$-continuous form $s$, there exists a
$\tau$-continuous state. For f\/inite dimensional $E$ such forms
$s$ are easily constructed. However, for inf\/inite dimensions
there may exist pathological (pre-) symplectic spaces
$(E,\sigma)$, for which no $\tau$-continuous form $s$, or even
worse, no $\tau$-continuous characteristic function exists at all,
excluding  $\tau$-continuous states. For such ``symplectic
pathology'' we refer to \cite{Robinson93}. For the standard
example of a symplectic space (see beginning of
Subsection~\ref{WQ-s2.2}) one may take, however, the
$\norm{\cdot}$-continuous form $s(f,g)=\betr{\hbar}\Re\skal{f}{g}$
in order to show the existence of $\tau$-continuous states for
every $\tau$ f\/iner than this norm.

Hence for our subsequent investigations, we may assume a
pre-symplectic space $(E,\sigma)$ for which $\cW(E,\hbar\sigma)$
has $\tau$-continuous states for many $\tau\in\cT(E,\sigma)$.
(This carries through to the quotient $E/\ker_\sigma$, which is
needed in the proof of Lemma~\ref{WQ:9-5:lemm:faithful-tau-cont}.)

For $\hbar=0$, where $\cC(E,0)$ consists of all
positive-def\/inite, normalized functions, the existence question
of $\tau$-continuous states is answered easily. Every character on
the vector group $E$ is an element of $\cC(E,0)$ and may be
interpreted as the characteristic function of a unique state on
the commutative Weyl algebra $\cW(E,0)$. Consequently, if $\tau$
is an arbitrary locally convex topology on $E$, then each
$\tau$-continuous character $E\ni f\mapsto\ex{iF(f)}$, where $F\in
E_\tau'$, def\/ines a $\tau$-continuous state.

It is shown in \cite{Hon97, Hon96a,Hon98} that for each topology
$\tau\in\cT(E,\sigma)$ the set of states
\begin{gather}
\label{WQ:eq:9-5:tau-states} \cF^\tau_\hbar=
\{\omega\in\cS(\cW(E,\hbar\sigma))\mid \text{$\omega$ is
$\tau$-continuous}\}
\end{gather}
constitutes a folium in the state space $\cS(\cW(E,\hbar\sigma))$.
The representation
$\Pi^\tau_\hbar\in\operatorname{rep}(\cW(E,\hbar\sigma))$,
associated with $\cF^\tau_\hbar$, is the largest $\tau$-continuous
representation of the Weyl algebra $\cW(E,\hbar\sigma)$, that is,
each $\tau$-continuous
$\Pi_\hbar\in\operatorname{rep}(\cW(E,\hbar\sigma))$ is a
subrepresentation of $\Pi_\hbar^\tau$. By
Lemma~\ref{WQ:9-5:lemm:faithful-tau-cont} of the Appendix,
$\Pi_\hbar^\tau$ is faithful.

The partial ordering on $\cT(E,\sigma)$ carries over to the
associated folia of $\cW(E,\hbar\sigma)$
\begin{gather}
\label{WQB:eq:ordering:tau-F} \tau_1\leq\tau_2\leq d
\quad\Longrightarrow\quad \cF^{\tau_1}_\hbar \subseteq
\cF^{\tau_2}_\hbar \subseteq \cF^d_\hbar =\cS(\cW(E,\hbar\sigma))
\end{gather}
($d$ indicates the discrete topology), which leads with
equation~\eqref{WQB:eq:ordering:F-Pi} to the partial ordering of
the associated (quasi-equivalence classes of) representations
according to their continuity properties.

\subsection{Folium of regular states of the Weyl algebra}
\label{WQ-s4.2}

A representation
$\Pi_\hbar\in\operatorname{rep}(\cW(E,\hbar\sigma))$ is called
\emph{regular}, if for each $f\in E$ the mapping $\dR\ni t\mapsto
\Pi_\hbar(W^\hbar(tf))$ is $\sigma$-strongly continuous in a
neighbourhood of the origin, or equivalently, if $E_\alpha\ni
f\mapsto \Pi_\hbar(W^\hbar(f))$ is $\sigma$-strongly continuous on
every f\/inite dimensional subspace $E_\alpha$ of $E$ with respect
to the unique vector space topology on $E_\alpha$. A state
$\omega$ on $\cW(E,\hbar\sigma)$ is termed \emph{regular}, if
$\dR\ni t\mapsto \dual{\omega}{W^\hbar(tf)}$ is continuous in a
neighbourhood of the origin for each $f\in E$. This is equivalent
to its GNS representation being regular
\cite[Section~5.2.3]{BratteliRobinson2}. The set of regular states
\begin{gather}
\label{WQ:eq:9-5:reg-states} \cF^{\text{reg}}_\hbar:=
\{\omega\in\cS(\cW(E,\hbar\sigma))\mid \text{$\omega$ is
regular}\}
\end{gather}
again constitutes a folium of $\cW(E,\hbar\sigma)$. The
representation associated with $\cF^{\text{reg}}_\hbar$, denoted
by
$\Pi^{\text{reg}}_\hbar\in\operatorname{rep}(\cW(E,\hbar\sigma))$,
is the largest regular representation of $\cW(E,\hbar\sigma)$.

Regularity suggests that $E$ carries a vector space topology which
is induced by its f\/inite dimensional subspaces, namely the
f\/inest locally convex topology $\tau_{\text{il}}$ (introduced at
the beginning of Section~\ref{WQ-s5}), which already has been
discussed in \cite{Segal58}. But for every locally convex topology
on $E$ its restriction to $E_\alpha$ just gives the natural unique
vector space topology on $E_\alpha$.

Since every linear map from $E$ into a locally convex space is
continuous with respect to $\tau_{\text{il}}$, it follows that the
pre-symplectic form $\sigma$ is at least separately
$\tau_{\text{il}}$-continuous, and consequently,
$\tau_{\text{il}}\in\cT(E,\sigma)$, what demonstrates the
existence of a locally convex element in $\cT(E,\sigma)$. In
general, $\sigma$ may not be jointly $\tau_{\text{il}}$-continuous
on $E\times E$, but only bounded and
hypocontinuous~\cite{Hoermann97}. In any case, the standard
example of a symplectic space, $\sigma=\Im\skal{\cdot}{\cdot}$, is
jointly $\tau_{\text{il}}$-continuous (since $\tau_{\text{il}}$ is
f\/iner than the norm topology arising from
$\skal{\cdot}{\cdot}$).

The folium of all $\tau_{\text{il}}$-continuous states on
$\cW(E,\hbar\sigma)$ is denoted by $\cF^{\text{il}}_\hbar$, and
the associated largest $\tau_{\text{il}}$-continuous
representation by
$\Pi^{\text{il}}_\hbar\in\operatorname{rep}(\cW(E,\hbar\sigma))$.
Since $\tau_{\text{il}}$ is the f\/inest locally convex topology
in $\cT(E,\sigma)$ we know the ordering relations
\begin{gather}
\label{WQB:eq:9-5:tau-reg-folia} \cF^\tau_\hbar\subseteq
\cF^{\text{il}}_\hbar\subseteq \cF^{\text{reg}}_\hbar
\quad\Longleftrightarrow\quad \Pi^\tau_\hbar\leq
\Pi^{\text{il}}_\hbar\leq \Pi^{\text{reg}}_\hbar,
\end{gather}
for each locally convex $\tau\in\cT(E,\sigma)$. Because the
characteristic functions~\eqref{WQB4:eq:defi:CF} are non-linear,
it is for inf\/inite dimensional $E$ not possible to show that
regularity implies $\tau_{\text{il}}$-continuity. In general one
knows that $\cF^{\text{il}}_\hbar$  is a proper subfolium of
$\cF^{\text{reg}}_\hbar$. In \cite{Hoermann97} an example of
a~test function space $(E,\sigma)$ is given, for which the two
folia coincide.

For inf\/inite dimensional $E$ there exists  a continuum of
inequivalent regular representations of~$\cW(E,\hbar\sigma)$ and,
for given locally convex $\tau$, one has even a continuum of
inequivalent $\tau$-continuous representations.

\subsection[Representations of the measure Banach-$*$-algebras]{Representations of the measure Banach-$\boldsymbol{*}$-algebras}
\label{WQ-s5.1xtra}

We f\/ix here an $\hbar\in\dR$ and consider a certain type of
Hilbert space representations of the Banach-$*$-algebra
$(\cM(E),\star_{\!\hbar})$. First, however, we restrict ourselves
to $(\cM(E_\alpha),\star_{\!\hbar})$ for the f\/inite dimensional,
and thus locally compact, subspaces $E_\alpha$ of $E$, referring
some facts known from the literature (e.g.~\cite{HewittRoss,
Pedersen79, Grundling97}).

There exists a 1:1:1:1-correspondence between the
$\sigma$-strongly continuous unitary representations
$(\pi_\hbar,\cH_\hbar)$ of the additive group $E_\alpha$, the
non-degenerate regular (=\,continuous) representations
$(\Pi_\hbar,\cH_\hbar)$ of $(\cM(E_\alpha)_d,\star_{\!\hbar})$
resp.~of $\cW(E_\alpha,\hbar\sigma)$, the non-degenerate
representations $(\Pi_\hbar,\cH_\hbar)$ of
$(\cM(E_\alpha)_a,\star_{\!\hbar})$, and those representations
$(\Pi_\hbar,\cH_\hbar)$ of $(\cM(E_\alpha),\star_{\!\hbar})$ for
which $\Pi_\hbar(\cM(E_\alpha)_a)\cH_\hbar$ is dense in the
representation Hilbert space $\cH_\hbar$. This correspondence is
given by $\Pi_\hbar(\mu)=\int_{E_\alpha} d\mu(f)\, \pi_\hbar(f)$,
$\mu\in\cM(E_\alpha)$. It preserves irreducibility. Such a
$(\Pi_\hbar,\cH_\hbar)$ is called \emph{regular} in the
literature, since it extends the original regular representation
of $(\cM(E_\alpha)_d,\star_{\!\hbar})$ resp.~of
$\cW(E_\alpha,\hbar\sigma)$ to a representation of the larger
algebra $(\cM(E_\alpha),\star_{\!\hbar})$
resp.~$C^*(\cM(E_\alpha),\star_{\!\hbar})$.

Since a regular representation $(\Pi_\hbar,\cH_\hbar)$ of
$(\cM(E_\alpha),\star_{\!\hbar})$ acts non-degenerately on the
closed $*$-ideal of the absolutely continuous measures
$\cM(E_\alpha)_a$, one has
\begin{gather*}
\norm{\Pi_\hbar(\mu)} = \sup\{
\norm{\Pi_\hbar(\mu\star_{\!\hbar}\gamma)}
        \mid \gamma\in\cM(E_\alpha)_a,\;\norm{\gamma}_1\leq1  \}
=
\lim_{n\rightarrow\infty}\norm{\Pi_\hbar(\mu\star_{\!\hbar}\gamma_n)},
\end{gather*}
where $(\gamma_n)_{n\in\dN}$, with $\norm{\gamma_n}_1=1$, is an
approximate identity for $(\cM(E_\alpha)_a,\star_{\!\hbar})$
(because
$\Pi_\hbar(\nu)\xi=\lim_n\Pi_\hbar(\gamma_n)\Pi_\hbar(\nu)\xi$ for
$\nu\in \cM(E_\alpha)_a$ and $\xi\in\cH_\Pi$, and
$\Pi_\hbar(\cM(E_\alpha)_a)\cH_\Pi$ is dense in $\cH_\Pi$, where
$\norm{\Pi_\hbar(\mu)}=\sup_{\norm{\eta}=1}\norm{\Pi_\hbar(\mu)\eta}$).
For $\cM(E_\alpha)_a$ there exist countable approximate identities
(since $E_\alpha$ is f\/inite dimensional), as e.g. such
$\gamma_n\in \cM(E_\alpha)_a^+$ which are concentrated in the ball
of radius $n^{-1}$ around the origin.

For an arbitrary representation $(\Psi_\hbar,\cH_\hbar)$ of
$(\cM(E_\alpha),\star_{\!\hbar})$ one may separate out a regular
part: $\cM(E_\alpha)_a$ being a closed $*$-ideal ensures that the
orthogonal projection $P$ onto the closure of
$\Psi_\hbar(\cM(E_\alpha)_a)\cH_\hbar$ commutes with each
$\Psi_\hbar(\mu)$, $\mu\in\cM(E_\alpha)$. Hence
$(\Psi_\hbar,\cH_\hbar)$ decomposes uniquely into the direct sum
$\Psi_\hbar=\Pi_\hbar\oplus\Pi_\hbar^N$, where
$\Pi_\hbar:=P\Psi_\hbar$ is an regular representation of
$(\cM(E_\alpha),\star_{\!\hbar})$, and where
$\Pi_\hbar^N:=(\1-P)\Psi_\hbar$ is a representation of
$(\cM(E_\alpha),\star_{\!\hbar})$ which vanishes on the
$*$-ideal~$\cM(E_\alpha)_a$, which we simply call a
\emph{non-regular} representation.

The Banach-$*$-algebra $(\cM(E_\alpha),\star_{\!\hbar})$ is a
dense sub-$*$-algebra of its enveloping $C^*$-algebra
$C^*(\cM(E_\alpha),\star_{\!\hbar})$, the $C^*$-norm of the latter
being written as $\norm{\cdot}_\hbar$. One f\/inds that
$C^*(\cM(E_\alpha)_d,\star_{\!\hbar})$
$=\cW(E_\alpha,\hbar\sigma)$ is a sub-$C^*$-algebra, and both
$C^*(\cM(E_\alpha)_a,\star_{\!\hbar})$ and
$C^*(\cM(E_\alpha)_c,\star_{\!\hbar})$ are closed $*$-ideals of
$C^*(\cM(E_\alpha),\star_{\!\hbar})$. The algebras
$(\cM(E_\alpha)_a,\star_{\!\hbar})$ and
$C^*(\cM(E_\alpha)_a,\star_{\!\hbar})$ are denoted \emph{twisted
group Banach-$*$-algebra} resp.~\emph{twisted group $C^*$-algebra}
of the vector group $E_\alpha$ with respect to the
multiplier~\eqref{WQB:eq:multiplier-hbar}.

The $\mathrm{L}^2$-representation $\Pi_\hbar^{L^2}$, given by
$\Pi_\hbar^{L^2}(\mu)\phi:=\mu\star_{\!\hbar}\phi$ for
$\phi\in\mathrm{L}^2(E_\alpha)$, is an example of an injective
regular representation of $(\cM(E_\alpha),\star_{\!\hbar})$ (e.g.
\cite{HewittRoss}, \cite{Pedersen79}). The operator-norm-closure
of $\Pi_\hbar^{L^2}(\cM(E_\alpha)_a)$ is called \emph{restricted
twisted group $C^*$-algebra}. If $\sigma$ is non-degenerate on
$E_\alpha$ and $\hbar\neq0$, then it follows that these two
twisted group $C^*$-algebras coincide (by
Lemma~\ref{WQB5:lemm-faithrep-E}\ref{WQB5:lemm-faithrep-E-a}
below), and are $*$-isomorphic to the $C^*$-algebra of compact
operators on a separable Hilbert space \cite{Segal67,Folland89},
cf.~also \cite{Grundling97}. 

$C^*(\cM(E_\alpha)_a,\star_{\!\hbar})$ being a closed $*$-ideal of
$C^*(\cM(E_\alpha),\star_{\!\hbar})$ ensures the existence of a
canonical $*$-homomorphism $m_\hbar$ from
$C^*(\cM(E_\alpha),\star_{\!\hbar})$ into the multiplier
$C^*$-algebra $\dM^\alpha_\hbar$ of
$C^*(\cM(E_\alpha)_a,\star_{\!\hbar})$ (e.g.
\cite[VIII.1]{FellDoran88}, \cite[III.6]{Takesaki79}). The above
1:1:1:1-correspondence shows that the
 extension of the largest regular
representation $\Pi_\hbar^{\text{reg}}$ of
$\cW(E_\alpha,\hbar\sigma)$ resp.~of
$(\cM(E_\alpha)_d,\star_{\!\hbar})$ leads just to the universal
representation of the twisted group $C^*$-algebra
$C^*(\cM(E_\alpha)_a,\star_{\!\hbar})$. Hence $\dM^\alpha_\hbar$
may be realized as a sub-$C^*$-algebra of the W$*$-algebra
$\ovl{\Pi_\hbar^{\text{reg}}(\cM(E_\alpha)_a)}^w$, in which case
$m_\hbar$ coincides with $\Pi_\hbar^{\text{reg}}$. By the
subsequent Lemma (also by
Lemma~\ref{WQ:9-5:lemm:faithful-tau-cont} of the Appendix)
$\Pi_\hbar^{\text{reg}}$ in addition acts faithfully on the
discrete twisted group algebra
$C^*(\cM(E_\alpha)_d,\star_{\!\hbar})=\cW(E_\alpha,\hbar\sigma)$.
Consequently, also $m_\hbar$ acts faithfully on the
sub-$C^*$-algebra $C^*(\cM(E_\alpha)_d,\star_{\!\hbar})$, and the
latter may be considered as sub-$C^*$-algebra of
$\dM^\alpha_\hbar$, too.

\begin{lemma}
\label{WQB5:lemm-faithrep-Eal} Let $\Pi_\hbar$ be a regular
representation of $(\cM(E_\alpha),\star_{\!\hbar})$. If the
representation $\Pi_\hbar$ acts faithfully on
$C^*(\cM(E_\alpha)_a,\star_{\!\hbar})$, then it is faithful on
$C^*(\cM(E_\alpha)_d,\star_{\!\hbar})=\cW(E_\alpha,\hbar\sigma)$,
too.
\end{lemma}

{\samepage
\begin{proof}
$C^*(\cM(E_\alpha)_a,\star_{\!\hbar})$ is a thick $*$-ideal in its
multiplier algebra $\dM^\alpha_\hbar$. So every faithful
representation of $C^*(\cM(E_\alpha)_a,\star_{\!\hbar})$ extends
uniquely to a faithful representation of $\dM^\alpha_\hbar$ (e.g.
\cite[Proposition~III.6.25]{Takesaki79} and its proof). Now
restrict to $C^*(\cM(E_\alpha)_d,\star_{\!\hbar})$.
\end{proof}

}

Because of the above decomposition theorem of representations
$\Psi_\hbar=\Pi_\hbar\oplus\Pi_\hbar^N$, $m_\hbar$ possibly may
not act faithfully on the $C^*$-algebra
$C^*(\cM(E_\alpha),\star_{\!\hbar})$, nevertheless $m_\hbar$ acts
injectively on the measure space $\cM(E_\alpha)$ itself
($\mu\star_{\!\hbar}\nu=0$ (resp.~$\nu\star_{\!\hbar}\mu=0$) for
all $\nu\in \cM(E_\alpha)_a$ yields $\mu=0$). In
Subsection~\ref{WQ-s6.3} we show the lower semicontinuity of
$\dR\ni\hbar\mapsto\norm{m_\hbar(\mu)}=\norm{\Pi_\hbar^{\text{reg}}(\mu)}$
for all $\mu\in\cM(E_\alpha)$, but continuity follows only for
those measures $\mu$ satisfying
$\norm{\mu}_\hbar=\norm{m_\hbar(\mu)}$ for all $\hbar\in\dR$. By
the above arguments, however, $m_\hbar$ acts faithfully on the two
sub-$C^*$-algebras $C^*(\cM(E_\alpha)_a,\star_{\!\hbar})$ and
$C^*(\cM(E_\alpha)_d,\star_{\!\hbar})$ of
$C^*(\cM(E_\alpha),\star_{\!\hbar})$, and hence the above proper
continuity is valid for each $\mu\in
\cM(E_\alpha)_a\bigcup\cM(E_\alpha)_d$.

Taking the inductive limit of the $C^*$-algebras
$C^*(\cM(E_\alpha),\star_{\!\hbar})$, $\alpha\in I$, we arrive at
the enveloping $C^*$-algebra $C^*(\cM(E),\star_{\!\hbar})$ of the
whole Banach-$*$-algebra $(\cM(E),\star_{\!\hbar})$, the
$C^*$-norm of which is denoted by $\norm{\cdot}_\hbar$, too. Also
here we have that
$C^*(\cM(E)_d,\star_{\!\hbar})=\cW(E,\hbar\sigma)$ is a
sub-$C^*$-algebra, and $C^*(\cM(E)_c,\star_{\!\hbar})$ is a closed
$*$-ideal of $C^*(\cM(E),\star_{\!\hbar})$. But analogously as for
the Banach-$*$-algebras $(\cM(E_\alpha)_a,\star_{\!\hbar})$,
$\alpha\in I$, there  is no inductive limit of the absolutely
continuous measure $C^*$-algebras
$C^*(\cM(E_\alpha)_a,\star_{\!\hbar})$, $\alpha\in I$. A certain
generalization of the above 1:1:1:1-correspondence to inf\/inite
dimensional $(E,\sigma)$ is found in \cite{Grundling97}.

We now turn to representations of our measure Banach-$*$-algebra
$(\cM(E),\star_{\!\hbar})$. By the Subsections~\ref{WQ-s2.2}
and~\ref{WQ-s4.1}, especially by
equation~\eqref{WQB:eq:pi-Pi-quasiequi}, each representation
(class)
\begin{gather*}
\Pi_\hbar\in \operatorname{rep}(\cM(E)_d,\star_{\!\hbar})=
\operatorname{rep}(\cW(E,\hbar\sigma)) \qquad\text{with}\quad
\Pi_\hbar\leq\Pi_\hbar^{\text{reg}}
\end{gather*}
gives rise to the projective unitary group representation (class)
\begin{gather*}
E\ni f\longmapsto
\pi_\hbar(f)=\Pi_\hbar(W^\hbar(f))=\Pi_\hbar(\delta(f)),
\end{gather*}
which is continuous on each f\/inite dimensional subspace
$E_\alpha$ of $E$ with respect to the $\sigma$-strong
(equivalently $\sigma$-weak) topology on the W$*$-algebra
$\ovl{\Pi_\hbar(\cW(E,\hbar\sigma))}^w$. Since $\sigma$ is jointly
continuous on each $E_\alpha$, it is obvious that $\Pi_\hbar$ may
be extended by integration in a weak operator topology, like
\begin{gather}
\label{WQB5:eq:Pi(mu)} \mu \longmapsto \Pi_\hbar(\mu):= \int_E
d\mu(f)\; \pi_\hbar(f)\; \in
\ovl{\Pi_\hbar(\cW(E,\hbar\sigma))}^w,
\end{gather}
to a representation of each Banach-$*$-algebra
$(\cM(E_\alpha),\star_{\!\hbar})$, $\alpha\in I$. Finally, by
taking the inductive limit over $\alpha\in I$ and by using a
$\norm{\cdot}_1$-density argument, one arrives at a representation
of the whole Banach-$*$-algebra $(\cM(E),\star_{\!\hbar})$,
denoted by the same symbol. If $\Pi_\hbar\leq\Pi_\hbar^\tau$
(requiring $\Pi_\hbar$ to be $\tau$-continuous), for some locally
convex $\tau\in\cT(E,\sigma)$, then the extension in
equation~\eqref{WQB5:eq:Pi(mu)} may be performed for all Borel
measures $\mu \in M^\tau(E)$, in which case $\Pi_\hbar$ is only a
linear mapping; only its restriction to $\cM(E)$ is a
$*$-homomorphism. Note that for an only regular $\Pi_\hbar$ the
linear extension to $M^\tau(E)$ does not work.

Since every representation $\Pi_\hbar$ from the sub-$C^*$-algebra
$C^*(\cM(E)_d,\star_{\!\hbar})=\cW(E,\hbar\sigma)$ may be extended
to a representation on $C^*(\cM(E),\star_{\!\hbar})$ with  in
general a larger representation Hilbert space, we may regard
subsequently only representations (resp.~equivalence classes of
representations) of $(\cM(E),\star_{\!\hbar})$. So, when writing
$\Pi_\hbar$, we henceforth mean an element of
$\operatorname{rep}(\cM(E),\star_{\!\hbar})
=\operatorname{rep}(C^*(\cM(E),\star_{\!\hbar}))$.

To avoid the abuse of ``regular'' and to indicate the construction
method, we make the following convention for inf\/inite
dimensional $E$. 
\begin{definition}[Integration type representations of $\boldsymbol{(\cM(E),\star_{\!\hbar})}$]
\label{wqb5-defITR} Under an \emph{integration type
representation} $\Pi_\hbar$ of $(\cM(E),\star_{\!\hbar})$ (or of
$(C^*(\cM(E),\star_{\!\hbar})$) we understand a representation,
which extends either a regular or a $\tau$-continuous
representation $(\Pi_\hbar,\cH_\hbar)$ of the $C^*$-Weyl algebra
$\cW(E,\hbar\sigma)=C^*(\cM(E)_d,\star_{\!\hbar})$, using
equation~\eqref{WQB5:eq:Pi(mu)}. By construction, the
representation Hilbert space is still $\cH_\hbar$.

(Recall that for inf\/inite dimensional $E$ ``regularity'' of a
representation of the Weyl algebra
$\cW(E,\hbar\sigma)=C^*(\cM(E)_d,\star_{\!\hbar})$ is weaker than
``$\tau$-continuity'' (in contradistinction to the f\/inite
dimensional case), but both properties allow to extend the weak
integration over test functions from discrete to more general
measures.)

The set of (quasi-equivalence classes of) integration type
representations is denoted nevertheless by
$\operatorname{rep}_{\text{reg}}(\cM(E),\star_{\!\hbar})
=\operatorname{rep}_{\text{reg}}(C^*(\cM(E),\star_{\!\hbar}))$ (to
have a short subscript). The integrated extensions of the largest
regular resp.~$\tau$-continuous representation, namely of
$\Pi_\hbar^{\text{reg}}$ resp.~of $\Pi_\hbar^\tau$,  are denoted
by the old symbols.

Especially, when writing $\Pi_\hbar\leq\Pi_\hbar^{\text{reg}}$ or
$\Pi_\hbar\leq\Pi_\hbar^\tau$, we indicate that $\Pi_\hbar$ is of
integration type, too.
\end{definition}

\begin{lemma}
\label{WQB5:lemm-faithrep-E} For $\hbar\neq0$ the following
assertions are valid:
\begin{enumerate}\itemsep=0pt
\renewcommand{\labelenumi}{(\alph{enumi})}
\renewcommand{\theenumi}{(\alph{enumi})}
\item \label{WQB5:lemm-faithrep-E-a} Let $\alpha\in I_\sigma$
(i.e.~$\sigma$ is non-degenerate on $E_\alpha$, Section~{\rm
\ref{WQ-s5}}). Then $C^*(\cM(E_\alpha)_a,\star_{\!\hbar})$ is
simple (a statement implying the well-known simplicity of the Weyl
algebra $\cW(E_\alpha,\hbar\sigma)$ by  means of Lemma~{\rm
\ref{WQB5:lemm-faithrep-Eal}}). \item
\label{WQB5:lemm-faithrep-E-b} Let $\sigma$ be non-degenerate on
$E$. Then we have
$\norm{\Pi_\hbar(\mu)}=\norm{\Pi_\hbar^{\text{\rm reg}}(\mu)}$,
$\mu\in\cM(E)$, for every subrepresentation
$\Pi_\hbar\leq\Pi_\hbar^{\text{\rm reg}}$. This is also valid when
we replace $E$ by $E_\alpha$ with $\alpha\in I_\sigma$.
\end{enumerate}
\end{lemma}

\begin{proof}
\cite{Kastler65} ensures that the regular part $\Pi_\hbar$ of any
representation $\Psi_\hbar=\Pi_\hbar\oplus\Pi_\hbar^N$ of
$(\cM(E_\alpha),\star_{\!\hbar})$ is a direct sum of identical
copies of the irreducible Schr{\"{o}}dinger representation, a
generalized von Neumann uniqueness result. Thus
$\norm{\Pi_\hbar(\mu)}$ has the same value for every integration
type representation $\Pi_\hbar$. Moreover,
$\norm{\Pi_\hbar(\mu)}=\norm{\mu}_\hbar$ for all
$\mu\in\cM(E_\alpha)_a$, since $\Pi_\hbar^N(\cM(E_\alpha)_a)=0$.
For part~\ref{WQB5:lemm-faithrep-E-b} take the inductive limit
over $\alpha\in I_\sigma$.
\end{proof}

Let us f\/inally give the classical case ($\hbar=0$) a special
treatment. Recall that $\norm{\wh{\mu}}_0$ denotes the supremum
norm of the continuous phase space function $\wh{\mu}$.
$\norm{\wh{\mu}}_0\leq\norm{\mu}_1$ for all $\mu\in M^\tau(E)$,
which may be shown directly, and which is for $\mu\in\cM(E)$ a
consequence of the continuity of $*$-homomorphisms
\cite[Proposition~I.5.2]{Takesaki79}. 
\begin{proposition}
\label{WQ:9-5:prop-Pi0tau} The function algebra, given by the
Fourier transformed $(\cM(E),\star_{\!0})$, is $*$-isomorphic to
the largest $\tau$-continuous representation class
$\Pi_0^\tau\in\operatorname{rep}_{\text{reg}}(C^*(\cM(E),\star_{\!0}))$,
restricted to $(\cM(E),\star_{\!0})$, for an arbitrary locally
convex topology $\tau$ on $E$, what essentially amounts to the
equality of the $C^*$-norms
$\norm{\Pi_0^\tau(\mu)}=\norm{\wh{\mu}}_0$ for all $\mu\in\cM(E)$.
(For represented algebra elements we use the operator norm.)

Furthermore, $\Pi_0^\tau$ acts faithfully on the
sub-$C^*$-algebras $C^*(\cM(E_\alpha)_a,\star_{\!0})$, for every
$\alpha\in I$, and on $C^*(\cM(E)_d,\star_{\!0})$. In terms of the
$C^*$-norm $\norm{\cdot}_{0}$ of $C^*(\cM(E),\star_{\!0})$, that
means
\begin{gather*}
\textstyle
\norm{\mu}_{0}=\norm{\Pi_0^\tau(\mu)}=\norm{\wh{\mu}}_0,
\qquad\forall
\,\mu\in\cM(E)_d\bigcup\bigl(\bigcup_\alpha\cM(E_\alpha)_a\bigr).
\end{gather*}
\end{proposition}
\begin{proof}
Let $\mu\in\cM(E_\alpha)$. As mentioned in
Subsection~\ref{WQ-s4.1} the characteristic funtions $C_\omega$ of
the states $\omega$ from the folium $\cF_0^\tau$ associated with
$\Pi_0^\tau$ are $\tau$-continuous, positive-def\/inite functions
on~$E$. By Bochner's theorem (which is not valid for inf\/inite
dimensional $E$ because of the lack of local compactness) for the
restriction of $C_\omega$ to $E_\alpha$ there exists  a unique
(positive) probability measure $\rho_\omega$ on $E_\alpha'$ such
that $C_\omega(f)=\int_{E_\alpha'}d\rho_\omega[F]\ex{iF(f)}$ for
all $f\in E_\alpha$. We conclude that
\begin{gather*}
\dual{\omega}{\mu^*\star_{\!0}\mu} = \int_{E_\alpha} d\mu^*(f)
\int_{E_\alpha} d\mu(g) C_\omega(f+g) =
\int_{E_\alpha'}d\rho_\omega[F]  \betr{\wh{\mu}[F]}^2
\leq\norm{\wh{\mu}}_0^2.
\end{gather*}
Since for every $F\in E_\tau'$ there is a state in $\cF_0^\tau$
with characteristic function $E\ni f\mapsto\ex{iF(f)}$ (cf.~again
Subsection~\ref{WQ-s4.1}), we conclude that
$\norm{\Pi_0^\tau(\mu)}^2=
\sup\{\dual{\omega}{\mu^*\star_{\!0}\mu}\mid \omega\in\cF_0^\tau\}
=\norm{\wh{\mu}}_0^2$ (recall that $\cF_0^\tau$ consists of the
normal states on $\ovl{\Pi_\hbar(\cW(E,0))}^w$). Taking the
inductive limit over $\alpha\in I$ we get
$\norm{\Pi_0^\tau(\mu)}=\norm{\wh{\mu}}_0$ for all $\mu\in
\cM(E)$. The rest is immediate with
Lemma~\ref{WQ:9-5:lemm:faithful-tau-cont} of the Appendix.
\end{proof}
\begin{lemma}
\label{WQ:9-5:obse-positive-measure} For $\mu\in M^\tau(E)^+$ we
have $\norm{\mu}_1=\mu(E)=\wh{\mu}[0]=\norm{\wh{\mu}}_0$.
\end{lemma}
\begin{proof}
For $\mu\in M^\tau(E)^+$ it follows that $\wh{\mu}$ is a
continuous positive-def\/inite function on the additive group
$E_\tau'$. Hence $\betr{\wh{\mu}[F]}\leq\wh{\mu}[0]$ for all $F\in
E_\tau'$, e.g. \cite[(32.4)]{HewittRoss}, and thus
$\norm{\wh{\mu}}_0=\wh{\mu}[0]$. On the other side, $\mu$ being
positive ensures $\mu=\betr{\mu}$, and hence
$\norm{\mu}_1=\mu(E)=\wh{\mu}[0]$.
\end{proof}

\section[Extended Weyl quantization as strict deformation quantization (of $C^*$-type)]{Extended Weyl quantization\\ as strict deformation quantization (of $\boldsymbol{C^*}$-type)}
\label{WQ-s6}

\subsection{Families of representations, quantization maps}
\label{WQ-s6.1}

Up to now we have developed the Banach-$*$-algebra version
$(Q_\hbar^B)_{\hbar\in\dR}$ of strict deformation quantization (in
Theorem~\ref{WQ:9-5:coro:Neumann-Dirac-M(E)}) for the generalized
Weyl quantization. The transition to the original $C^*$-type
version from Def\/inition~\ref{WQ:defi:9-5:SQ} is carried through
in a second step by the selection of a family of
(quasi-equivalence classes of) representations
\begin{gather*}
\Pi\equiv(\Pi_\hbar)_\hbar:=\{\Pi_\hbar\in
\operatorname{rep}(\cM(E),\star_{\!\hbar})\mid \hbar\neq0\}.
\end{gather*}
This being given, we def\/ine for every $\hbar\neq0$  the family
of quantization maps by
\begin{gather}
\label{WQB:eq:quat-map} \wh{\mu} \;\longmapsto\;
Q_\hbar^\Pi(\wh{\mu})
:=\Pi_\hbar(\underbrace{Q_\hbar^B(\mu)}_{\mbox{$=\mu$}})=\Pi_\hbar(\mu).
\end{gather}
As is common for Weyl quantization, we start from phase space
functions $\wh{\mu}:\mathsf{P}\rightarrow\dC$ (with
$\mathsf{P}\equiv E_\tau'$) instead of the associated measures
$\mu\in\cM(E)$ resp.~$M^\tau(E)$.

Since Fourier transformation is a bijection, the quantization map
$Q_\hbar^\Pi$ is well-def\/ined on $\wh{\cM}(E_\tau')$, even on
$\wh{M}(E_\tau')$, provided $\Pi_\hbar\leq\Pi_\hbar^\tau$. Recall
from Def\/inition~\ref{wqb5-defITR}, that $\Pi_\hbar$ may be an
integration type representation
\begin{gather}
\label{WQB:eq:quat-map01} \Pi_\hbar(\mu) =\int_E d\mu(f)\;
\underbrace{\Pi_\hbar(W^\hbar(f))}_{\mbox{$=\pi_\hbar(f)$}}\; \in
\ovl{\Pi_\hbar(\cW(E,\hbar\sigma))}^w .
\end{gather}

For a true strict deformation quantization one has to restrict the
domain of def\/inition of the quantization maps $Q_\hbar^\Pi$,
$\hbar\neq0$, to a Poisson algebra $\wh{\cP}=\dF\cP$ of phase
space functions. We may choose $\cP$ as any measure Poisson
algebra from Theorem~\ref{WQ:9-5:theo:PoissonBracket-M(E)}, whose
Fourier transformations are treated in
Theorem~\ref{WQ:9-5:theo:PA-B}.

For each $\hbar$ the images $Q_\hbar^\Pi(\wh{\mu})=\Pi_\hbar(\mu)$
have to be operators in a $C^*$-algebra, for which we take e.g.
the smallest $C^*$-algebra $\cA^\hbar$ comprising the represented
Banach-$*$-algebra $\Pi_\hbar(\cM(E),\star_{\!\hbar})$. We
elaborate this for selected families $\Pi\equiv(\Pi_\hbar)_\hbar$
of representations.

As mentioned previously, we have the norm estimations
\begin{gather}
\label{wqb6-est2} \norm{\Pi_\hbar(\mu)}\leq\norm{\mu}_1,
\qquad\forall\,\hbar\in\dR,\qquad \mu\in \cM(E),
\end{gather}
where $\norm{\cdot}$ are the $C^*$-norms of the dif\/ferent
$\cA^\hbar$. From these estimations the Dirac and the von Neumann
conditions follow easily, see Subsection~\ref{WQ-s6.2}. But,
whereas Rief\/fel's condition has been trivially fulf\/illed in
the Banach-$*$-deformation quantization
$(Q_\hbar^B)_{\hbar\in\dR}$ of
Theorem~\ref{WQ:9-5:coro:Neumann-Dirac-M(E)}, relation
\eqref{wqb6-est2} does not provide a simple deduction of the
continuity of $\dR\ni
\hbar\mapsto\|Q_\hbar^\Pi(\wh{\mu})\|=\|\Pi_\hbar(\mu)\|$ with the
appropriate $C^*$-norms for the dif\/ferent values of $\hbar$
(Rief\/fel's condition).

For the representation family $\Pi=(\Pi_\hbar)_\hbar$ we consider
$\Pi^\#:=(\Pi_\hbar^\#)_\hbar$ with
$\#\in\{\text{reg},\;\text{il},\;\tau\}$, $\tau\in\cT(E,\sigma)$.
We write $\Pi\leq\Pi^\#$, if $\Pi_\hbar\leq\Pi_\hbar^\#$ for every
$\hbar\neq0$. A family $\Pi\leq\Pi^\tau$ may be linearly extended
from the universal $(\cM(E),\star_{\!\hbar})$ to all of
$M^\tau(E)$, by using the integration type method
equation~\eqref{WQB5:eq:Pi(mu)}. Thus,
$\norm{\Pi_\hbar(\mu)}\leq\norm{\mu}_1$ in any case. If in
$\Pi=(\Pi_\hbar)_\hbar$ each $\Pi_\hbar$ acts faithfully on a
sub-$C^*$-algebra $\cB^\hbar$ of the enveloping $C^*$-algebra
$C^*(\cM(E),\star_{\!\hbar})$, then we call $\Pi$  ``faithful
on~$(\cB^\hbar)_{\hbar\neq0}$''.

Up to now, a family $\Pi=(\Pi_\hbar)_\hbar$ of quasi-equivalence
classes of representations was def\/ined for the values
$\hbar\neq0$, only. We add for the classical case $\hbar=0$ the
Fourier transformation $\dF$ as the representation $\Pi_0$, which
by Proposition~\ref{WQ:9-5:prop-Pi0tau} is $*$-isomorphic to the
integration type representation
$\Pi_0^\tau\in\operatorname{rep}_{\text{reg}}(\cM(E),\star_{\!0})$,
for each $\tau$. Let us write then
\begin{gather*}
Q_0^\Pi(\wh{\mu}):=\Pi_0(\mu)=\wh{\mu} ,\qquad\forall \, \mu\in
\cM(E)\quad \text{resp.} \ \ M^\tau(E).
\end{gather*}
We subsequently make the identif\/ication
$\|Q_0^\Pi(\wh{\mu})\|=\norm{\wh{\mu}}_0$, with the sup-norm
$\norm{\cdot}_0$, dropping occasionally the subscript ``$0$''.

\subsection{Dirac's and von Neumann's conditions}
\label{WQ-s6.2}

As an immediate consequence of inequality \eqref{wqb6-est2} we
obtain from Theorem~\ref{WQ:9-5:coro:Neumann-Dirac-M(E)} the main
part of a strict quantization.

\begin{theorem}
\label{WQ:9-5:theo:pre-SDQ} For each family
$\Pi=(\Pi_\hbar)_\hbar$ of representations, not necessarily of
integration type:
\begin{enumerate}\itemsep=0pt
\renewcommand{\labelenumi}{(\alph{enumi})}
\renewcommand{\theenumi}{(\alph{enumi})}
\item \label{WQ:9-5:theo:pre-SDQ-a} \textbf{\itshape [Dirac]}
$\lim\limits_{\hbar\rightarrow0} \norm{
\{Q_\hbar^\Pi(\wh{\mu}),Q_\hbar^\Pi(\wh{\nu})\}_\hbar -
Q_\hbar^\Pi(\{\wh{\mu},\wh{\nu}\})
                      }  =   0$
for all $\mu,\nu\in \cM^1_\varsigma(E)$, with the Poisson bracket
$\{\cdot,\cdot\}$ from equation~\eqref{WQ:9-5:eq:PA-B-Fou}, and
with the $\hbar$-scaled commutators $\{\cdot,\cdot\}_\hbar$
from~\eqref{WQ:eq:9-5:hbar-scaled-commut}. \item
\label{WQ:9-5:theo:pre-SDQ-b} \textbf{\itshape [von Neumann]}
$\lim\limits_{\hbar\rightarrow0} \norm{
Q_\hbar^\Pi(\wh{\mu})Q_\hbar^\Pi(\wh{\nu})
        -  Q_\hbar^\Pi(\wh{\mu}\cdot_{\!0}\wh{\nu})
      }  =  0$
for all $\mu,\nu\in \cM(E)$.
\end{enumerate}
\end{theorem}

That means that the Dirac and von Neumann conditions are valid for
all $\wh{\mu}$ and $\wh{\nu}$ contained in any sub-Poisson algebra
$\tilde{\cP}$ of
$(\wh{\cM}_\varsigma^\infty(E_\tau'),\cdot_{\!0},\{\cdot,\cdot\})$,
where we think especially on the Fourier transformed measure
Poisson algebras $\tilde{\cP}:=\wh{\cP}=\dF\cP$, with $\cP$ from
Theorem~\ref{WQ:9-5:theo:PoissonBracket-M(E)} (including the case
$\cP=\cM_\varsigma^\infty(E)$).

\subsection[On Rieffel's continuity condition]{On Rief\/fel's continuity condition}
\label{WQ-s6.3} 
In Rief\/fel's condition the norms of the representations
$\Pi_\hbar$ of $(\cM(E),\star_{\!\hbar})$ are compared with each
other for dif\/ferent $\hbar\in\dR$, which causes some
complications. This problem does not arise for the usual
$C^*$-Weyl algebras, corresponding to the discrete measures. We
include this previously discussed case into the present method of
dealing with representations. 
\begin{example}[Discrete case]
\label{wqb6-exam} For each family $\Pi=(\Pi_\hbar)_\hbar$, which
is faithful on
$(C^*(\cM(E)_d,\star_{\!\hbar})=\cW(E,\hbar\sigma))_{\hbar\neq0}$,
the present quantization maps $(Q_\hbar^\Pi)_{\hbar\in\dR}$ are
$*$-isomorphic to the previous $C^*$-algebraic quantization maps
of  the strict deformation quantization in
Theorem~\ref{WQ:theo:9-5:SDQ-Weylalgebra} (use
Proposition~\ref{WQ:9-5:prop-Pi0tau} for $\hbar=0$). Candidates
$\cP$ for a measure Poisson algebra are given by $\cM(E)_{df}
\cong \Delta(E,0)$, and by
$\cM(E)_d\bigcap\cM_\varsigma^\infty(E)$ from
Theorem~\ref{WQ:9-5:theo:PoissonBracket-M(E)} (already used
in~\cite{BinzHonRie03b}).
\end{example}

The present, representation dependent quantization procedure, aims
however at a larger class of $C^*$-algebraic quantizations,
founded on the enlarged measure space $\cM(E)\supset\cM(E)_d$.
Before stating (and proving) the results in detail, let us give a
brieve overview. 
\begin{summary}
\label{wqb6-sum} Let $\Pi$ be a family of integration type
representations of $(\cM(E),\star_{\!\hbar})$. Under certain
circumstances ($\sigma$ non-degenerate, or $\Pi$ suitably
well-matched), we achieve by our own efforts lower semicontinuity
of $\dR\ni
\hbar\mapsto\|Q_\hbar^\Pi(\wh{\mu})\|=\|\Pi_\hbar(\mu)\|$, for all
$\mu\in\cM(E)$. Using some results of Rieffel one even gets true
continuity for non-degenerate $\sigma$.

Thus, if Rieffel's condition is weakened to lower semicontinuity,
then $(Q_\hbar^\Pi)_{\hbar\in \dR}$ constitutes a strict
deformation quantization of each classical Poisson algebra
$\wh{\cP}=\dF\cP$ of phase space functions, which is mentioned in
Theorem~{\rm \ref{WQ:9-5:theo:PoissonBracket-M(E)}}.
\end{summary}

In virtue of
Lemma~\ref{WQB5:lemm-faithrep-E}\ref{WQB5:lemm-faithrep-E-b} the
case of a non-degenerate $\sigma$ is simpler than that of a
degenerate one, since only for degenerate $\sigma$ one may have
that $\norm{\Pi_\hbar(\mu)}<\norm{\Pi_\hbar^{\text{reg}}(\mu)}$,
for some $\mu\in\cM(E)$, in a proper subrepresentation
$\Pi_\hbar<\Pi_\hbar^{\text{reg}}$. The following two Subsections
treat technical questions of this nature.

\subsubsection[Non-degenerate symplectic form $\sigma$]{Non-degenerate symplectic form $\boldsymbol{\sigma}$}
\label{WQ-s6.3non}

Let us assume $\sigma$ non-degenerate on the inf\/inite
dimensional test function space $E$. The proof of the next result
is deferred to Appendix~\ref{WQ-appB}. 
\begin{proposition}[Lower semicontinuity]
\label{WQ6:prop:non} Let $\Pi\leq\Pi^{\text{\rm reg}}$ be an
arbitrary family of integration type representations of
$(\cM(E),\star_{\!\hbar})$ (formed with a non-degenerate
$\sigma$). Then
$\dR\ni\hbar\mapsto\|Q_\hbar^\Pi(\wh{\mu})\|=\|\Pi_\hbar(\mu)\|$
is lower semicontinuous for every $\mu\in\cM(E)$.
\end{proposition}

We want to relate our investigation to Rief\/fel's works
\cite{Rieffel93a, Rieffel94a}. This may be carried out for
non-degenerate $\sigma$, since only in this case the matrix $J$ in
equation~\eqref{eq:wqb6:rieffel} below may be related to $\sigma$
via a transformation of the variables.

Let $\alpha\in I_\sigma$ (i.e.~$\sigma$ acts non-degenerately on
the f\/inite dimensional subspaces $E_\alpha$, cf.~beginning of
Section~\ref{WQ-s5}). In
equation~\eqref{WQ:eq:9-5:deformedproducts} we see how the Fourier
transformation of the twisted convolution products leads to the
deformed or Moyal products for functions \cite{GarciaBondia88}.
The latter products $\cdot_{\!\hbar}$ may be formulated for
measures $\mu$ and $\nu$ on $E_\alpha$ in terms of oscillatory
integrals as follows
\begin{gather}
\label{eq:wqb6:rieffel} \wh{\mu}\cdot_{\!\hbar}\wh{\nu}[F] =
\int_{E_\alpha'}dG \int_{E_\alpha'}dH \;\wh{\mu}[F+\hbar JG] \:
\wh{\nu}[F+H] \,\ex{iG\cdot H},
\end{gather}
where $J$ is an anti-symmetric matrix arising from $\sigma$ and
$G\cdot H$ is an inner product on $E_\alpha'$. Rief\/fel shows the
continuity of
$\dR\ni\hbar\mapsto\|\wh{\Pi}_\hbar^{L^2}(\wh{\mu})\|$ for
$\mu\in\cM_\varsigma^\infty(E_\alpha)$ for the Fourier transforms
$\wh{\Pi}_\hbar^{L^2}$ of the $\mathrm{L}^2$-representations
$\Pi_\hbar^{L^2}$ from Subsection~\ref{WQ-s5.1xtra}, a specif\/ic,
well-matched $\mathrm{L}^2$-representation family. For $\hbar=0$
observe that $\wh{\mu}$ acts as multiplicator on the Fourier
transformed $\mathrm{L}^2$-space, the operator norm of which is
just given by its sup-norm
$\norm{\wh{\mu}}_0=\|\wh{\Pi}_0^{L^2}(\wh{\mu})\|=
\|\Pi_0^{L^2}(\mu)\|$. Thus in virtue of
Lemma~\ref{WQB5:lemm-faithrep-E}\ref{WQB5:lemm-faithrep-E-b} and
by taking the inductive limit over $\alpha\in I_\sigma$ one
arrives at the continuity of
$\dR\ni\hbar\mapsto\|\Pi_\hbar(\mu)\|$ for all $\mu\in\cM(E)$, for
arbitrary families $\Pi\leq\Pi^{\text{reg}}$ of integration type
representation classes.

\subsubsection[Degenerate $\sigma$, well-matched representation families]{Degenerate $\boldsymbol{\sigma}$, well-matched representation families}
\label{WQ-s6.3deg}

Let us now allow for a degenerate, non-trivial $\sigma$ on $E$.
Because of
Lemma~\ref{WQB5:lemm-faithrep-E}\ref{WQB5:lemm-faithrep-E-b} a
norm def\/icit
$\|\Pi_\hbar(\mu)\|<\|\Pi_\hbar^{\text{reg}}(\mu)\|, \,
\mu\in\cM(E)$, is possible in a proper subrepresentation
$\Pi_\hbar<\Pi_\hbar^{\text{reg}}$. Thus we need a certain
compatibility condition between representations $\Pi_\hbar$, with
dif\/ferent $\hbar\neq0$, within a~family $\Pi=(\Pi_\hbar)_\hbar$.
Otherwise, one could not expect continuity properties for
$\dR\ni\hbar\mapsto\|\Pi_\hbar(\mu)\|$. 
\begin{definition}[Well-matched families of representation classes]
\label{WQB:defi:comparing} Let $\tau\in\cT(E,\sigma)$ be a~locally
convex topology. A family $\Pi=(\Pi_\hbar)_\hbar$ is called
\emph{well-matched} (with respect to $\tau$), if the following
assertions are valid:
\begin{enumerate}\itemsep=0pt
\renewcommand{\labelenumi}{(\Alph{enumi})}
\renewcommand{\theenumi}{(\Alph{enumi})}
\item \label{WQB:defi:comparing-A} For each $\hbar\neq0$ there is
an $\dR$-linear $\tau$-homeomorphism $T_\hbar$ on $E$ satisfying
$\sigma(T_\hbar f,T_\hbar g)=\hbar\sigma(f,g)$ for all $f,g\in E$.
(By Lemma~\ref{WQ:9-5:lemm:beta-hbar-*iso} then there exists a
unique $*$-isomorphism $\beta_\hbar$ from $\cW(E,\hbar\sigma)$
onto $\cW(E,\sigma)$ with $\beta_\hbar(W^\hbar(f))=W^1(T_\hbar
f)$, for all $f\in E$.) \item \label{WQB:defi:comparing-C} For
$\hbar=1$ we have $\Pi_1\leq\Pi_1^\tau$ and
$\Pi_\hbar=\Pi_1\circ\beta_\hbar$ for all $\hbar\neq0$, with the
representations acting on $\cW(E,\hbar\sigma)$ (relations which
are extensible via the integration type construction to
$C^*(\cM(E),\star_{\!\hbar})$ according to
Def\/inition~\ref{wqb5-defITR}). \item
\label{WQB:defi:comparing-B} Moreover, $\hbar\mapsto T_\hbar f$ is
assumed $\tau$-continuous with $\lim\limits_{\hbar\rightarrow
0}T_\hbar f=0$, for all $f\in E$.
\end{enumerate}

Instead of $\tau$-continuous representations one may also treat
merely \emph{regular} ones. Then the $\dR$-linear bijections
$T_\hbar$ on $E$ are demanded to leave each f\/inite dimensional
subspace invariant, so that part~\ref{WQB:defi:comparing-B} is
reasonable, and one assumes $\Pi_1\leq\Pi_1^{\text{reg}}$.
\end{definition}
The dual mapping $\beta_\hbar^*$ is an af\/f\/ine bijection from
the folium $\cF_{\hbar=1}^\tau$ onto the folium $\cF_\hbar^\tau$,
and consequently $\beta_\hbar$ extends $\sigma$-strong
continuously to a $*$-isomorphism from the W$*$-algebra
$\ovl{\Pi_\hbar^\tau(\cW(E,\hbar\sigma))}^w$ onto the W$*$-algebra
$\ovl{\Pi_{\hbar=1}^\tau(\cW(E,\sigma))}^w$, which in addition
maps $\Pi_\hbar^\tau(\cM(E),\star_{\!\hbar})$ onto
$\Pi_{\hbar=1}^\tau(\cM(E),\star_{\!1})$ and
$\Pi_\hbar^\tau(C^*(\cM(E),\star_{\!\hbar}))$ onto
$\Pi_{\hbar=1}^\tau(C^*(\cM(E),\star_{\!1}))$. Some examples of
$T_\hbar$ fulf\/illing the above Def\/inition are given in
Subsection~\ref{WQ-s4.3}. (If in concrete situations such
$T_\hbar$ are only given for $\hbar$ in a subset $J\subset\dR$ --
as in Subsection~\ref{WQ-s2.1} --, then the subsequent results
remain valid by restricting $\hbar\in\dR$ to $\hbar\in J$.)

The notion of $E_\rho'$-invariance, used in our next result, is
introduced in Appendix~\ref{WQ-appA}, the proof is found in
Appendix~\ref{WQ-appB}. 
\begin{proposition}[Lower semicontinuity]
\label{WQ:9-5:prop:pre-SDQ-R} Let $\Pi\leq\Pi^{\text{\rm reg}}$ be
a family of integration type representations.
\begin{enumerate}\itemsep=0pt
\renewcommand{\labelenumi}{(\alph{enumi})}
\renewcommand{\theenumi}{(\alph{enumi})}
\item \label{WQ:9-5:prop:pre-SDQ-R-b} Assume $\Pi$ well-matched.
Then $\hbar\mapsto\|Q_\hbar^\Pi(\wh{\mu})\|$ is lower
semicontinuous on $\dR\!\setminus\!\{0\}$, for every $\mu\in
\cM(E)$, and we have $\lim\limits_{\hbar\rightarrow0} \|
Q_\hbar^\Pi(\wh{\mu})\|
= \norm{\wh{\mu}}_0$ for all $\mu\in \cM(E)^+$.\\
Suppose in addition a locally convex topology $\rho$ on $E$ so
that either each $\Pi_\hbar$ is $E_\rho'$-invariant, or that
$\Pi_{\hbar=1}$ is partially $E_\rho'$-invariant and $E_\rho'\circ
T_\hbar^{-1}=E_\rho'$, for each $\hbar\neq0$. Then the above lower
semicontinuity is valid on all of $\dR$ (now including the
origin). \item \label{WQ:9-5:prop:pre-SDQ-R-c} Let $\Pi$ be
well-matched with reference to $\tau$ (thus
especially~$\Pi\leq\Pi^\tau$), and let $\rho\leq\tau$ with
$\tau\in\cT(E,\sigma)$ a locally convex topology. Then the
assertions of part~\ref{WQ:9-5:prop:pre-SDQ-R-b} are even valid
for $\mu\in M^\tau(E)$.
\end{enumerate}

Examples are provided, if $\Pi^{\text{\rm reg}}$ is well-matched
and is $E_\rho'$-invariant, for every locally convex~$\rho$ on
$E$, and also $\Pi^\tau$ is well-matched and $E_\tau'$-invariant.
\end{proposition}

Let us now investigate some circumstances under which the lower
semicontinuity may be strengthend to proper continuity of
$\dR\ni\hbar\mapsto\|Q_\hbar^\Pi(\wh{\mu})\|$. We suppose a
$\tau$-well-matched family $\Pi$ with $E_\rho'$-invariance, as in
the above Proposition. It is immediately checked that
$\mu\mapsto\limsup\limits_{\hbar\rightarrow\hbar_0}\norm{\Pi_\hbar(\mu)}$
def\/ines a semi-$C^*$-norm on the Banach-$*$-algebra
$(\cM(E),\star_{\!\hbar_0})$. Consequently,
$\limsup\limits_{\hbar\rightarrow\hbar_0}\norm{\Pi_\hbar(\mu)}
\leq\norm{\mu}_{\hbar_0}$ for each $\hbar_0\in\dR$, since
$\norm{\mu}_{\hbar_0}$ is the $C^*$-norm of the enveloping
$C^*$-algebra $C^*(\cM(E),\star_{\!\hbar_0})$ as in
Subsection~\ref{WQ-s5.1xtra}. Now
equation~\eqref{WQB:9-5:eq:lower-semicont} yields: 
\begin{example}
\label{WQB6:obse:=} Let $\mu\in\cM(E)$. If one is able to show
that $\norm{\Pi_\hbar(\mu)}=\norm{\mu}_{\hbar}$ for all
$\hbar\in\dR$, then $\dR\ni\hbar\mapsto\|Q_\hbar^\Pi(\wh{\mu})\|$
is continuous, that is, Rief\/fel's condition is satisf\/ied for
this $\mu$.
\end{example}
This observation leads to the question, if there exist
well-matched families $\Pi=(\Pi_\hbar)_\hbar$, which are faithful
on the enveloping $C^*$-algebras $C^*(\cM(E),\star_{\!\hbar})$,
$\hbar\in\dR$ (here including $\hbar=0$), resp.~on parts of it? We
know already from Proposition~\ref{WQ:9-5:prop-Pi0tau} that the
representation of the classical obser\-vables $\Pi_0$ is faithful
on $\cW(E,0)=C^*(\cM(E)_d,\star_{\!0})$ and on each
$C^*(\cM(E_\alpha)_a,\star_{\!0})$, $\alpha\in I$. In
\cite{BinzHonRie03b} are used similar arguments for proving the
referred Theorem~\ref{WQ:theo:9-5:SDQ-Weylalgebra}.

Let $\alpha\in I$ be f\/ixed, and suppose $\Pi_{\hbar=1}$ faithful
on $C^*(\cM(E_\alpha)_a,\star_{\!1})$. Here we suppose the family
$\Pi=(\Pi_\hbar)_\hbar$ to be well-matched with respect to the
regularity condition or with respect to  $\tau$, but with
$T_\hbar(E_\alpha)=E_\alpha$ for all $\hbar\neq0$. Then
$\Pi=(\Pi_\hbar)_\hbar$ is faithful on the $C^*$-algebras
$(C^*(\cM(E_\alpha)_a,\star_{\!\hbar}))_{\hbar\in\dR}$ and also on
the $C^*$-Weyl algebras
$\cW(E_\alpha,\hbar\sigma)=C^*(\cM(E_\alpha)_d,\star_{\!\hbar})$,
$\hbar\in\dR$ (for $\hbar\neq0$ by
Lemma~\ref{WQB5:lemm-faithrep-Eal}). Consequently, for all
$\hbar\in\dR$ it is $\norm{\Pi_\hbar(\mu)}=\norm{\mu}_{\hbar}$
($\;=\norm{\wh{\mu}}_0$ in case of $\hbar=0$), and by
Example~\ref{WQB6:obse:=} Rief\/fel's condition is valid for every
measure $\mu\in\cM(E_\alpha)_a\bigcup\cM(E_\alpha)_d$.

If $\Pi_{\hbar=1}$ acts faithfully on each
$C^*(\cM(E_\alpha)_a,\star_{\!1})$, $\alpha\in I$, then
$\Pi=(\Pi_\hbar)_\hbar$ may be $\tau$-well-matched with possibly
$T_\hbar(E_\alpha)=E_\beta$ for $\alpha\neq\beta$, in order to
obtain Rief\/fel's condition for all
$\mu\in\bigcup\limits_{\alpha\in
I}\cM(E_\alpha)_a\bigcup\cM(E)_d$.

For f\/inite dimensional $E_\alpha$ (its dual $E_\alpha'$ may be
treated as part of $E_\tau'$) the Fourier transformed measures
$\wh{\cM}(E_\alpha')$ form a proper subspace of the bounded,
$\dC$-valued, uniformly continuous functions on $E_\alpha'$. After
all, one knows that $\wh{\cM}(E_\alpha')_a$ is a
$\norm{\cdot}_0$-dense sub-$*$-algebra of the $C^*$-algebra
$(\operatorname{C}_\infty(E_\alpha'),\cdot_{\!0})$ of the
continuous functions on $E_\alpha'$ vanishing at inf\/inity (cf.
e.g. \cite[(31.5)]{HewittRoss}).

Especially, for an arbitrary norm $\kappa$ on $E_\alpha$ every
$\wh{\mu}\in\wh{\cM}_\kappa^\infty(E_\alpha')$ is a smooth bounded
function on $E_\alpha'$, all of whose derivatives of all orders
are bounded
 (and so is similar to an element in Rief\/fel's function class $\cB$ \cite{Rieffel94a}).
However the Poisson algebra
$\wh{\cM}_\varsigma^\infty(E_\alpha')_a=\wh{\cM}_\varsigma^\infty(E_\alpha')
\bigcap\wh{\cM}(E_\alpha')_a$ covers the inf\/initely
dif\/ferentiable functions with compact support, resp.~the
functions of rapid decrease, which are rather dif\/ferent from the
almost periodic functions occurring in the usual $C^*$-Weyl
quantization. Nevertheless we have proved them to be strictly
deformation quantizable (including the full Rief\/fel condition),
also when $\sigma$ acts degenerately on $E_\alpha$. 
Note, if~$\sigma$ is  degenerate on $E_\alpha$,
the elements of $\wh{\cM}_\varsigma^\infty(E_\alpha')_a$
are not dif\/ferentiable
in all directions, since $\ker(\varsigma)\bigcap
E_\alpha$ is non-trivial.

But for
degenerate~$\sigma$ the elements of
$\wh{\cM}_\varsigma^\infty(E_\alpha')_a$ are not dif\/ferentiable
in all directions for non-trivial $\ker(\varsigma)\bigcap
E_\alpha$ (the latter being possible only for degenerate $\sigma$
on $E_\alpha$).

One apparently has to use mathematical techniques dif\/ferent from
ours for ef\/f\/iciently investigating  Rief\/fel's condition,
when one wants to quantize all phase space functions, which are
the Fourier transforms of  $\mu\in\cM(E)$, where even the results
concerned with measures $\mu$ on f\/inite dimensional test
functions spaces $E_\alpha$ have to be completed.

\appendix 

\section[$E_\rho'$-invariant folia and representations]{$\boldsymbol{E_\rho'}$-invariant folia and representations}
\label{WQ-appA}

Let $\rho$ be a locally convex topology on $E$ with dual space
$E_\rho'$. For each $F\in E_\rho'$ there exists a unique
$*$-automorphism $\gamma_\hbar^F$ on the Weyl algebra
$\cW(E,\hbar\sigma)$, called gauge transformation of the second
kind \cite{MSTV73,BinzHonRie99b}, such that
\begin{gather*}
\gamma_\hbar^F(W^\hbar(f))=\ex{iF(f)}W^\hbar(f) ,\qquad\forall\,
f\in E.
\end{gather*}
\begin{definition}[$\boldsymbol{E_\rho'}$-invariance]
\label{WQB:def:Erho-inv} A folium
$\cF_\hbar\in\operatorname{fol}(\cW(E,\hbar\sigma))$ respectively
the associated representation
$\Pi_\hbar\in\operatorname{rep}(\cW(E,\hbar\sigma))$ are called
partially $E_\rho'$-invariant, if there exists at least one state
$\omega\in\cF_\hbar$ such that $\omega\circ\gamma^F_\hbar\in
\cF_\hbar$ for all $F\in E_\rho'$.

$\cF_\hbar$ resp.~$\Pi_\hbar$ are called (globally)
$E_\rho'$-invariant, if $\cF_\hbar=\cF_\hbar\circ\gamma^F_\hbar$,
or equivalently, if $\Pi_\hbar=\Pi_\hbar\circ\gamma^F_\hbar$ for
all $F\in E_\rho'$.
\end{definition}
Note that for non-degenerate $\sigma$ and $\hbar\neq0$ every
$\cF_\hbar\in\operatorname{fol}(\cW(E,\hbar\sigma))$ is
$E_\rho'$-invariant for the locally convex topology $\rho$ arising
from the semi-norms $E\ni f\mapsto\betr{\sigma(g,f)}$, $g\in E$
(since for $\omega\in \cF_\hbar$ one has
$\dual{\omega}{W^\hbar(-g).\,W^\hbar(g)}\in\cF_\hbar$ by the
def\/inition of a folium), especially $\rho\in\cT(E,\sigma)$.

As mentioned already, the $C^*$-Weyl algebra $\cW(E,\hbar\sigma)$
is simple, if and only if $\sigma$ is non-degenerate and
$\hbar\neq0$, in which case every representation is faithful. But
also for degenerate $\sigma$ we have: 
\begin{lemma}
\label{WQ:9-5:lemm:faithful-tau-cont} Let $\rho\leq\tau$ for
locally convex $\rho$ and $\tau\in\cT(E,\sigma)$. Then
$\cF_\hbar^\tau$ is $E_\rho'$-invariant, and the associated
representation $\Pi_\hbar^\tau$ is faithful on
$\cW(E,\hbar\sigma)$.
\end{lemma}
\begin{proof}
The $E_\rho'$-invariance is immediate. For the proof of the case
$\hbar=0$ one may assume that $\sigma=0$. So suppose $\hbar\neq0$.
By construction the quotient symplectic form
$\sigma_q(q(f),q(g)):=\sigma(f,g)$ on the quotient $E/\ker_\sigma$
is non-degenerate ($\ker_\sigma$ is the null space of $\sigma$
from equation~\eqref{WQ:eq:9-5:nullspace-sigma}, and
$q(f):=f+\ker_\sigma$ the quotient map). Hence the $C^*$-Weyl
algebra $\cW(E/\ker_\sigma,\hbar\sigma_q)$ is simple, and so every
non-degenerate representation of it is faithful.

Let $\Pi'$ be a non-degenerate representation of
$\cW(E/\ker_\sigma,\hbar\sigma_q)$. Then
$\pi(f):=\Pi'(W^\hbar(q(f)))$, $f\in E$, def\/ines a projective
unitary representation of the additive group $E$, and hence by
Subsection~\ref{WQ-s2.2} there exists a unique representation
$\Pi$ of $\cW(E,\hbar\sigma)$ with $\Pi(W^\hbar(f))=\pi(f)$. The
separate $\tau$-continuity of $\sigma$ implies that $\ker_\sigma$
is closed. Thus on the quotient $E/\ker_\sigma$ there exists a
canonical locally convex topology $\tau_q$ arising from $\tau$,
for which the quotient map $q:E\rightarrow E/\ker_\sigma$ is
continuous and $\sigma_q$ is separately continuous (e.g.
\cite[Proposition~V.2.1]{Conway85}). Consequently, if $\Pi'$ is
$\tau_q$-continuous, then $\Pi$ is $\tau$-continuous. Especially,
by taking GNS representations every $\omega\in
\cF_\hbar^{\tau_q}\in\operatorname{fol}(\cW(E/\ker_\sigma,\hbar\sigma_q))$
extends to a unique state $\varphi\in\cF_\hbar^\tau$ with
$\dual{\varphi}{W^\hbar(f)}=\dual{\omega}{W^\hbar(q(f))}$ for all
$f\in E$.

Let us f\/ix a state $\omega\in\cF_\hbar^{\tau_q}$ with associated
$\varphi\in\cF_\hbar^\tau$. Then
$\varphi_F:=\varphi\circ\gamma^\hbar_F\in \cF_\hbar^\tau$ for all
$F\in E_\rho'$. By construction it follows that
$\norm{A}_\hbar\geq \bigl\|\Pi^\tau_\hbar(A)\bigr\|\geq \sup_{F\in
E_\rho'}\betr{\dual{\varphi_F}{A}}$ for all
$A\in\cW(E,\hbar\sigma)$. Now we restrict ourselves to arbitrary
elements $A=\textstyle\sum_k z_k W^\hbar(f_k)$ from the
commutative sub-$*$-algebra $\Delta(\ker_\sigma,0)$ (note that
$\sigma$ restricted to $\ker_\sigma$ vanishes), i.e., with
$f_k\in\ker_\sigma$. Since $q(f)=0$
$\dual{\varphi}{W^\hbar(f)}=\dual{\omega}{W^\hbar(0)}=1$ for all
$f\in\ker_\sigma$, yielding, for all $F\in E_\rho'$,
\begin{gather*}
\dual{\varphi_F}{ \sum_k\! z_k W^\hbar(f_k)} = \sum_k\!
z_k\ex{iF(f_k)}\dual{\varphi}{W^\hbar(f_k)}
\\
\phantom{\dual{\varphi_F}{\textstyle\sum_k\! z_k W^\hbar(f_k)}}{}
= \sum_k\! z_k\ex{iF(f_k)} = \bigl( \sum_k\! z_k W_c(f_k)\bigr)[F]
\end{gather*}
with the Weyl functions $W_c(f)$ from
equation~\eqref{WQ:eq:9-5:xi(f)}. By the above estimation we get
\begin{gather*}
\left\| \sum_k z_k W^\hbar(f_k)\right\|_\hbar
 \geq
\left\| \Pi^\tau_\hbar\left( \sum_k z_k W^\hbar(f_k)\right)
\right\|
\\
 \phantom{\left\| \sum_k z_k W^\hbar(f_k)\right\|_\hbar}{} \geq
\sup\limits_{F\in E_\rho'} \left|  \sum_k z_k\ex{iF(f_k)}  \right|
= \left\|   \sum_k z_k W_c(f_k)  \right\|_0.
\end{gather*}
But from \cite{BinzHonRie99b} it is known that
$\cW(\ker_\sigma,0)$ is a sub-$C^*$-algebra of
$\cW(E,\hbar\sigma)$, which means that the norm on
$\cW(\ker_\sigma,0)$ is just the restriction of the norm on
$\cW(E,\hbar\sigma)$. With the help of
Proposition~\ref{WQ:prop:9-5:hbar=0-*iso} we conclude that
$\|\sum_k z_k W^\hbar(f_k)\|_\hbar =\|\sum_k z_k W_c(f_k)\|_0$.
Consequently, $\norm{A}=\|\Pi^\tau_\hbar(A)\|$ for all
$A\in\Delta(\ker_\sigma,0)$, and thus,
$\ker(\Pi^\tau_\hbar)\bigcap\cW(\ker_\sigma,0)=\{0\}$. Finally,
\cite[(4.21.iii)]{MSTV73} implies that the closed $*$-ideal
$\ker(\Pi^\tau_\hbar)$ vanishs, that is, $\Pi^\tau_\hbar$ is
faithful.

Note, the proof ensures that also the direct sum representation
$\textstyle\bigoplus_F
\Pi_\varphi\circ\gamma_F^\hbar\leq\Pi^\tau_\hbar$ is faithful, for
every $\varphi$ constructed from an $\omega\in\cF_\hbar^{\tau_q}$.
Hence there exist many proper subrepresentations
of~$\Pi^\tau_\hbar$, which are faithful.
\end{proof}

\section{Proofs of the Propositions~\ref{WQ:9-5:prop:pre-SDQ-R}
                and~\ref{WQ6:prop:non}}
\label{WQ-appB}

Let us f\/irst treat Proposition~\ref{WQ:9-5:prop:pre-SDQ-R}. We
demonstrate part \ref{WQ:9-5:prop:pre-SDQ-R-c},
part~\ref{WQ:9-5:prop:pre-SDQ-R-b} then is obvious. For
$\hbar\neq0$ let
$\cF_\hbar\in\operatorname{fol}(\cW(E,\hbar\sigma))$ be the folium
associated with
$\Pi_\hbar\in\operatorname{rep}(\cW(E,\hbar\sigma))$. Then
$\beta_\hbar^*(\cF_1)=\cF_\hbar$, and $\beta_\hbar$ extends
$\sigma$-strongly to the associated W$*$-algebras.
$\omega_\hbar:=\omega\circ\beta_\hbar$ is $\tau$-continuous for
$\omega\in\cF_1\subseteq\cF_1^\tau$, and thus
\begin{gather}
\label{WQB:eq:theo:cont-hbar-om} \hbar\mapsto
\dual{\omega_\hbar}{W^\hbar(f)}=
\dual{\omega}{\beta_\hbar(W^\hbar(f))}= \dual{\omega}{W^1(T_\hbar
f)}
\end{gather}
is continuous on $\dR\!\setminus\!\{0\}$ with
$\lim\limits_{\hbar\rightarrow0}\dual{\omega_\hbar}{W^\hbar(f)}=0$
for each $f\in E$. For $\hbar_0\neq0$ and $\mu\in M^\tau(E)$ we
conclude that
\begin{gather*}
\dual{\omega_\hbar}
     {Q_\hbar^\Pi(\wh{\mu})^*Q_\hbar^\Pi(\wh{\mu})}
= \int_E d\mu^*(f) \int_E d\mu(g)
\ex{-\tfrac{i}{2}\hbar\sigma(f,g)}
\dual{\omega_\hbar}{W^\hbar(f+g)}
\\
\quad{} \stackrel{\hbar\rightarrow\hbar_0}{\longrightarrow} \int_E
d\mu^*(f) \int_E d\mu(g) \ex{-\tfrac{i}{2}\hbar_0\sigma(f,g)}
\dual{\omega_{\hbar_0}}{W^{\hbar_0}(f+g)} =
\dual{\omega_{\hbar_0}}
     {Q_{\hbar_0}^\Pi(\wh{\mu})^*Q_{\hbar_0}^\Pi(\wh{\mu})}.
\end{gather*}
Consequently $\dual{\omega_{\hbar_0}}
     {Q_{\hbar_0}^\Pi(\wh{\mu})^*Q_{\hbar_0}^\Pi(\wh{\mu})}
= \lim\limits_{\hbar\rightarrow \hbar_0} \dual{\omega_\hbar}
     {Q_\hbar^\Pi(\wh{\mu})^*Q_\hbar^\Pi(\wh{\mu})}
\leq \liminf\limits_{\hbar\rightarrow\hbar_0}
\|Q_\hbar^\Pi(\wh{\mu})\|^2$. Since $\cF_\hbar$ consists of the
normal states on $\ovl{\Pi_\hbar(\cW(E,\hbar\sigma))}^w$, we have
$\norm{A}^2=\sup\{\dual{\varphi}{A^*A}\mid \varphi \in
\cF_\hbar\}$ for all $A\in\ovl{\Pi_\hbar(\cW(E,\hbar\sigma))}^w$.
Hence, taking the supremum over all states $\omega\in\cF_1$, or
equivalently over the states $\omega_{\hbar_0}\in\cF_{\hbar_0}$,
we conclude the lower semicontinuity at $\hbar_0\neq0$, more
precisely
\begin{gather}
\label{WQB:9-5:eq:lower-semicont} \|Q_{\hbar_0}^\Pi(\wh{\mu})\|
\leq
\liminf\limits_{\hbar\rightarrow\hbar_0}\|Q_\hbar^\Pi(\wh{\mu})\|
\leq
\limsup\limits_{\hbar\rightarrow\hbar_0}\|Q_\hbar^\Pi(\wh{\mu})\|
\leq \norm{\mu}_1, \qquad\forall \, \mu\in M^\tau(E).
\end{gather}
Similarly we obtain for $\hbar_0=0$ that
\begin{gather*}
\dual{\omega_\hbar}{Q_\hbar^\Pi(\wh{\mu})^*Q_\hbar^\Pi(\wh{\mu})}
\;\stackrel{\hbar\rightarrow0}{\longrightarrow}\; \int_E d\mu^*(f)
\int_E d\mu(g) =\wh{\mu}^*[0]\wh{\mu}[0] =\betr{\wh{\mu}[0]}^2,
\end{gather*}
and hence $\betr{\wh{\mu}[0]} \leq
\liminf\limits_{\hbar\rightarrow0}\|Q_\hbar^\Pi(\wh{\mu})\| \leq
\limsup\limits_{\hbar\rightarrow0}\|Q_\hbar^\Pi(\wh{\mu})\| \leq
\norm{\mu}_1$ for all $\mu\in M^\tau(E)$. But
Lemma~\ref{WQ:9-5:obse-positive-measure} tells that
$\norm{\mu}_1=\wh{\mu}[0]$ for all $\mu\in M^\tau(E)^+$.

Suppose now the additional property of (partial)
$E_\rho'$-invariance. Then there exists a state $\varphi\in\cF_1$
so that
$\varphi_\hbar^F:=\beta_\hbar^*(\varphi)\circ\gamma_F^\hbar\in\cF_\hbar$
for all $F\in E_\rho'$ for each $\hbar\neq0$. We have
$\dual{\varphi_\hbar^F}{W^\hbar(f)}=\ex{iF(f)}\dual{\varphi}{W^1(T_\hbar
f)}$ for each $f\in E$. Similarly to the above reasoning we get
\begin{gather*}
\dual{\varphi_\hbar^F}
     {Q_\hbar^\Pi(\wh{\mu})^*Q_\hbar^\Pi(\wh{\mu})}
\\
\qquad {}= \int_E d\mu^*(f) \int_E d\mu(g)
\ex{i(F(f)+F(g)-\textstyle\frac{1}{2}\hbar\sigma(f,g))}
\dual{\varphi}{W^1(T_\hbar(f+g))}
\\
\qquad{} \stackrel{\hbar\rightarrow0}{\longrightarrow}\; \int_E
d\mu^*(f) \int_E d\mu(g) \ex{iF(f)}\ex{iF(g)}
=\wh{\mu}^*[F]\wh{\mu}[F] =\betr{\wh{\mu}[F]}^2
\end{gather*}
for all $F\in E_\rho'$. Taking the sup-norm for $\wh{\mu}$ yields
$\|\wh{\mu}\|_0\leq
\liminf\limits_{\hbar\rightarrow0}\|Q_\hbar^\Pi(\wh{\mu})\|$ for
all $\mu\in M^\tau(E)$.

The proof of Proposition~\ref{WQ6:prop:non} now is easy. For
non-degenerate $\sigma$ it follows from
Lemma~\ref{WQB5:lemm-faithrep-E}\ref{WQB5:lemm-faithrep-E-b} that
the well matching properties are not necessary, since the norms
for regular families $\Pi$ don't dif\/fer from those of the well
matched families $\Pi^\tau$, resp.~$\Pi^{\text{reg}}$. The result
follows with help of the previous proof.

\pdfbookmark[1]{References}{ref}
\LastPageEnding

\end{document}